\documentstyle[12pt,epsf,epsfig]{article}
\input epsf
\epsfverbosetrue
\def\beq{\begin{equation}}
\def\eeq{\end{equation}}
\def\bea{\begin{eqnarray}}
\def\eea{\end{eqnarray}}
\def\bq{\begin{quote}}
\def\eq{\end{quote}}

\def\gappeq{\mathrel{\rlap {\raise.5ex\hbox{$>$}}
{\lower.5ex\hbox{$\sim$}}}}

\def\lappeq{\mathrel{\rlap{\raise.5ex\hbox{$<$}}
{\lower.5ex\hbox{$\sim$}}}}

\begin{document}
\topmargin -0.5cm
\oddsidemargin -0.3cm
\pagestyle{empty}
\begin{flushright}
CERN-TH/96-371\\
hep-th/9701069\\
\end{flushright}
\vspace*{1cm}
\begin{center}
  {\bf\Large Introduction to $S$-Duality in $N=2$ Supersymmetric} \\
\vspace{.5cm}
{\bf\Large Gauge Theories} \\
\vspace*{1cm}
{\large (A Pedagogical Review of the Work of Seiberg and Witten)} \\
\vspace*{2cm}
{\bf Luis Alvarez-Gaum\'e}{\footnote{\tt e-mail:
    luis.ALVAREZ-GAUME@cern.ch}} and    
{\bf S. F. Hassan}{\footnote{\tt e-mail: fawad@mail.cern.ch}} \\
\vspace{.5cm}
Theoretical Physics Division, CERN \\
CH - 1211 Geneva 23 \\
\vspace*{2cm}
{\bf Abstract}
\end{center}
\begin{quote}
In these notes we attempt to give a pedagogical introduction to the
work of Seiberg and Witten on $S$-duality and the exact results of
$N=2$ supersymmetric gauge theories with and without matter. The first
half is devoted to a review of monopoles in gauge theories and the
construction of supersymmetric gauge theories. In the second half, we
describe the work of Seiberg and Witten. 
\end{quote}
\vfill
\begin{flushleft}
CERN-TH/96-371\\
December 1996\\
\end{flushleft}
\vfill\eject

\setcounter{page}{1}
\pagestyle{plain}

\noindent{\Large\bf Introduction}

These notes are the very late written version of a series of lectures
given at the Trieste Summer School in 1995.  The aim was to provide an
elementary introduction to the work of Seiberg and Witten on exact
results concerning $N=2$ supersymmetric extensions of Quantum
Chromodynamics.  We wanted to provide, in a single place, all the
background material necessary to study their work in detail.  We had
in mind graduate students who have already gone through their Quantum
Field Theory course, but we do not expect much more background to
follow these lectures. We have done our best to make the treatment
pedagogical.  In some sections, we have heavily drawn on previous
reviews, for instance in the treatment of supersymmetry we have
followed Bagger and Wess \cite{WB}, and in the theory of monopoles we
have used the reviews by Goddard and Olive \cite{GO} and by S. Coleman
\cite{coleman1,coleman2}.  These sources provide excellent
presentations of these topics, and we had no compelling reason to try
to make improvements on their presentation.  It is also quite obvious
that we have drawn heavily on the original papers of Seiberg and
Witten \cite{SW-I,SW2}, but we have tried to provide the necessary
tools to make their reading more accessible to interested students
and/or researchers. Recently there have been other lecture notes
published at a more advanced level, where one can find more details
and also the connection with String Theory (see, for instance, the
notes by W. Lerche \cite{Lerche}).  We would also like to mention that
we have not tried to be exhaustive in quoting all the literature on
the subject.  A more complete reference list can be found in
\cite{Lerche}.  The reference list is intented to provide a guidance
to enter the vastly growing literature on duality in String Theory and
Field Theory.  We apologize to those authors who may be offended by
not finding their works referenced.

These notes are divided into four sections, with each section further
subdivided into several subsections. Section 1 is devoted to an
introduction to monopoles in gauge theories. We start with a
discussion of the Dirac monopole and the idea of charge quantization,
and then describe the 't Hooft-Polyakov monopole in gauge theories
with spontaneous symmetry breaking. Then we introduce the notion of
Bogomol'nyi bound and the BPS states. After this, we describe the
topological classification of monopoles and then describe the
Montonen-Olive conjecture of electric-magnetic duality. We end this
section with a description of how, in the presence of a $\theta$-term
in the Lagrangian, the electric charge of a monopole is shifted by its
magnetic charge. Section 2 is devoted to an introduction to
supersymmetric gauge theories. First we describe the supersymmetry
algebra and its representations without and with central charges and
discuss its local realizations in terms of superfields. Then we
construct $N=1$ Lagrangians and finally, $N=2$ supersymmetric
Lagrangians with both gauge multiplets and matter multiplets
(hypermultiplets). At the end, we calculate the $N=2$ central charge
both in the pure gauge theory, as well as in the theory with matter
and establish its relation to the BPS bound. Having built the
foundations in the first two sections, in section 3 we describe the
Seiberg-Witten analysis of the $N=2$ pure gauge theory with gauge
group $SU(2)$. In the first two subsections, we discuss the
parametrization of the moduli space and the breaking of R-symmetries.
Then we describe how the chiral $U(1)$ anomaly of the theory can be
used to obtain the one-loop form of the low-energy effective action.
The rest of the section is devoted to finding the exact low-energy
effective action by using duality and the singularity structure on the
moduli space of the theory. We express the exact solution in terms of
complete elliptic integrals. In section 4, we briefly describe the
Seiberg-Witten analysis of $N=2$ $SU(2)$ gauge theory with $N_f$
matter fields. After a discussion of the general features of these
theories, we describe how the duality group is no longer pure
$SL(2,Z)$. Then we describe the singularity structure on the moduli
space of these theories and sketch the procedure for obtaining the
exact solutions. Our aim in section 4 is to give a flavour of the
analysis of theories with matter and, for a deeper understanding, the
interested reader is referred to the original work of Seiberg and
Witten.

\section{Magnetic Monopoles in Gauge Theories}

In this section, we begin by reviewing the proporties of the Dirac
monopole and the idea of charge quantization. Then we describe the
magnetic monopoles and dyons which arise in non-Abelian gauge theories
with spontaneous symmetry breaking and discuss their general
properties. We also introduce the notion of the Bogomol'nyi bound and
BPS states. In the last two subsections, we describe the
Montonen-Olive conjecture of electric-magnetic duality and Witten's 
argument about how the presence of the $\theta$-term in the Lagrangian
modifies the monopole and dyon electric charges. For a more detailed
discussion of most of the material in this section, the reader is
referred to the review article by Goddard and Olive \cite{GO}, and to
the lecture notes by Coleman \cite{coleman1,coleman2}.

\subsection{Conventions and Preliminaries}

Let us start by stating our conventions: we always take $c=1$, and
almost always $\hbar=1$, except when it is important to make a
distinction between classical and quantum effects. For index
manipulations, we use the flat Minkowsky metric $\eta$ of signature
$\{ +,-,-,-\}$.  Moreover, we choose units in which Maxwell's
equations take the form:
\beq
\begin{array}{l}
\vec{\nabla}\cdot\vec{E}=\rho\,,\qquad\vec{\nabla}\times\vec{B}-
\partial \vec{E}/\partial t =\vec{j}\,,\\
\vec{\nabla}\cdot\vec{B}=0\,,\qquad \vec{\nabla}\times\vec{E} +
\partial\vec{B}/\partial t = 0\,.
\label{I}
\end{array}
\eeq
In these units, a factor of $(4 \pi)^{-1}$ appears in Coulomb's law:
For a static point-like charge $q$ at the origin, we have $\rho = q 
\delta^{3}(\vec{r})$. Integrating the first Maxwell equation over a
sphere of radius $r$ and using spherical symmetry, we get
$\int_{S^2} \vec{E} \cdot d \vec{s}=4\pi r^2 E(r)=q$. Hence,
$\vec{E}=q\vec{r}/4\pi r^{3}$, as stated. The electrostatic
potential $\phi$ defined by $\vec{E}=-\vec{\nabla}\phi$ is given by 
$\phi = q / 4 \pi r$. 

In relativistic notation, one introduces the four-potential 
$A^{\mu}= \{\phi, \vec{A}\}$. The electric and magnetic fields are
defined as components of the corresponding field strength tensor
$F_{\mu\nu}$ as follows:
\bea
F_{\mu \nu} & = &\partial_{\mu}A_{\nu} - \partial_{\nu}A_{\mu}\,,
\nonumber \\
F_{0i} & = &\partial_{0}A_{i} - \partial_{i}A_{0} =
- \partial_{0}A^{i} - \partial_{i}A^{0} = E^{i}\,,\nonumber \\
F_{ij} & = & \partial_{i}A_{j} - \partial_{j}A_{i} =
- (\partial_{i}A^{j} - \partial^{j}A_{i}) =
-\epsilon_{ijk} B^{k}\,,
\label{III}
\eea
so that $B^i=-\frac{1}{2}\epsilon^{ijk}F_{jk}=-\frac{1}{2}
\epsilon^{oi\mu\nu}F_{\mu\nu}$. The dual field strength tensor is
given by   
$$
\widetilde F^{\mu \nu} = \frac{1}{2} \varepsilon^{\mu \nu \alpha
\beta} F_{\alpha \beta}\,,
$$
with $\varepsilon^{0123}=+1$. In component notation, we have 
\beq
F^{\mu \nu} =
\left(
\begin{array}{cccc}
0 & -E_{x} & -E_{y} & -E_{z}  \\
E_{x} & 0 & -B_{z} & B_{y} \\
E_{y} & B_{z} & 0 & -B_{x} \\
E_{z} & -B_{x} & B_{y} & 0
\end{array}
\right)\,,
\qquad
\widetilde F^{\mu \nu} =
\left(
\begin{array}{cccc}
0 & -B_{x} & -B_{y} & -B_{z} \\
B_{x} & 0 & E_{z} & -E_{y} \\
B_{y} & -E_{z} & 0 & E_{x} \\
B_{z} & E_{y} & -E_{x} & 0
\end{array}
\right)\,.
\label{IV}
\eeq

In terms of the electric four-current $j^{\mu}=\{\rho,\vec{j}\}$, the
Maxwell's equations take the compact form
\beq
\partial_\nu F^{\mu\nu}=-j^\mu\,,\qquad\partial_\nu\,\widetilde
F^{\mu\nu}=0\,. \label{V}
\eeq
Note that when $j^\mu =0$, the above equations are invariant under the
replacement $\vec{E} \rightarrow \vec{B}, \vec{B} \rightarrow
-\vec{E}$. This is referred to as the electric-magnetic duality
transformation. In the presence of electric sources, this
transformation is no longer a symmetry of Maxwell's equations. In
order to restore the duality invariance of these equations for
non-zero $j^{\mu}$, Dirac \cite{Dirac} introduced the magnetic
four-current $k^{\mu}=\{\sigma ,\vec{k}\}$ and modified Maxwell's
equations to
\beq
\partial_\nu F^{\mu\nu}=-j^\mu\,,\qquad\partial_\nu\,
\widetilde F^{\mu\nu}= -k^\mu \,.   
\label{VI}
\eeq
The above equations are now invariant under a combined duality
transformation of the fields and the currents which can be written as 
\beq
F\rightarrow \widetilde F\,,\,\widetilde F\rightarrow-F\,;\qquad 
j^\mu\rightarrow k^\mu\,,\, k^\mu\rightarrow -j^\mu\,.
\label{emduality}
\eeq
Note that the full invariance group of the equations (\ref{VI}) is
larger than this discrete duality. In fact, they are invariant under a
continuous $SO(2)$ group which rotates the electric and magnetic
quantities into each other. 

For point-like electric and magnetic sources, the current densities 
can be written as 
\bea
j^{\mu}(x) & = & \sum_{a} q_{a} \, \int \, dx_{a}^{\mu}
\delta^{4} (x - x_{a})\,, \nonumber \\
k^{\mu}(x) & = & \sum_{a} g_{a} \, \int \, dx_{a}^{\mu}
\delta^{4} (x - x_{a})\,.\nonumber
\eea
A particle of electric charge $q$ and magnetic charge $g$ experiences
a Lorentz force given by
$$
m\,\frac{d^2x^\mu}{d\tau^2}=(q F^{\mu\nu}+g\, \widetilde F^{\mu\nu})
\frac{dx_\nu}{d\tau}\,.
$$

Although, at the level of quations of motion, Dirac's modification of
Maxwell's theory seems trivial, it has highly non-trivial consequences
in quantum theory. One way of realizing the problem is to note that
the vector potential $\vec{A}$ is indispensable in the quantum
formulation of the theory. On the other hand, equations
$\vec{B}=\vec{\nabla}\times \vec{A}$ and $\vec{\nabla}\cdot\vec{B}\ne
0$ are not compatible unless the vector potential $\vec{A}$ has
singularities and, hence, is not globally well-defined. It turns out
that these singularities are gauge dependent and, therefore, their
presence should not be experimentally  detectable. In the classical
theory, which can be formulated in terms of $\vec B$ alone, this
requirement is trivially satisfied. However, in quantum theory, it
leads to the important phenomenon of the quantization of electric
charge. In the following, we will first give a semiclassical
derivation of this effect and then describe a derivation based on the
notion of the Dirac string. 
 
\subsection{A Semiclassical Derivation of Charge Quantization}

Consider a non-relativistic charge $q$ in the vicinity of a magnetic
monopole of strength $g$, situated at the origin. The charge $q$
experiences a force $m\ddot{\vec{r}}=q\dot{\vec{r}}\times\vec{B}$,
where $\vec{B}$ is the monopole field given by $\vec{B}= 
g\vec{r}/4\pi r^3$. The change in the orbital angular momentum of the
electric charge under the effect of this force is given by 
$$
\frac{d}{dt}\left(\,m\vec{r}\times\dot{\vec{r}}\right) =
m\vec{r}\times\ddot{\vec{r}}= \frac{qg}{4 \pi r^{3}}\,
\vec{r}\times\left(\dot{\vec{r}}\times\vec{r}\right)= 
\frac{d}{dt}\left(\frac{qg}{4\pi}\,\frac{\vec{r}}{r}\right)\,.
$$
Hence, the total conserved angular momentum of the system is
\beq
\vec{J}=\vec{r}\times m\dot{\vec{r}}-\frac{qg}{4\pi}\,
\frac{\vec{r}}{r}\,.
\label{XI}
\eeq
The second term on the right hand side (henceforth denoted by
$\vec{J}_{em}$) is the contribution coming from the elecromagnetic
field.  This term can be directly computed by using the fact that the
momentum density of an electromagnetic field is given by its Poynting
vector, $\vec{E}\times\vec{B}$, and hence its contribution to the
angular momentum is given by
$$
\vec{J}_{em} = \int \, d^{3}x \, \vec{r} \times
(\vec{E} \times \vec{B}) = \frac{g}{4 \pi} \,
\int \, d^{3} x \, \vec{r} \times
\left( \vec{E} \times \frac{\vec{r}}{r^{3}} \right)\,.
$$
In components,
\bea
J^i_{em} & = & \frac{g}{4 \pi} \, \int \, d^{3}x E^{j}
\left( \delta_{ij} - \frac{x_i x_j}{r^{2}} \right )
\frac{1}{r} = \frac{g}{4 \pi} \, \int \, d^{3}x E^{j}
\partial_{j} (\hat{x}^{i}) \nonumber \\
& = & \frac{g}{4 \pi} \int_{S^2} \hat{x}^i \vec{E}\cdot \vec{ds} 
- \frac{g}{4 \pi} \, \int \, d^{3}x \vec{\nabla} \cdot
\vec{E} \hat{x}^{i}\,. 
\label{XIV}
\eea
When the separation between the electric and magnetic charges is
negligible compared to their distance from the boundary $S^2$, the
contribution of the first integral to $\vec{J}_{em}$ vanishes by
spherical symmetry. We are therefore left with   
\beq
\vec{J}_{em}= -\frac{gq}{4 \pi} \hat{r}
\label{Jem}\,.
\eeq

Returning to equation (\ref{XI}), if we assume that orbital angular
momentum is quantized. Then it follows that 
\beq
\frac{qg}{4 \pi} = \frac{1}{2} n \hbar\,,
\label{XII}
\eeq
where $n$ is an integer. Note that in the above we have assumed the
total angular momentum of the charge-monopole system to be quantized
in half-integral units. This is a strange assumption considering that
we did not have to treat the electrically charged particle or the
monopole as fermios.  both of the components are bosonic. However, it
turns out that this actually is the case and that the situation does
not contradict the spin-statistics theorem
\cite{Goldhaber,Wilczek}. We will not discuss this issue further but
remark that the derivation of the same equation presented in the next
subsection does not depend on this assumption.

Equation (\ref{XII}) is the Dirac charge quantization
condition. It implies that if there exists a magnetic monopole of
charge $g$ somewhere in the universe, then all electric charges are
quantized in units of $2\pi\hbar/g$. If we have a number of purely
electric charges $q_i$ and purely magnetic charges $g_j$, then any
pair of them will satisfy a quantization condition:
\beq
\frac{q_{i}g_{j}}{4 \pi \hbar} = \frac{1}{2} n_{ij}
\label{XIII}
\eeq
Thus, any electric charge is an integral multiple of $2\pi\hbar/g_j$.
For a given $g_j$, let these charges have $n_{0j}$ as the highest
common factor. Then, all the electric charges are multiples
of $q_{0}=n_{0j}2\pi\hbar /g_{j}$. Note that $q_0$ itself may not
exist in the spectrum. Similar considerations apply to the
quantization of magnetic charge. 

Till now, we have only dealt with particles that carry either an
electric or a magnetic charge. Let us now consider dyons, {\it i.e.,}
particles that carry both electric and magnetic charges. Consider two
dyons of charges $(q_1,g_1)$ and $(q_2,g_2)$. For this system, we can
repeat the calculation of $\vec{J}_{em}$ by following the steps in
(\ref{XIV}) where now the electromagnetic fields are split as
$\vec{E}=\vec{E}_1+\vec{E}_2$ and $\vec{B}=\vec{B}_1+ \vec{B}_2$.  The
answer is easily found to be

\beq
\vec{J}_{em}=-\frac{1}{4\pi}\left(q_1 g_2 - q_2 g_1\right)\hat{r} 
\label{Jemd}
\eeq
The charge quantization condition is thus generalized to 
\beq
\frac{q_1 g_2 - q_2 g_1}{4\pi\hbar}=\frac{1}{2} n_{12}
\label{DSZ}
\eeq
This is referred to as the Dirac-Schwinger-Zwanziger condition
\cite{Schwinger,Zwanziger}. This condition is invariant under the
$SO(2)$ transformation $(q+ig)\rightarrow e^{i\phi}(q+ig)$ which is
also a symmetry of (\ref{VI}). 

\subsection{The Dirac String}

In the following, we present a more rigorous derivation of the Dirac
quantization condition which is based on the notion of a Dirac
string. Let $\vec{B}_{mon}$ denote the magnetic field around a
monopole. Since $\vec{\nabla}\cdot\vec{B}_{mon}\ne 0$, it is not
possible to construct a well-defined $\vec{A}_{mon}$ such that
$\vec{B}_{mon}=\vec{\nabla} \vec{A}_{mon}$. To overcome this problem,
Dirac intruduced a  semi-infinite solenoid (or string) running from
$(0,0,-\infty)$ to the monopole position, $(0,0,0)$. This solenoid
carries a magnetic field $\vec{B}_{sol}=g\theta (-z)\delta(x)
\delta(y)\hat{z}$ which also is not divergence free. However, the
total magnetic field,   
\beq
\vec{B}=\vec{B}_{mon}+\vec{B}_{sol}=\frac{g}{4\pi r^{2}}+g\theta(-z)
\delta(x)\delta(y)\hat{z}\,,
\eeq
satisfies $\vec{\nabla}\cdot\vec{B}=g\delta(\vec{r})-g\delta(\vec{r}) 
=0$. It is therefore possible to construct a non-singular $\vec{A}=
\vec{A}_{mon}+\vec{A}_{sol}$ corresponding to the monopole-solenoid
system. In fact, the singular $\vec{A}_{sol}$ associated with the
Dirac string is used to cancel the singularity in $\vec{A}_{mon}$. The
position of the singularity in $\vec{A}_{mon}$, and therefore the
position of the Dirac string, can be shifted by singular gauge
transformations. Since the Dirac string is an artificial construct, it
should be unobservable and should not contribute to any physical
process. However, in an Aharonov-Bohm experiment, the presence of the
string can affect the wavefunction of an electric charge by
contributing to its phase. This resulting Aharonov-Bohm phase along a
contour $\Gamma$ encircling the string and enclosing an area $S$ can
be easily computed to be
$$
e \oint_{\Gamma} \vec{A}_{sol} \cdot d \vec{l} = e \int_S
\vec{B}_{sol} \cdot d\vec{s} = e g\,.
$$
Here, $e=q/\hbar$ is the electromagnetic coupling constant. This
phase, and therefore the Dirac string, is unobservable provided 
$$
e g = 2\pi n\,,
$$
which is again the Dirac charge quantization condition.

\epsfxsize=8cm
\epsfysize=6cm
\centerline{\epsffile{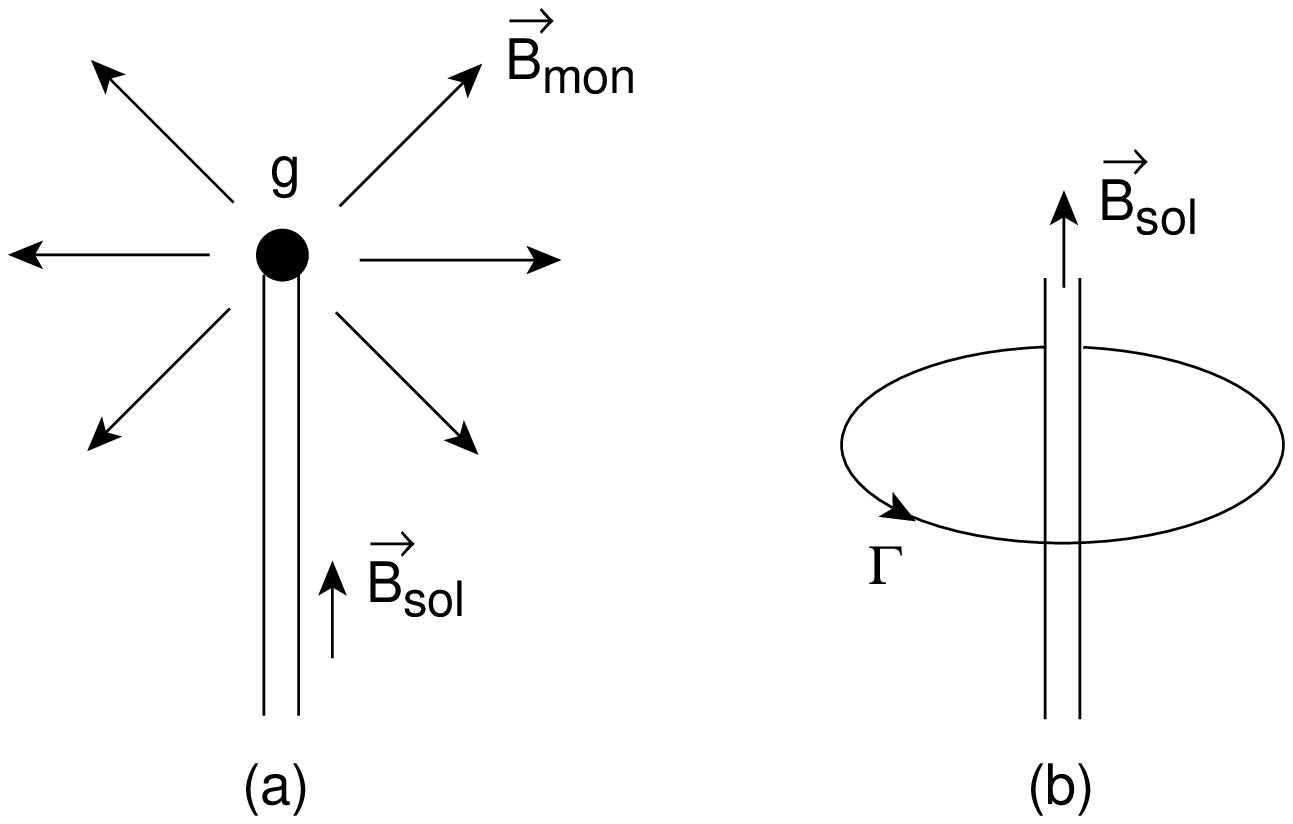}}
\begin{center}
Figure 1\end{center}
\subsection{The Georgi-Glashow Model: A Simple Theory with Monopoles}

Till now we have been working in the framework of particle mechanics
where both electric and magnetic charges are point-like objects and
are introduced by hand into the theory. However, in field theory,
these objects can also arise as solitons which are non-trivial
solutions of the field equations with localized energy density. If the
gauge field configuration associated with a soliton solution has a
magnetic character, the soliton can be identified as a magnetic
monopole. The simplest solitons are found in the Sine-Gordon theory,
which is a scalar field theory in $1+1$ dimensions. However, by
Derrick's theorem, a scalar field theory in more than two dimensions
cannot support static finite energy solutions. This is basically due
to the fact that because of the non-trivial structure of the soliton
at large distances, the total energy of the configuration
diverges. This situation can be cured by the addition of gauge fields
to the theory. Thus, a scalar theory with gauge interactions in four 
dimensions can admit static finite energy field configurations. The
stability of such solitonic configurations are often related to the
fact that they are characterized by conserved topological charges. 
For a more detailed discussion of these issues, the reader is referred
to \cite{GO, coleman1, coleman2}. In the following, we describe the
Georgi-Glashow model which is a simple theory in $3+1$ dimensions with
monopole solutions.  

The Georgi-Glashow model is a Yang-Mills-Higgs system which contains 
a Higgs multiplet $\phi^a\,(a=1,2,3)$ transforming as a vector in the
adjoint representation of the gauge group $SO(3)$, and the gauge
fields $W_\mu=W^a_\mu T^a$. Here, $T^a$ are the hermitian generators
of $SO(3)$ satifying $[T^a,T^b]=if^{abc}T^c$. In the adjoint
representaion, we have $(T^a)_{bc}=-if^a_{bc}$ and, for $SO(3)$,
$f^{abc}=\epsilon^{abc}$. The field strength of $W_\mu$ and the
cavariant derivative on $\phi^a$ are defined by 
\bea
G_{\mu \nu} &=& \partial_{\mu}W_{\nu} - \partial_{\nu} W_{\mu} +
ie [W_{\mu},W_{\nu}]\,,
\nonumber\\
D_\mu\phi^a &=& \partial_\mu\phi^a - e\epsilon^{abc} W^b_\mu \phi^c\,.
\eea
The minimal Lagrangian is then given by
\beq
{\cal L}  =  - \frac{1}{4} G^{a}_{\mu \nu} G^{a \mu \nu} +
\frac{1}{2} D^{\mu}\phi^a D_{\mu}\phi^a - V(\phi)\,,
\label{GGmodel}
\eeq
where,
\beq
V(\phi)=\frac{\lambda}{4}\left(\phi^2 - a^2 \right)^2\,.
\label{GGpot}
\eeq
The equations of motion following from this Lagrangian are 
\beq
(D_\nu G^{\mu \nu})^a=-e\,\epsilon^{abc}\,\phi^b\,(D^\mu\phi)^c,
\qquad D^\mu D_\mu \phi^a =-\lambda\phi^a (\phi^{2} - a^{2})\,. 
\label{XXIX}
\eeq
The field strength also satifies the Bianchi identity 
\beq
D_\nu\,\widetilde G^{\mu \nu a} = 0\,.
\label{bianchi}
\eeq

Let us find the vacuum configurations in this theory. Using the
notations $G_a^{0i}=-{\cal E}_a^i$ and $G_a^{ij} = -\epsilon^{ij}_
{\hphantom{ij}k}{\cal B}_a^k$, the energy density is written as
\beq
\theta_{00}=\frac{1}{2}\left(({\cal E}_a^i)^2+({\cal B}_a^i)^2 +
  (D^0\phi_a)^2 + ( D^i \phi_a)^2 \right) + V (\phi)\,.
\label{XXX}
\eeq 
Note that $\theta_{00} \geq 0$, and it vanishes only if 
\beq
G_{a}^{\mu \nu} = 0, \quad D_{\mu} \phi = 0, \quad V(\phi) = 0\,.
\label{XXXI}
\eeq 
The first equation implies that in the vacuum, $W^a_\mu$ is pure
gauge and the last two equations define the Higgs vacuum. The
structure of the space of vacua is determined by $V(\phi)=0$ which
solves to $\phi^a= \phi^a_{vac}$ such that $|\phi_{vac}|=a$. The space
of Higgs vacua is therefore a two-sphere ($S^2$) of radius $a$ in the
field space. To formulate a perturbation theory, we have to choose one
of these vacua and hence, break the gauge symmetry spontaneously (this
is the usual Higgs mechanism). The part of the symmetry which keeps
this vacuum invariant, still survives and the corresponding unbroken 
generator is $\phi^c_{vac} T^c/a$. The gauge boson associated with
this generator is $A_\mu=\phi^c_{vac}W^c_\mu/a$ and the electric
charge operator for this surviving $U(1)$ is given by  
\beq 
Q = \hbar e \frac{\phi^c_{vac}T^c}{a}\,.
\label{XXXIII}
\eeq
If the group is compact, this charge is quantized. The perturbative
spectrum of the theory can be found by expanding $\phi^a$ around the
chosen vacuum as 
$$
\phi^{a} = \phi^{a}_{vac} + \phi '^{a}\,.
$$
A convenient choice is $\phi^c_{vac}=\delta^{c3}a$. The perturbative 
spectrum (which becomes manifest after choosing an appropriate gauge)
consists of a massive Higgs ($H$), a massless photon ($\gamma$) and
two charged massive bosons ($W^\pm$):  
$$
\begin{tabular}{|c|c|c|c|}
\hline
& Mass & Spin & Charge \\
\hline
$H$ & $a(2\lambda)^{\frac{1}{2}} \hbar$ & $0$ &$0$ \\
\hline
$\gamma$ & $0$ & $\hbar$ & $0$ \\
\hline
$W^{\pm}$ & $ae\hbar = aq$ & $\hbar$ & $\pm q = \pm e \hbar$ \\
\hline
\end{tabular}
$$
In the next section, we investigate the existence of monopoles
(non-perturbative states) in the Georgi-Glashow model.

\subsection{The 't Hooft - Polyakov Monopole}

Let us look for time-independent, finite energy solutions in the
Georgi-Glashow model. Finiteness of energy requires that as
$r\rightarrow \infty$, the energy density $\theta_{00}$ given by
(\ref{XXX}) must approach zero faster than $1/r^3$. This means that
as $r\rightarrow \infty$, our solution must go over to a Higgs vacuum
defined by (\ref{XXXI}).  In the following, we will first assume that
such a finite energy solution exists and show that it can have a
monopole charge related to its soliton number which is, in turn,
determined by the associated Higgs vacuum. This result is proven
without having to deal with any particular solution explicitly. Next,
we will describe the 't Hooft-Polyakov ansatz for explicitly
constructing one such monopole solution. We will also comment on the
existence of Dyonic solutions. For convenience, in this section we
will use the vector notation for the $SO(3)$ gauge group indices and
not for the spatial indices. 

\noindent\underline{The Topological Nature of Magnetic Charge:}
\noindent Let $\vec{\phi}_{vac}$ denote the field $\vec{\phi}$ in a
Higgs vacuum. It then satisfies the equations
\bea
&\vec{\phi}_{vac}\cdot\vec{\phi}_{vac} = a^2\,, &\nonumber \\
&\partial_{\mu} \vec{\phi}_{vac}-e\,\vec{W}_{\mu} \times
\vec{\phi}_{vac} = 0\,,& 
\label{vec-vac}
\eea
which can be solved for $\vec{W}_\mu$. The most general solution is
given by 
\beq
\vec{W}_{\mu} = \frac{1}{ea^{2}}\, \vec{\phi}_{vac} \times
\partial_\mu\vec{\phi}_{vac}+\frac{1}{a}\vec{\phi}_{vac}A_{\mu}\,.
\label{sol-vac}
\eeq
To see that this actually solves (\ref{vec-vac}), note that 
$\partial_\mu\vec{\phi}_{vac}\cdot\vec{\phi}_{vac}=0$, so that
$$
\frac{1}{ea^2}(\vec{\phi}_{vac}\times\partial_{\mu}\vec{\phi}_{vac})
\times \vec{\phi}_{vac}= \frac{1}{ea^2}\left(\partial_\mu\vec{\phi}_{vac}a^2
-\vec{\phi}_{vac} (\vec{\phi}_{vac}\cdot\partial_\mu\phi_{vac})\right)= 
\frac{1}{e}\partial_\mu\vec{\phi}_{vac}\,.
$$
The first term on the right-hand side of Eq. (\ref{sol-vac}) is the
particular solution, and $\vec{\phi}_{vac} A_\mu$ is the general
solution to the homogeneous equation. Using this solution, we can now
compute the field strength tensor $\vec{G}_{\mu\nu}$. The field
strength $F_{\mu\nu}$ corresponding to the unbroken part of the gauge
group can be identified as  
\bea
F_{\mu\nu}&=&\frac{1}{a}\vec{\phi}_{vac}\cdot\vec{G}_{\mu \nu}
\nonumber\\
&=&\partial_\mu A_\nu -\partial_\nu A_\mu +
\frac{1}{a^3 e}\vec{\phi}_{vac}\cdot (\partial_\mu\vec{\phi}_{vac}
\times\partial_\nu\vec{\phi}_{vac})\,.
\label{F-vac}
\eea
Using the equations of motion in the Higgs vacuum it follows that  
$$
\partial_\mu F^{\mu\nu}=0\,,\qquad \partial_\mu\,\widetilde F^{\mu\nu}=0\,.
$$
This confirms that $F_{\mu\nu}$ is a valid $U(1)$ field strength
tensor. The magnetic field is given by
$B^i=-\frac{1}{2}\epsilon^{ijk}F_{jk}$. Let us now consider a static,
finite energy solution and a surface $\Sigma$ enclosing the core of
the solution. We take $\Sigma$ to be far enough so that, on it, the
solution is already in the Higgs vacuum. We can now use the magnetic
field in the Higgs vacuum to calculate the magnetic charge $g_\Sigma$
associated with our solution: 
\beq
g_\Sigma=\int_\Sigma\,B^i ds^i=-\frac{1}{2 ea^3} \int_\Sigma\, 
\epsilon_{ijk}\,\vec{\phi}_{vac}\cdot \left(\partial^j\vec{\phi}_{vac}
\times\partial^k\vec{\phi}_{vac}\right)ds^i\,.
\label{XL}
\eeq
It turns out that the expression on the right hand side is a
topological quantity as we explain below: Since $\phi^2=a$; the
manifold of Higgs vacua (${\cal M}_0$) has the topology of $S^2$. 
The field $\vec{\phi}_{vac}$ defines a map from $\Sigma$ into 
${\cal M}_0$. Since $\Sigma$ is also an $S^2$, the map 
$\phi_{vac}:\Sigma\rightarrow{\cal M}_0$ is characterized by its
homotopy group $\pi_2 (S^2)$. In other words, $\phi_{vac}$ is
characterized by an integer $\nu$ (the winding number) which counts the
number of times it wraps $\Sigma$ around ${\cal M}_0$. In terms of
the map $\phi_{vac}$, this integer is given by
\beq
\nu = \frac{1}{4\pi a^{3}} \, \int_\Sigma\,\frac{1}{2}
\epsilon_{ijk}\vec{\phi}_{vac}\cdot\left(\partial^j\vec{\phi}_{vac}
\times \partial^k\vec{\phi}_{vac}\right)\,d s^i\,.
\label{XLI}
\eeq
Comparing this with the expression for magnetic charge, we get the
important result
\beq
g_{\Sigma} = - \frac{4\pi \nu}{e}\,.
\label{XLII}
\eeq
Hence, the winding number of the soliton determines its monopole
charge. Note that the above equation differs from the Dirac
quantization condition by a factor of $2$. This is because the
smallest electric charge which could exist in our model is 
$q_0=e\hbar/2$ in terms of which, (\ref{XLII}) reduces to the Dirac
condition. 

\noindent\underline{An Ansatz for Monopoles:}
\noindent Now we describe an ansatz proposed by 't Hooft \cite{thooft}
and Polyakov \cite{polya} for constructing a monopole solution in the 
Georgi-Glashow model. For a spherically symmetric, parity-invariant,
static solution of finite energy, they proposed:
\bea
\phi^a & = & \frac{x^a}{er^2}\,H(aer)\,,\nonumber\\
W^a_i&=&-\epsilon^a_{ij}\frac{x^j}{er^2}\left(1 - K(aer)\right)\,,
\qquad W^a_0= 0\,.
\label{tp-ansatz}
\eea
For the non-trivial Higgs vacuum at $r\rightarrow\infty$, they chose
$\phi^c_{vac}=ax^c/r= a\hat{x}^c$. Note that this maps an $S^2$ at
spatial infinity onto the vacuum manifold with a unit winding number.
The asymptotic behaviour of the functions $H(aer)$ and $K(aer)$ are
determined by the Higgs vacuum as $r\rightarrow\infty$ and regularity
at $r=0$. Explicitly, defining $\xi=aer$, we have: as
$\xi\rightarrow\infty,\,H\sim\xi,\, K\rightarrow 0$ and as
$\xi\rightarrow 0,\, H\sim\xi,\,(K-1)\sim\xi$. The mass of this
solution can be parametrized as
$$
M = \frac{4 \pi a}{e} f \, (\lambda /e^{2})
$$
For this ansatz, the equations of motion reduce to two coupled
equations for $K$ and $H$ which have been solved exactly only in
certain limits.  
For $r\rightarrow 0$, one gets $H\rightarrow ec_1 r^2$ and
$K=1+ec_2r^2$ which shows that the fields are non-singular at
$r=0$. For $r\rightarrow\infty$, we get $H\rightarrow\xi+c_3
exp(-a\sqrt{2\lambda}r)$ and $K\rightarrow c_4\xi exp(-\xi)$ which
leads to $W^a_i\approx-\epsilon^a_{ij}x^j/er^2$. Once again, defining
$F_{ij}=\phi^c G^c_{ij}/a$, the magnetic field turns out to be  
$B^{i} = - x^i/e r^3$. The associated monopole charge is $g=-4\pi/e$,
as expected from the unit winding number of the solution. It should be
mentioned that 't Hooft's definition of the Abelian field strength
tensor is slightly different but, at large distances, it reduces to
the form given above. 

Note that in the above monopole solution, the presence of the Dirac
string is not obvious. To extract the Dirac string, we have to perform
a singular gauge transformation on this solution which rotates the
non-trivial Higgs vacuum $\phi^c_{vac}=a\hat{x}^c$ into the trivial
vacuum $\phi^c_{vac}= a \delta^{c3}$. In the process,the gauge field
develops a Dirac string singularity which now serves as the source of
the magnetic charge \cite{thooft}.  

\noindent\underline{The Julia-Zee Dyons:}

The 't Hooft-Polyakov monopole carries one unit of magnetic charge and
no electric charge. The Georgi-Glashow model also admits solutions
which carry both magnetic as well as electric charges. An ansatz for
constructing such a solution was proposed by Julia and Zee
\cite{Julia-Zee}. In this ansatz, $\phi^a$ and $W^a_i$ have exactly
the same form as in the 't Hooft-Polyakov ansatz, but $W^a_0$ is no
longer zero: $W^a_0=x^a J(aer)/er^2$. This serves as the source for
the electric charge of the dyon.  It turns out that the dyon electric
charge depends of a continuous parameter and, at the classical level,
does not satisfy the quantization condition. However, semiclassical
arguments \cite{TW1,GSW} show that, in CP invariant theories, and at
the quantum level, the dyon electric charge is quantized as $q=n\hbar
e$. This can be easiy understood if we recognize that a monopole is
not invariant under a guage transformation which is, of course, a
symmetry of the equations of motion. To treate the associated
zero-mode properly, the gauge degree of freedom should be regarded as
a collective coordinate. Upon quantization, this collective coordinate
leads to the existence of electrically charged states for the monopole
with discrete charges.  In the presence of a CP violating term in the
Lagrangian, the situation is more subtle as we will discuss later. In
the next subsection, we describe a limit in which the equations of
motion can be solved exactly for the 'tHooft-Polyakov and the
Julia-Zee ansatz. This is the limit in which the soliton mass
saturates the Bogomol'nyi bound.

\subsection{The Bogomol'nyi Bound and the BPS States}

In this subsection, we derive the Bogomol'nyi bound \cite{bogomol} on
the mass of a dyon in term of its electric and magnetic charges which
are the sources for $F^{\mu\nu}=\vec{\phi}\cdot \vec{G}^{\mu\nu}/a$. 
Using the Bianchi identity (\ref{bianchi}) and the first equation in
(\ref{XXIX}), we can write the charges as 
\bea 
g&\equiv&\int_{S^2_\infty}B_i dS^i=\frac{1}{a}\int{\cal B}^a_i\phi^a
dS^i = \frac{1}{a} \int {\cal B}^a_i (D^i \phi)^a d^3x\,,
\nonumber\\
q&\equiv&\int_{S^2_\infty}E_i dS^i=\frac{1}{a}\int{\cal E}^a_i\phi^a
dS^i = \frac{1}{a} \int {\cal E}^a_i (D^i \phi)^a d^3x\,. 
\label{gq}
\eea
Now, in the center of mass frame, the dyon mass is given by 
$$
M\equiv \int d^3x \theta_{00}= \int d^3x \left(\,\frac{1}{2}
\left[({\cal E}^a_k)^2+({\cal B}^a_k)^2 + (D_k\phi^a)^2 + 
(D_0\phi^a)^2 \right]+ V(\phi) \right)\,,
$$
where, $\theta_{\mu\nu}$ is the energy momentum tensor. After a little
manipulation, and using the expressions for the electric and magnetic
charges given in (\ref{gq}), this can be written 
as 
\bea
M&=&\int d^3x \left(\,\frac{1}{2}\left[({\cal E}^a_k -
D_k\phi^a\sin\theta)^2 + ({\cal B}^a_k-D_k\phi^a\cos\theta )^2
+(D_0\phi^a)^2\right] + V(\phi) \right)
\nonumber\\
&+& a (q \sin \theta + g \cos \theta )\,,
\label{M-bog}
\eea
where $\theta$ is an arbitrary angle.  Since the terms in the first
line are positive, we can write $M\geq (q \sin \theta + g \cos \theta
)$. This bound is maximum for $\tan\theta = q/g$. Thus we get the
Bogomol'nyi bound on the dyon mass as
$$
M \geq a\, \sqrt{g^2 + q^2}\,.
$$
For the 't Hooft-Polyakov solution, we have $q = 0$, and thus, $M \geq
a|g|$. But $|g|=4\pi/e$ and $M_W=ae\hbar=aq$, so that 
$$
M \geq a \frac{4 \pi}{e} = \frac{4 \pi}{e^{2}\hbar} M_{W} =
\frac{4 \pi \hbar}{q^{2}} M_{W} = \frac{\nu}{\alpha} M_{W}\,.
$$
Here, $\alpha$ is the fine structure constant and $\nu=1$ or $1/4$,
depending on whether the electron charge is $q$ or $q/2$. Since
$\alpha$ is a very small number ($\sim 1/137$ for electromagnetism),
the above relation implies that the monopole is much heavier than the
W-bosons associated with the symmetry breaking. 

From (\ref{M-bog}) it is clear that the bound is not saturated unless
$\lambda\rightarrow 0$, so that $V(\phi)=0$. This is the 
Bogomol'nyi-Prasad-Sommerfield (BPS) limit of the theory
\cite{bogomol,PS}. Note that in this limit,  $\phi_{vac}^2 =a^2$ is no
longer determined by the theory and, therefore, has to be imposed as
a boundary condition on the Higgs field. Moreover, in this limit, the
Higgs scalar becomes massless. Now, to saturate the bound we have to set 
\beq
D_0 \phi^a=0\,,\qquad {\cal E}^a_k=(D_k\phi)^a\sin\theta\,,
\qquad{\cal B}^a_k = (D_k\phi)^a\cos\theta\,,
\label{L}
\eeq
where, $\tan\theta=q/g$. In the BPS limit, one can use the 
't Hooft-Polyakov (or the Julia-Zee) ansatz either in (\ref{XXIX}), or
in (\ref{L}) to obtain the exact monopole (or dyon) solutions
\cite{bogomol, PS}. These solutions automatically saturate the
Bogomol'nyi bound and are referred to as the BPS states. Also, note
that in the BPS limit, all the perturbative excitations of the theory
saturate this bound and, therefore, belong to the BPS spectrum. As we
will see later, the BPS bound appears in a very natural way in
theories with $N=2$ supersymmetry. 

\subsection{Monopoles from a Distance}

Till now, we have described the monopoles arising in the
Georgi-Glashow model in terms of the structure of the  Higgs vacuum of
the theory. In this section, we will consider monopoles in a general
Yang-Mills-Higgs system and relate the Higgs vacuum description to a
description in terms of the unbroken gauge fields. These are the gauge
fields which remain massless and are relevant for the study of
monopoles at large distances. This formulation is convenient for
describing non-abelian monopoles. 

Let $\phi$ transform as a vector in a given representation of a gauge
group $G$. For convenience of notation, in the following we do not
distinguish between the group element $g$ and a given realization of
it. Writing the gauge fields as $W_\mu=T^aW^a_\mu$, we can construct
the covariant derivative of $\phi$ and the curvature tensor of $W_\mu$
as
\bea
D_\mu \phi &=& \partial_{\mu} \phi + ie W_{\mu} \phi\,,
\nonumber \\
G_{\mu\nu}&=&\partial_\mu W_\nu -\partial_\nu W_\mu +
ie [W_\mu, W_\nu] \,.
\nonumber
\eea
The Lagrangian density ${\cal L}$, the stress-energy tensor  
$\theta_{\mu\nu}$ and the gauge current $j^a_\mu$ are then given by  
\bea
{\cal L}&=&-\frac{1}{4}(G^a_{\mu\nu})^2+(D^\mu\phi)^\dagger D_\mu\phi
-V(\phi) \,,
\nonumber\\
\theta_{\mu\nu}&=&-G^a_{\mu\lambda}G^{a\lambda}_\nu +
(D_\mu\phi)^\dagger(D_\nu\phi) + (D_\nu\phi)^\dagger(D_\mu\phi) 
- g_{\mu\nu} {\cal L} \,,
\nonumber \\
\theta_{00}&=& \frac{1}{2}{\cal E}^{ia}{\cal E}^a_i +
\frac{1}{2}{\cal B}^{ia}{\cal B}^a_i +D_0\phi^\dagger D_0\phi +
D_i\phi^\dagger D_i\phi + V(\phi)\,,
\nonumber \\
j^a_\mu &=& ie \phi^\dagger T^a D_\mu\phi -
ie (D_\mu\phi)^\dagger T^a \phi\,.
\label{LIII}
\eea
Here, the Higgs potential is gauge invariant: $V(g\phi)=V(\phi)$. 
The equations of motion following from the above Lagrangian are 
\beq
D^\mu D_\mu\phi^a=-\frac{\partial V}{\partial \phi^a}\,,\qquad
D^\nu G^a_{\mu\nu} = - j_\mu^a\,.
\label{LIV}
\eeq
When the gauge group is $SO(3)$ spontaneously broken to $U(1)$, we can
work out the Bogomol'nyi bound exactly as in the previous section and
the outcome is 
\beq
M\geq \sqrt 2 \, a \sqrt{g^2 + q^2}\,.
\label{Bbound}
\eeq
For a general gauge group $G$, the Higgs vacuum, as in the
Georgi-Glashow model, is defined by  
$$
V(\phi)=0\,,\qquad D_\mu\phi=0 \,.
$$
The first equation defines the vacuum manifold ${\cal M}_0\equiv 
\{\phi\,: V(\phi)=0\}$, and the second equation leads to 
$$
[D_\mu, D_\nu] \phi =G_{\mu\nu} \phi = 0\,. 
$$
Thus, in the Higgs vacuum, $G_{\mu\nu}$ takes values in a subgroup of
the gauge group $G$ which keeps the Higgs vacuum invariant. We denote
this unbroken subgroup of $G$ by $H$. The generators of $G$ which do
not keep the Higgs vacuum invariant are of course broken and the
corresponding gauge bosons become massive.  If $V(\phi)$ does not have
extra global symmetries and accidental minima, then it is reasonable
to assume that the action of $G$ on ${\cal M}_0$ is transitive. This
means that any point in ${\cal M}_0$ is related to any other point
(and, in particular, to a reference point $\phi_0$) by some element of
$G$. Therefore, the little group or the invariance group, $H\subset
G$, of any point in ${\cal M}_0$ is isomorphic to the the little group
of any other point. Hence, the structure of ${\cal M}_0$ is described
by the right coset $G/H$.

For a solution  to have finite energy, at sufficiently large distances
from the core of the solution the field $\phi$ must take values in
the Higgs vacuum. Let $\Sigma$ be a 2-dimensional surface around the
core such that, on this surface, $\phi$ is already in ${\cal M}_0$. 
On this surface, $\phi$ describes a map from $\Sigma$ (with the
topology of $S^2$) into ${\cal M}_0$. This map is characterised by its
homotopy class which has to be an element of $\pi_2({\cal M}_0)\simeq
\pi_2(G/H)$. As described before, the associated topological number is
the magnetic charge of the solution. As long as no monopole crosses
the surface $\Sigma$, $\phi$ remains a continuous function of time and its
homotopy class does 
not change. To show that the map $\phi$ satisfies the group properties
of $\pi_2(G/H)$, one has to consider several widely separated
monopoles and study how their magnetic charges combine. For a
discussion of this issue, see \cite{coleman1}. 

The above discussion of the topological characterisation of the
monopole is in terms of the structure of the Higgs vacuum. However, it
is more natural to have a description in terms of the unbroken gauge
fields. The relationship  between these two descriptions is contained
in the equation $D_\mu\phi =0$ which is valid on the $S^2$ surface
$\Sigma$. Let us parametrise $S^2$ by a square $\{ 0 \leq s, t \leq 1
\}$. The map $r(s,t)$ from this square to the sphare is single valued
everywhere except on the boundary of the square which is identified
with a single point on the sphere: 
$$
r(0,t) = r(1,t) =r(s,0) =r(s,1) = r_0\,.
$$
For fixed $s$, as $t$ varies from $0$ to $1$, $r(s,t)$ describes a
closed path on $S^2$. 

\epsfxsize=11cm
\centerline{\epsffile{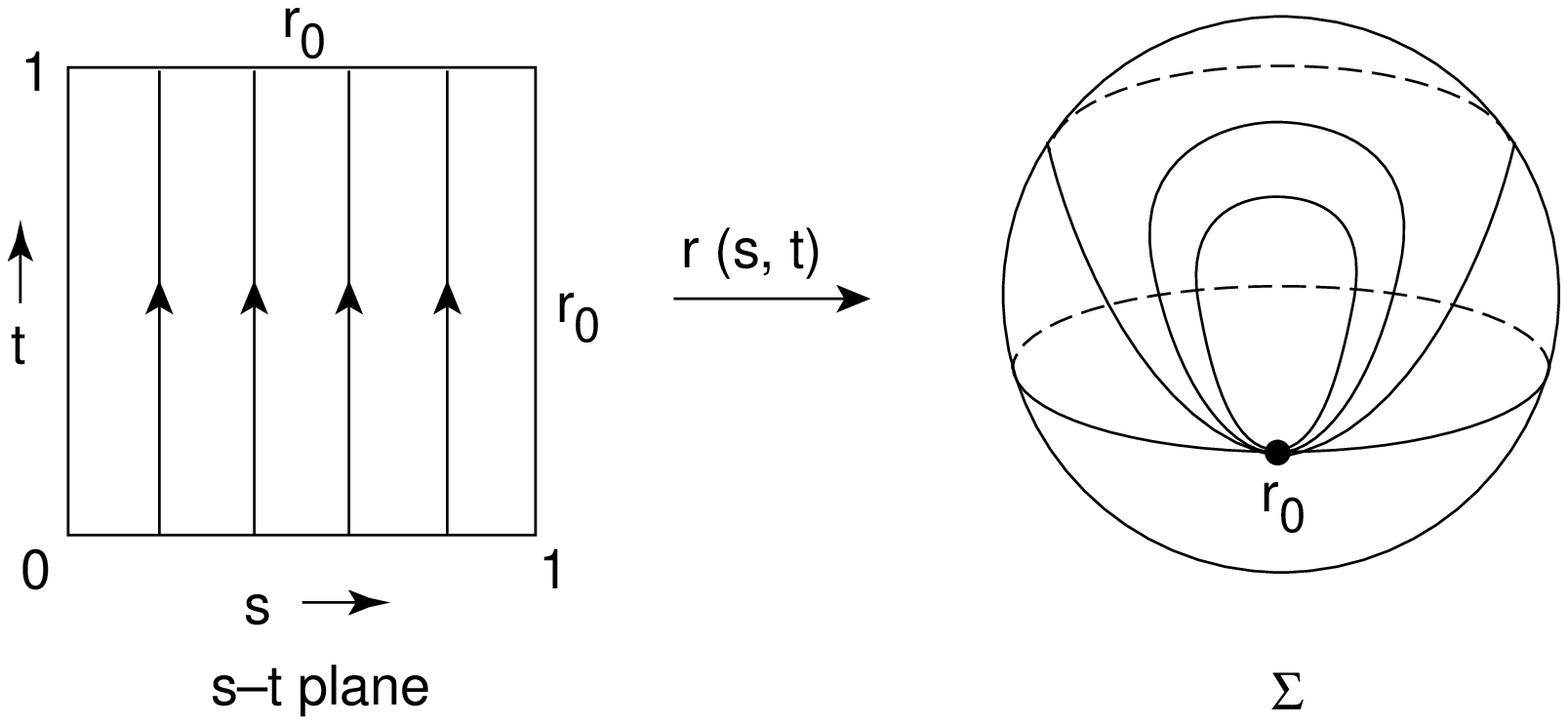}}
\begin{center}Figure 2\end{center}
The covariant derivative along $t$ becomes
$$
D_t \phi(s,t) =\frac{\partial r^i}{\partial t}D_i \phi(s,t)=0\,.
$$
For constant $s$, this can be solved as 
$$
\phi(s,t)=g(s,t)\phi(s,0)\,, 
$$
where, $g(s,t)$ satisfies $D_t g=0$, $g(s,0)=1$ and is given by the
path-ordered integral 
\beq
g (s,t) =  P\left(\,\exp \left(\,ie\int^{t}_{0}
\vec{W}\cdot\frac{\partial \vec{r}}{\partial t} \, dt\right)\right)\,. 
\label{LVIII}
\eeq
Clearly, $g(s,t)$ is an element of $G$ which gauge transforms any
$\phi(s,t)$ into a reference $\phi(s,0)$ and  $\phi(s,0)=\phi_0$ is
independent of $s$. Therefore, $g(s,t)$ contains the same topological
information as $\phi(s,t)$ and is characterised by a homotopy class
in $\pi_2(G/H)$. Since at $s=0,1$ we have $\partial r/ \partial t =0$,
then, $g(0,t)=g(1,t)=1$. For $t=1$, $\phi(s,1)=g(s,1)\phi_0$. But
since $\phi(s,1)=\phi_0$, we conclude that $g(s,1)$ must be an element
of the unbroken gauge group $H$ and denote it by $h(s)$:
\beq
h(s)\equiv g(s,1)=P\,\exp\left(ie\int^1_0\,\vec{W}\cdot\frac{\partial
\vec{r}}{\partial t}\,dt\right)\,. 
\label{LIX}
\eeq
Since $h(1)=h(0)=1$, as we move along $s$, $h(s)$ describes a closed
path in $H$ and is thus, an element of $\pi_1(H)$. However, note that
as we vary $t$ from $1$ to $0$, $g(s,t)$ continuously interpolates
between $h(s)$ and the identity. Therefore, $h(s)$ can describe only
those closed paths in $H$ which are homotopic to the trivial path when
$H$ is embedded in $G$. We denote the subgroup of $\pi_1(H)$ which 
corresponds to such paths by $\pi_1(H)_G$. If the homotopy class of
the map $\phi(s,t)$ is changed in $\pi_2(G/H)$, the homotopy class of
$h(s)$ changes in $\pi_1(H)_G$. In fact, there is a one-to-one
correspondence between $\pi_2(G/H)$ and $\pi_1(H)_G$ and the two
groups are isomorphic \cite{coleman1}. Thus, we can equally well
characterise the monopole by its homotopy class in $\pi_1(H)_G$.   

Let us discuss this in some more detail. Since any closed path in $H$
is also a closed path in $G$, there is a 
natural homomorphism from $\pi_1(H)$ into $\pi_1(G)$. As the
discussion above shows, $\pi_1(H)_G$ is in the kernel of this
homomorphism. Moreover, for a compact connected group $G$,
$\pi_2(G)=0$, which implies that $\pi_2(G)$ can be embedded in
$\pi_2(G/H)$ as its identity element. What we have done above,
basically, is to construct part of the following exact sequence:
$$
\pi_2(G)\rightarrow\pi_2(G/H)\rightarrow\pi_1(H)\rightarrow 
\pi_1(G)\rightarrow\pi_1(G/H)\rightarrow\pi_0(H)
$$
Note that if $G$ is simply connected then, $\pi_1(G)=0$ and 
$\pi_1(H)_G=\pi_1(H)$. So that the full $\pi_1(H)$ enters the physical
description of the monopole. For non-simply connected $G$, this
possibility can be realised in the presence of a Dirac string. Such a
string appears as a singular point on the $(s,t)$ plane, in the
presence of which, it is no longer possible to continuously deform 
$h(s)$ to the identity map. Therefore, the homotopy class of $h(s)$ is
no longer restricted to $\pi_1(H)_G$. However, 
for non-simply connected $G$, it is still possible to have a
description in terms of an unrestricted $\pi_1$ group provided we
embed $G$ in its universal covering group $\tilde G$. Let us denote
the little group of $\phi$ by $\hat H$. Since $\tilde G$ is
necessarily connected, $\pi_1(\tilde G)=0$ and $\pi_1(\hat H)= 
\pi_1(H)_G$. As an example, let us consider the Georgi-Glashow
model. Here, $G = SO(3)$ (with $(Ta)_{ij}=-i\epsilon_{aij}$) and  
$\tilde{G} = SU(2)$ (with $T_a=\frac{1}{2}\sigma_a$). The
homotopically distinct paths in $H$ (or $\hat H$) are: 
$$
h(t)=\exp(it\,\vec{\phi}\cdot\vec{T}\,4\pi N/a),\quad {\rm for}\,\,\, 
0 \leq t \leq 1\,.
$$
The integer $N$, which  characterises the elements of $\pi_1$, is
determined as follows: 
$$
\begin{array}{llll}
\tilde G=SU(2),& \tilde H=U(1),& \vec{\phi}\cdot\vec{T}\in 
\frac{1}{2}{\bf Z},& \Rightarrow\,\, N \in {\bf Z}\,, \cr
G=SO(3),& H=SO(2),& \vec{\phi}\cdot\vec{T}\in {\bf Z},&
\Rightarrow\,\, N \in \frac{1}{2}{\bf Z}\,.
\end{array}
$$
Since $SO(3)\sim S^2/{\bf Z}_2$, only paths with integer $N$
contribute in both cases. Later we show that $g=4\pi N/e$. For $SU(2)$,
$q_0=e\hbar/2$ and $gq_0/4\pi\hbar=N/2$. For $SO(3)$, $q_0=e\hbar$ and
$gq_0/4\pi\hbar=N$.

For Glashow-Weinberg-Salam model, $G=SU(2)\times U(1)$ and $H=U(1)$ is
a linear combination of $SU(2)$ and $U(1)$.  Although $\pi_{1} (H) =
{\bf Z}$, $\pi_{1} (H)_{G} = 0$. Therefore, in this model, any
non-trivial monopole must have a Dirac string.

We set out to describe the monopole in terms of the unbroken gauge
fields (the $H$-fields). Although, we have obtained a description in
terms of $\pi_1(H)_G$, it is not manifest that $h(s)$, as given by
(\ref{LIX}), involves only the $H$-fields. We show this in the
following: The quantity $g^{-1}(s,t)D_s g(s,t)$ is invariant under a
$t$-dependent gauge transformation. Moreover, by construction, $D_t
g(s,t)=0$. Hence, we can write
\beq
\partial_t(g^{-1}D_s g)= D_t(g^{-1}D_s g)=g^{-1}[D_t,D_s] g =
ie g^{-1} G_{ij} g \frac{\partial r^i}{\partial t} \frac{\partial r^j}
{\partial s}\,.
\label{LX}
\eeq
Let us integrate the first and the last terms above from $t=0$ to
$t=1$. Since $g^{-1}D_s g=0$ at $t=0$ and $g^{-1}D_s g=h^{-1}dh/ds$ at
$t=1$, we get
\beq
h^{-1} \, \frac{dh}{ds} = ie \, \int^{1}_{0} dt \,
g^{-1} G_{ij}\,g \,\frac{\partial r^i}{\partial t} \,
\frac{\partial r^j}{\partial s}\,.
\label{LXI}
\eeq
Since $G_{ij}$ was calculated on $\Sigma$, it involves only the
$H$-gauge fields. The conjugation by $g$ does not bring in a
dependence on the massive gauge fields as is evident from the
left-hand side of the equation. Hence the map is given entirely in
terms of the $H$-fields without any reference to the Higgs field. As a
simple application, consider the Dirac monopole. Integrating
(\ref{LXI}) from $s=0$ to $s=1$, we get $h(1)=exp(ie\int {\vec
  B}\cdot{\vec ds})$. Since $h(1)=1$, this leads to the Dirac
quantization condition $eg=2\pi n$ or $qg/4\pi\hbar=n/2$. Another
interesting consequence of (\ref{LXI}) is a possible explanation for
the fractional charges of quarks. We describe this in the next
section. 

\subsection{The Monopole and Fractional Charges}

So far, we have seen how the existence of a monopole can quantize the
electric charge in integer units. In the physical world, however, we
also come across fractionally charged quarks. In the following, we see
how the existence of a monopole can also account for these fractional
charges \cite{CO,GO}.

Let us represent our adjoint Higgs by $\phi=\phi^a T^a$, where $T^a$
are the fundamental representation matrices. Moreover, we only
consider $\phi$ on the surface $\Sigma$ as described in the previous
subsection. With $\phi$ in the Higgs vacuum, a generator $T^a$ belongs
to the unbroken subgroup $H$ of $G$ provided $[T^a,\phi]=0$. This
implies that $\phi$ itself is a generator of $H$ and commutes with its
other generators. Thus the Lie Algebra of $H$ is of the form
$L(H)=u(1)\oplus L(K)$ and we choose $L(K)$ to be orthogonal to
$u(1)$: ${\rm Tr}(\phi K^a)=0$ for $K^a\in K$.  Locally, $H$ has the
structure $U(1)\times K$, though, this is not necessarily the global
structure. We refer to $K$ as the colour group and identify the $U(1)$
as corresponding to electromagnetism. The gauge fields in $H$ can
be decomposed as $W^\mu=A^\mu\phi/a + X^\mu$, with ${\rm Tr}(\phi
X^\mu)=0$. Expanding the covariant derivative $\partial_\mu+ ie W_\mu$
we can identify the electric charge operator which couples to $A_\mu$
as $Q=(e\hbar/a)\phi$.

Since $h^{-1} dh/ds$ is a generator of $H$, we may write
\beq
h^{-1} \frac{dh}{ds}=ie\alpha (s)\,\frac{\phi_0}{a} +
i \beta_{a} K^{a} \,.
\label{hab}
\eeq
Using ${\rm Tr}(T_a T_b)=\delta_{ab}$, $\phi(r)=g\phi_0 g^{-1}$ and
equation (\ref{LXI}), we get 
\bea
\alpha (s)&=&-\frac{i}{ae}{\rm Tr}\left(\phi_0 h^{-1}\frac{dh}{ds} 
\right) \nonumber \\
&=&\frac{1}{a} \int^{1}_{0}{\rm Tr} \left(\varphi (r) G_{ij}\right) 
\frac{\partial r^i}{\partial t}\frac{\partial r^j}{\partial s}\,dt\,.
\nonumber
\eea
Identifying  the electromagnetic field strength tensor as
$F^{\mu\nu}={\rm Tr}\,(\phi G^{\mu\nu}/a)$, we get $\alpha(s)=
d\Omega/ds$, where, $\Omega(s)$ is the magnetic flux in a solid angle
subtended at the origin by the path $0\leq t\leq 1$ at fixed $s$ on
$S^2$. Substituting this back in (\ref{hab}) and integrating from
$s=0$  to $s$, gives
$$
h(s)= k(s) e^{i Q \Omega(s) / \hbar}\,.
$$
Since $h(1)=1$ and $\Omega (1) = g$ (where $g$ is the total magnetic
charge inside $\Sigma$), the quantization condition is replaced by 
$$
e^{igQ/\hbar} = k(1)^{-1}=k \in K \,.
$$
The left-hand side is invariant under $K$, therefore, we can at most
have $k\in {\bf Z}(K)$, the center of $K$. If we take $K = SU(N)$,
then, $k=e^{2\pi in/N}$ with $n=1,2,\dots N$. If all values of $n$ are
allowed, then $U(1)\cap K\subset Z_{N}$. This corresponds to the fact
that globally, $H$ cannot be decomposed as $U(1)\times K$. Now, let 
$|s>$ be a colour singlet. Then $k|s> = |s>$ and $exp(igq_s/\hbar)=1$,
which is again the Dirac quantization condition. Thus, if $q_0$ and
$g_0$ are the units of electric and magnetic charges for colour
singlets, then $q_s=n_sq_0$ and $g=mg_0$ with $g_0q_0=2\pi\hbar$. 
The colour non-singlet states $|c>$ can be classified according to
their behaviour under the center of the colour group $K$:
$$
k|c> = e^{2\pi i \, t(c)/N} |c> =e^{igQ / \hbar} |c>\,,
$$
where, $t(c)$ is an integer mod $N$. For a minimal monopole $g_{0}$,
we obtain $g_0 q_c /\hbar=2\pi(m + t(c)/N) $, hence,
$$
q_c = q_0 (m + \frac{t(c)}{N})\,. 
$$
If we set $N=3$, as for QCD, and $m=0$, then $q_c=q_0/3, 2q_0/3,q_0$. 

\subsection{Non-Abelian Magnetic Charge and the Montonen-Olive
  Conjecture} 

In this subsection we first consider the generalization of charge
quantisation to non-abelian monopoles \cite{GNO}, and then describe
the electric-magnetic duality conjecture of Montonen and Olive 
\cite{MO}.

Goddard, Nuyts and Olive \cite{GNO} attempted to classify all
$H$-monopole configurations. To describe such a monopole, we consider
a static configuration and choose the gauge $\vec{r}\cdot\vec{W}^a =0$.
At large distances, it is reasonable to write the magnetic componenets
of the field strength as
$$
G_{ij} = \frac{1}{4 \pi r^2} \, \epsilon_{ijk} \, \hat{r}^k G(r)\,, 
$$
where $D_\mu G(r)=0$. Since $G(r)$ transforms in the adjoint
representation, we can write $G(r)=g(s,t)G_{0}g(s,t)^{-1}$. 
Substituting the above expression in (\ref{LXI}), and integrating over
$s$, we get
\bea
{\rm ln}\left(h(s)\right) & = & \frac{ie}{4\pi r^2} G_0  
\int^s_0 ds\int^1_0 dt\,\epsilon_{ijk}\hat{r}^k \,
\frac{\partial r^i}{\partial t}\frac{\partial r^j}{\partial s}
\nonumber \\
& = & \frac{ie}{4\pi} G_0\,\Omega (s)\,.
\label{LXX}
\eea
Here, $\Omega (s)$ is the solid angle subtended at the origin by the
loop $0\leq t\leq 1,\,s={\rm const}$ on $S^2$. The elements in
$\pi_1(H)_G$ are, therefore, given by  
$$
h(s)= e^{\frac{ie}{4 \pi} G_0\Omega(s)}\,. 
$$
Since, $h(1)=1$, the above equation implies that
\beq
e^{ie G_0} = 1\,,
\label{LXXI} 
\eeq
which is the generalized charge quantization condition.
Clearly, $G_0$ is arbitrary upto a conjugation in the gauge
group. This freedom can be used to solve the charge quantization
condition as follows: Assume that $H$ is compact and connected and let
$T$ denote an Abelian subgroup of $H$ generated by its Cartan
subalgebra. Then, any element of $H$ is conjugate to at least one
element of $T$. Thus, it is always possible to find a frame in which
$G_0 = \beta^a T^a$. The coefficients $\beta^a$ are still not unique
as they transform under the Weyl group of $H$ which keeps this
parametrization of $G_0$ unchanged. Therefore, the equivalence classes
of $\beta^a$, related by the action of the Weyl group, are the gauge
invariant objects which characterise the non-abelian magnetic
charges. The $\beta^a$ are determined by the quantization
condition $exp(ie \beta^a T^a)=1$. To solve this, let $\omega^a$
denote a weight vector of $H$ in the given representation, and let
$\Lambda(H)$ denote the weight lattice. Then the quantization
condition implies that
$$
e \beta^a \omega^a \in 2\pi{\bf Z}\,,\qquad {\rm for\, all}\,\, 
\omega\in \Lambda(H)\,.
$$      
Note that the factors of $e$ and $2\pi$ are convention dependent.  
Thus, $e\beta$ lies on a lattice dual to $\Lambda(H)$: $e\beta \in
\Lambda^*(H)$. This dual lattice can by itself be regarded as the
weight lattice of a dual group $H^v$ which has $e\beta$'s as its
weight vectors (For details, see \cite{GNO}). Moreover, $(H^v)^v=H$.
$H$ is referred to as the electric group and $H^v$ as the magnetic
group. The magnetic charges are related to $H^v$ in the same way that
electric charges are related to $H$. A simple example of a dual pair
of groups is provided by $SO(3)$ and $SU(2)$. In this case, $G_0=\beta
T^3$. $T^3$ has integral eigenvalues for $SO(3)$ and half-integral
eigenvalues for $SU(2)$. The quantization condition $exp(ie\beta
T^3)=1$ gives:  
\bea
{\rm for}\,\,\,H=SO(3)\,:\quad e\beta &=& 4\pi \frac{n}{2}=4\pi \times 
({\rm weight \,\,of\,\,} SU(2))\,, \nonumber\\ 
{\rm for}\,\,\,H=SU(2)\,:\quad e\beta &=& 4\pi n = 4\pi \times 
({\rm weight\,\,of\,\,}SO(3))\,. \nonumber
\eea
For a general $SU(N)$ group, the dual relation is given by 
$(SU(NM)/{\bf Z}_N)^v = (SU(NM)/{\bf Z}_M)$ and, in particular,
$SU(N)^v = SU(N)/Z_N$.

Now, we will briefly describe the Montonen-Olive conjecture \cite{MO}
which is based on the above results. This  conjecture states that a
gauge theory is characterized by $H \times H^{v}$, and that we have
two equivalent descriptions of the theory: One in terms of $H$-gauge
fields with normal charged particles in the perturbative spectrum and
another, in terms of $H^{v}$-gauge fields with monopoles in the
perturbative spectrum. The Noether currents (associated with electric
charges) also get interchanged with topological currents (associated
with magnetic charges). Hence, the coupling constant $q/\hbar$ of the
$H$-theory is replaced by $g/\hbar$ in the in the $H^v$-theory. Since,
$g\sim 1/e$, this conjecture relates a strongly coupled theory to a
weekly coupled one, and vice-versa. As a result of this, it is
not easy to either prove or disprove this conjecture. Montonen and
Olive provided some semiclassical evidence in favour of this
conjecture in the BPS limit of the Georgi-Glashow model. This model
contains a Higgs boson, a photon and two massive charged vector bosons
in its perturbative spectrum, and magnetic monopoles as solitonic
classical solutions. The unbroken gauge group is self dual,
$H=U(1)=H^v$, therefore, the dual theory has the same form as the
original one with the monoples as elementary states and the massive
gauge bosons as solitonic solutions. The gauge boson mass in the dual
theory (where it appears as a soliton) can be computed using the BPS
formula and turns out to have the right value. Moreover, the
long-range force between two monopoles as obtained by Manton
\cite{manton}, can be obtained by calculating the potential between
$W$-bosons in the dual theory and turns out to be the same. 

Since the Montonen-Olive duality is non-perturbative in nature, it
cannot be verified in a perturbative framework unless we have some
kind of control over the perturbative and non-perturbative aspects of
the theory. Such a control is provided by superysymmetry. In fact, in 
the $N=4$ super Yang-Mills theory, some very non-trivial predictions
of this duality were verified in \cite{sen}. In later parts, we will
consider in detail the analogue of the Montonen-Olive duality in $N=2$
supersymmetric gauge theories. A prerequisite for this, however, is
the introduction of the $\theta$-term in the Yang-Mills action which
affects the electric charges of dyons.    

\subsection{The $\theta$-Parameter and the Monopole Charge}

In this section we will show, following Witten \cite{Witten79}, that
in the presence of a $\theta$-term in the Lagrangian, the magnetic
charge of a particle always contributes to its electric charge.

As shown by Schwinger and Zwanziger, for two dyons of charges
$(q_1,g_1)$ and $(q_2,g_2)$, the quantization condition takes the form
\beq
q_1g_2 - q_2g_1 = 2 \pi n \hbar
\label{LXXVI}
\eeq
For an electric charge $q_0$ and a dyon $(q_n,g_n)$, this gives
$q_0g_n = 2\pi n\hbar$. Thus, the smallest magnetic charge the dyon
can have is $g_0=2\pi\hbar/q_0$. For two dyons of the same magnetic
charge $g_0$ and electric charges $q_1$ and $q_2$, the quantization
condition implies $q_1-q_2= nq_0$. Therefore, although the difference
of electric charges is quantized, the individual charges are still
arbitrary. This arbitrariness in the electric charge of dyons can be
fixed if the theory is CP invariant as follows: Under a CP
transformation $(q,g)\rightarrow (-q,g)$. If the theory is CP
invariant, the existence of a state $(q,g_0)$ necessarily leads to the
existence of $(-q,g_0)$. Applying the quantization condition to this
pair, we get $2q =q_0\times integer$. This implies that $q=nq_0$ or
$q=(n+\frac{1}{2})q_0$, though at a time we can either have dyons of
integral or half-odd integral charge, and not both together.

In the above argument, it was essential to assume CP invariance to
obtain integral or half-odd inegral values for the electric charges of
dyons. However, in the real world, CP invariance is violated and there
is no reason to expect that the electric charge should be quantized as
above. To study the effect of CP violation, we consider the
Georgi-Glashow model with an additional $\theta$-term which is the
source of CP violation:
\beq
{\cal L}=-\frac{1}{4}\,F^a_{\mu\nu}F^{a\mu\nu}+\frac{1}{2}
(D_{\mu} \vec{\phi})^{2} - \lambda (\phi^{2} - a^{2} )^{2}
+\frac{\theta e^2}{32\pi^2}\,F^a_{\mu\nu}\tilde F^{a\mu\nu}\,. 
\label{LXXVIII}
\eeq
Here, $\tilde F^{a\mu\nu}=\frac{1}{2}\epsilon^{\mu\nu\rho\sigma} 
F^a_{\rho\sigma}$ and the vector notation is used to represent indices
in the gauge space. The presence of the $\theta$-term does not affect
the equations of motion but changes the physics since the theory is no
longer CP invariant. We want to construct the electric charge operator
in this theory. The theory has an $SO(3)$ gauge symmetry but the
electric charge is associated with an unbroken $U(1)$ which keeps the
Higgs vacuum invariant. Hence, we define an operator $N$ which
implements a gauge rotation around the $\hat{\phi}$ direction with
gauge parameter $\Lambda^a = \phi^a/a$. These transformations
correspond to the electric charge. Under $N$, a vector $\vec v $ and
the gauge fields $\vec A_\mu$ transform as    
$$
\delta \vec{v} = \frac{1}{a} \, \vec{\phi} \times \vec{v}\,, \quad
\delta \vec{A}_{\mu} = \frac{1}{ea} \, D_{\mu} \vec{\phi}\,.
$$
Clearly, $\vec{\phi}$ is kept invariant. At large distances where
$|\phi|=a$, the operator $e^{2\pi i N}$ is a $2 \pi$-rotation about
$\hat{\phi}$ and therefore $\exp\,(2\pi iN)=1$. Elsewhere, the
rotation angle is $2\pi |\phi|/a$. However, by Gauss' law, if the
gauge transformation is $1$ at $\infty$, it leaves the physical states
invariant. Thus, it is only the large distance behaviour of the
transformation which matters and the eigenvalues of $N$ are quantized
in integer units. Now, we use Noether's formula to compute $N$: 
$$
N=\int\,d^3x\left(\frac{\delta{\cal L}}{\delta\partial_0 A^a_i}\;
\delta A_{i}^{a} + \frac{\delta {\cal L}}{\delta \partial_{0}
\phi^{a}} \; \delta \phi^{a} \right )\,.
$$
Since $\delta \vec\phi= 0$, only the gauge part (which also includes
the $\theta$-term) contributes:
\bea
\frac{\delta}{\delta \partial_{0}A^{a}_{i}} \,
\left ( F^{a}_{\mu \nu} F^{a\mu\nu}\right) &=&
4F^{aoi} = -4 {\cal E}^{ai}\,, \nonumber\\
\frac{\delta}{\delta \partial_{0}A^{a}_{i}} \,
\left(\tilde{F}^a_{\mu\nu}F^{a\mu\nu}\right) &=&
2\epsilon^{ijk}F^a_{jk} =-4 {\cal B}^{ai}\,. \nonumber
\eea
Thus, we get
\bea
N &=&\frac{1}{ae} \int d^{3}x D_{i}\vec{\phi} \cdot \vec{\cal E}^i
- \frac{\theta e}{8\pi^{2}a}\,\int d^3x \, D_{i}\vec{\phi}
\cdot \vec{\cal B}^i \nonumber\\
&=& \frac{1}{e} Q - \frac{\theta e}{8 \pi^2} M \,,\nonumber
\eea
where, we have used equations (\ref{gq}). Here, $Q$ and $M$ are the
electric and magnetic charge operators with eigenvalues $q$ and $g$,
respectivly, and $N$ is quantized in integer units. This leads to the
following formula for the electric charge 
$$
q=ne + \frac{\theta e^2}{8\pi^2} g \,.
$$  
For the 't Hooft-Polyakov monopole, $n=1$, $g=-4\pi/e$, and
therefore, $q=e(1-\theta/2\pi)$. For a general dyonic solution we get
\beq
g= \frac{4\pi}{e} m,\qquad q = ne + \frac{\theta e}{2\pi}m\,.
\label{gqtheta}
\eeq
Thus, in the presence of a $\theta$-term, a magnetic monopole always
carries an electric charge which is not an integral multiple of some
basic unit.  

It is very useful to represent the charged states as points on the
complex plane, with electric charges along the real axis and magnetic
charges along the imaginary axis. A state can thus be represented as
\beq
q+ig=e(n+m\tau)\,,
\label{qig}
\eeq
where,
\beq
\tau=\frac{\theta}{2\pi} + \frac{4\pi i}{e^2}
\label{tau}
\eeq
In this parametrisation, the Bogomol'nyi bound (\ref{Bbound}) takes
the form 
\beq
M \geq {\sqrt 2}|ae(n+m\tau)|\,.
\label{BBtau}
\eeq
Note that (\ref{qig}) implies that all states lie on a two-dimensional
lattice with lattice parameter $\tau$ and (\ref{BBtau}) implies that
the BPS bound for a state is proportional to the distance of its
lattice point from the origin. These equations play a very important
role in the subsequent discussions.    

\section{Supersymmetric Gauge Theories}

In this section we will explain some aspects of supersymmetry and
supersymmetric field theories which are relevant to the work of Witten
and Seiberg. We start by explaining our conventions and then briefly
describe the representations of supersymmetry algebra with and without
central charges. We then discuss the representations of $N=1$
supersymmetry in terms of quantum fields and construct Lagrangians
with $N=1$ and $N=2$ supersymmetry. Most of the material in this
section is by now standard and can be found in \cite{WB,West,Sohnius}. 
Towards the end of this section, we will explicitly calculate the
central charges in $N=2$ theories with and without matter.
\subsection{Conventions}

We start by descrbing our conventions. We use the flat metric
$\eta_{ab} = {\rm diag}~ (1,-1,-1,-1)$. The spinors of the Lorentz
group $SL(2, C)\sim SU(2)_L\times SU(2)_R$ are written with dotted and
undotted components and, under $SL(2,C)$, transform as 
$$
\psi'_\alpha = M_{\alpha}^{~\beta} \psi_\beta\,, \qquad
\bar\psi'_{\dot{\alpha}} = M^{*~\dot{\beta}}_{\dot{\alpha}}
\bar{\psi}_{\dot{\beta}}\,.
$$
Spinor indices are raised or lowered with the $\epsilon$-tensor,
$$
\epsilon^{\alpha \beta} = \epsilon^{\dot{\alpha}\dot{\beta}} =
\left( \begin{array}{lr}
0 & 1 \\
-1 & 0
\end{array} \right) =
(i \sigma_{2})\,.
$$
By definition, this tensor is invariant under a $SL(2,C)$
transformation: $\epsilon^{\alpha \beta} = M_{~\gamma}^{\alpha}
\epsilon^{\gamma \delta}M_{\delta}^{~\beta}$. This can be written as 
$M^T\sigma_2 M=\sigma_2$ which implies $\sigma_2 M=(M^T)^{-1}\sigma_2$
Using this, we can write the transformations of the spinors with
raised indices as
$$
\psi'^\alpha =\psi^\beta (M^{-1})^{~\alpha}_{\beta}\,,\qquad
\bar{\psi}'^{\dot{\alpha}} = \bar{\psi}^{\dot{\beta}}
(M^*)^{-1~\dot{\alpha}}_{\dot{\beta}}\,.
$$
Now, let us define
$$
(\sigma^\mu)_{\alpha \dot{\alpha}} \equiv (1, \vec{\sigma})\,,
$$
then, 
$$
\sigma^\mu P_\mu = \left ( \begin{array}{lr}
P^{0} - P^{3} & -P^{1} + i P^{2} \\
-P^{1} - iP^{2} & P^{0} + P^{3}
\end{array} \right )\,, 
$$
and $\det(\sigma^\mu P_\mu)=P_\mu P^\mu$. We can raise the indices on 
$\sigma^\mu$ using the $\epsilon$-tensor and define $\bar\sigma$ as
$$
(\bar{\sigma}^\mu)^{\dot{\alpha}\alpha}=-(\sigma^\mu)^{\alpha\dot{\alpha}}
=\epsilon^{\dot{\alpha}\dot{\beta}}\epsilon^{\alpha \beta}
(\sigma^\mu)_{\beta \dot{\beta}}\,. 
$$
Numerically, this gives,
$$
(\bar{\sigma}^\mu) = (i\sigma_2) (\sigma^\mu)^T (i\sigma_2)^T =
\sigma_2 (\sigma^\mu)^T\sigma_2 = (1, - \vec{\sigma})\,.
$$
With these conventions, Lorentz transformations are generated by
\bea
(\sigma^{\mu\nu})_\alpha^{~\beta} & = & \frac{1}{4}
[\sigma^\mu_{\alpha \dot{\beta}} \bar{\sigma}^{\nu\dot{\beta}\beta}- 
(\mu \leftrightarrow \nu)]\,, \nonumber \\
(\bar{\sigma}^{\mu\nu})^{\dot{\alpha}}_{\dot{~\beta}}&=&\frac{1}{4}
[\bar{\sigma}^{\mu\dot{\alpha}\beta}\sigma^\nu_{\beta\dot{\beta}}-
(\mu\leftrightarrow \nu)]\,.\nonumber
\eea
For the scalar product of spinors, we use the following conventions 
\bea
\psi \chi &= & \psi^{\alpha} \chi_{\alpha} = - \psi_{\alpha}
\chi^{\alpha} = \chi^{\alpha} \psi_{\alpha} = \chi \psi\,,\nonumber\\ 
\bar{\psi} \bar{\chi} &=& \bar{\psi}_{\dot{\alpha}} 
\bar{\chi}^{\dot{\alpha}} =\bar{\chi} \bar{\psi}\,,\nonumber \\
(\psi\chi)^\dagger&=&\bar{\chi}_{\dot{\alpha}}\bar{\psi}^{\dot{\alpha}}
=\bar{\chi}\bar{\psi}=\bar{\psi}\bar{\chi}\,. \nonumber
\eea
We list some more spinor identities
$$
\begin{array}{rl}
\chi\sigma^\mu\bar{\psi} & =-\bar{\psi}\bar{\sigma}^\mu\chi\,, \\
(\chi\sigma^\mu\bar{\psi})^\dagger & =\psi\sigma^\mu\bar{\chi}\,, \\
\chi\sigma^\mu\bar{\sigma}^\nu\psi & =\psi\sigma^\nu\bar{\sigma}^\mu
\chi \,,\\ 
(\chi\sigma^\mu\bar{\sigma}^\nu\psi)^\dagger & =\bar{\psi} 
\bar{\sigma}^\nu \sigma^\mu \bar{\chi}\,.
\end{array}
$$
In the above basis, the Dirac matrices and Dirac and Majorana spinors
are given by
\beq
\gamma^\mu = \left( \begin{array}{cc}
0 & \sigma^\mu \\ \bar{\sigma}^\mu & 0 \end{array} \right)\,, \qquad
\psi_D=\left(\begin{array}{c}\psi_\alpha\\\bar{\chi}^{\dot\alpha}
\end{array}\right)\,,\qquad
\psi_M=\left(\begin{array}{c}\psi_\alpha\\{\bar{\psi}^{\dot{\alpha}}}
\end{array}\right)\,.
\label{sp-comp}
\eeq
As usual, one defines $\gamma_5=-i\gamma^0\gamma^1\gamma^2\gamma^3$.
Consider a massless fermion moving in the $z$-direction. Then, $P^\mu=
E(1,0,0,1)$, and the Dirac equation gives $(\gamma^0-\gamma^3)\psi=0$. 
Since the helicity operator is now $J_3=\frac{i}{2}\gamma^1\gamma^2$,
one gets, $J_3\psi=\frac{i}{2}(\gamma^0)^2 \gamma^1\gamma^2\psi =
\frac{i}{2}\gamma^0\gamma^1\gamma^2\gamma^3\psi=-\frac{1}{2}\gamma_5\psi$.
Hence, 
$$
\gamma_5 =+1 \Rightarrow -ve \;{\rm helicity}\,,
\gamma_5 =-1 \Rightarrow +ve \;{\rm helicity}\,.
$$

\subsection{Supersymmetry Algebra without Central Charges}

In the absence of central charges, the supersymmetry algebra is
written as  
\bea
&&\{ Q^{I}_{\alpha}, \bar{Q}_{\dot{\alpha} J} \} = 2
\sigma^\mu_{\alpha \dot{\alpha}} P_\mu \delta^{I}_{J}\,, 
\nonumber\\
&&\{Q^I_\alpha , Q^J_\beta\}=0\,,\,\{{\bar Q}_{\dot\alpha I}\,,\, 
{\bar Q}_{\dot\beta J}\}=0\,.
\label{susy-noC}
\eea
Here, $Q$ and $\bar Q$ are the supersymmetry generators and transform
as spin-half operators under the  angular momentum algebra. The
indices $I, J$ run from $1$ to $N$, where $N$ is the total number of
supersymmetries. Moreover, the supersymmetry generators commute with
the momentum operator $P_\mu$ and hence, with $P^2$. Therefore, all
states in a given representation of the algebra have the same mass.  
For a theory to be supersymmetric, it is necessary that its particle
content form a representation of the above algebra. The supersymmetry
algebra can be embedded in the super-Poincar\'{e} algebra and its
representations can be obtained systematically using Wigner's
method. In the following, we will give a brief description of the
representations of supersymmetry algebra. 

\noindent\underline{Massless Irreducible Representations}:
\noindent For massless states, we can always go to a frame where
$P^{\mu} = M (1,0,0,1)$. Then the supersymmetry algebra becomes 
$$
\{ Q^{I}_{\alpha}, \bar{Q}_{\dot{\alpha}J} \} =
\left ( \begin{array}{lr}
0 & 0 \\ 0 & 4M \end{array} \right )
\delta^{I}_{J}\,.
$$
Now, in a unitary theory the norm of a state is always positive
definite. Since $Q_\alpha$ and $\bar Q_{\dot\alpha}$ are conjugate to
each other, and $\{ Q_{1}, \bar{Q}_{\dot{1}}\}=0$, it follows that 
$Q_1|phys>= \bar{Q}_{\dot{1}}|phys> =0$. As for the other generators,
it is convenient to rescale them as  
$$
a^I=\frac{1}{2\sqrt{M}}Q^I_2\,,\qquad
(a^I)^\dagger=\frac{1}{2\sqrt{M}}\bar Q^I_{\dot 2}\,.
$$
Then, the supersymmetry algebra takes the form
$$
\{ a^I, (a^J)^\dagger \} = \delta^{IJ}\,,\quad  \{ a^I, a^J \}=0\,,
\quad \{ (a^I)^\dagger, (a^J)^\dagger \} = 0\,.
$$
This is a Clifford algebra with $2N$ generators and has a
$2^{N}$-dimensional representation. From the point of view of the
angular momentum algebra, $a^I$ is a rising operator and
$(a^I)^\dagger$ is a lowering operator for the helicity of massless
states. We choose the vacuum such that $J_3|\Omega_\lambda>=\lambda 
|\Omega_\lambda>$ and $a^I|\Omega_\lambda>=0$ for all $I$. Other
states are generated by the action of $(a^I)^\dagger$'s on the vacuum
state. From anti-symmetry it follows that a state with $m$
$(a^I)^\dagger$'s, and hence with helicity $\lambda -m/2$, will have a
degeneracy of $^N{\bf C}_m$. The helicity of all states so constructed
will span the range $\lambda$ to $\lambda-N/2$. Some examples are:
$$
\begin{array}{rrrrrr}
N = 1\,:& \quad |\lambda >,&    |\lambda-1/2>\hphantom{,}& & &\\
N = 2\,:& \quad |\lambda >,& 2\,|\lambda-1/2>,&|\lambda -1>\hphantom{,}&&\\ 
N = 4\,:& \quad |\lambda >,& 4\,|\lambda-1/2>,&
6\,|\lambda -1 >,&4\,|\lambda - 3/2>,& |\lambda -2 >
\end{array}
$$
The irreducible representations are not necessarily CPT
invariant. Therefore, if we want to assign physical states to these
representations, we have to suplement them with their CPT conjugates.
If a representation is CPT self-conjugate, it is left unchanged.  
Below, we list the representations after the addition of the CPT
conjugates and indicate the particle spectra which can be assiged to
them:
$$
\begin{array}{lrllllll}
N=1,&\lambda=1/2:&\,\,\, |1/2>\,,&|0>\,,   &|-1/2>\,,&|0>\hphantom{,}&& \\
    &\lambda=1:&\,\,\,|1> \,,&|1/2>\,, &|-1>\,,  &|-1/2>\hphantom{,}&& \\
N=2,&\lambda=1/2:&\,\,\,|1/2>\,,&2|0>\,,&|-1/2>\,,&|-1/2>\,,&2|0>\,,&|1/2>\\
    &\lambda=1:&\,\,\, |1> \,,&2|1/2>\,,&|0>\,,   &|-1>\,, &2|-1/2>\,,&|0>\\
N=4,&\lambda=1:&\,\,\, |1>\,,&4|1/2>\,,&6|0>\,,&4|-1/2>\,,&|-1>\hphantom{,}&
\end{array}
$$
Thus, for $N=1$, the representation contains a Majorana spinor and a
complex scalar if $\lambda=1/2$ (scalar multiplet), or a massless
vector and a Majorana spinor if $\lambda=1$ (vector multiplet). 
For $N=2$ and $\lambda=1/2$, we have two Majorana spinors (or one
Dirac spinor) with two complex scalars. This representation has the
same particle content as two copies of the $N=1$, $\lambda=1/2$
multiplet. For $N=2$ and $\lambda=1$, we have a massless vector, two
Majorana spinors and a complex scalar. Note that this multiplet has
the same particle content as the two $N=1$ multiplets for
$\lambda=1/2$ and $\lambda=1$ put together. For $N=4$, the
representation is self-conjugate and accommodates a massless vector,
two Dirac fermions and three complex scalars.   

\noindent\underline{Massive Irreducible Representations}:
\noindent For massive states, we can always go to the rest frame where
$P_\mu=(M,0,0,0)$ and define 
$$
a^I_\alpha=Q^I_\alpha/\sqrt{2M}\,,\quad (a^I_\alpha)^\dagger= 
{\bar Q}_{\dot\alpha I}/\sqrt{2M}\,.
$$
Then the supersymmetry algebra reduces to 
$$
\{a^I_1,(a_1^J)^\dagger\}=\delta^{IJ}\,,\qquad 
\{a^I_2,(a_2^J)^\dagger\}=\delta^{IJ}\,,
$$
with all other anti-commutators vanishing. The Clifford vacuum is
defined by $a^I_\alpha |\Omega>=0$ and the representation is
constructed by applying $(a^I_\alpha)^\dagger$'s on $\Omega$. Let
$|\Omega>$ be a spin singlet. Then there are $^{2N}{\rm C}_m$ states
at level $m$ and the dimension of the the representation is given by  
$\sum_{m=0}^{2N}{}^{2N}{\rm\bf C}_m=2^{2N}$. The maximum spin which can
be reached is $N/2$ and not $N$ as one might naively expect. This is
because $(a^I_1)^\dagger(a^I_2)^\dagger=\frac{1}{2}
\epsilon^{\alpha\beta} (a^I_\alpha)^\dagger(a^I_\beta)^\dagger$ is a
scalar. Thus the state with $m=2N$ has spin zero, as the vacuum. The
degeneracy of states with a given spin is labelled by the irreducible
representations of the group $USp(2N)$ which we will not discuss
here. Instead, let us consider the simplest example. For $N=1$, the
massive representation contains $2^2=4$ states, 
$$
|\Omega>,\quad a^\dagger_\alpha |\Omega>,\quad\frac{1}{\sqrt{2}}
\epsilon^{\alpha \beta} a^\dagger_\alpha a^\dagger_\beta |\Omega>\,,
$$
with spin content $(0)\oplus (1/2)\oplus (0)$. Here, $(j)$ denotes a
state of total spin $j$ and degeneracy $2j+1$. Thus, in the above
example, we have a Weyl (or Majorana) spinor and a complex scalar
$(\lambda , \phi)$. For $N=2$, the representation contains $2^4=16$
states which, under the $SU(2)$ of angular momentum, decompose as
$5(0)\oplus 4(1/2)\oplus 1(1)$. The $N=4$ massive multiplet has
$2^8=256$ states and inclueds a spin $2$ state.

Till now we have considered representations based on a singlet vacuum.
Let us consider a vacuum $|\Omega_j>$ of spin $j$ which is $2j+1$-fold
degenerate. The representation now contains $(2j+1)2^{2N}$ states. The
spectrum is worked out by combining the $j=0$ representation of the
Clifford algebra and a spin $j$, using the angular momentum addition
rules. For example, to obtain the $N=1$ representation based on
$|\Omega_j>$, we combine $(0)\oplus(1/2)\oplus (0)$ with $(j)$ to
obtain $(j)\oplus(j+1/2)\oplus(j-1/2)\oplus(j)$.  For $j=1/2$, we get
$(1/2)\oplus(1)\oplus(0)\oplus(1/2)$ which corresponds to a gauge
field, a Dirac fermion and a scalar field, all of the same mass.  Note
that in all cases we get the same number of bosonic and fermionic
degrees of freedom.

\subsection{Supersymmetry Algebra with Central Charges}
 
As shown by Haag, Lapuszanski and Sohnius \cite{HLS}, the
supersymmetry algebra (\ref{susy-noC}) admits a central extension and
can be generalised to 
\bea
&&\{ Q^{I}_{\alpha}, \bar{Q}_{\dot{\beta}J} \} =
2 \sigma^\mu_{\alpha \dot{\beta}} P_\mu \delta^{I}_{J}\,, 
\nonumber\\
&&\{ Q^I_\alpha, Q^J_\beta \}=2{\sqrt 2}\epsilon_{\alpha\beta} Z^{IJ}\,,
\nonumber \\
&&\{\bar{Q}_{\dot{\alpha}I},\bar{Q}_{\dot{\beta}J} \} =
2{\sqrt 2}\epsilon_{\dot{\alpha} \dot{\beta}} Z^{*}_{IJ}\,,
\label{susy-C}
\eea
where, $Z$ and $Z^*$ are the central charge metrices which are
antisymmetric in $I$ and $J$. Let us focus on the case of even $N$.
Using a unitary transformation, we can skew-diagonalize $Z$:
$\tilde{Z}^{IJ} = U^{I}_{A}U^{J}_{B}Z^{AB}$, so that it takes the form
$Z=\epsilon\otimes D$, where $D$ is an $N/2$-dimensional diagonal
matrix. Thus, the index $I$ which counts the number of supersymmetries
can be decomposed into $(a,m)$, with $a=1,2$ coming from the
antisymmetric tensor $\epsilon$, and $m=1,...,N/2$ coming from the
diagonal matrix $D$.  By a further chiral rotation, we may choose the
eigenvalues of $D$ to be real. Once we have skew-diagonalized, it is
sufficient to consider just the $N=2$ supersymmetry, for which the
algebra takes the form
\bea
&&\{Q^a_\alpha,{\bar Q}_{\dot\beta b}\}=2(\sigma^\mu)_{\alpha\dot\beta}  
P_\mu \delta^{a}_{b}\,,\nonumber \\
&&\{ Q^a_\alpha,Q^b_\beta\}=2{\sqrt 2}\epsilon_{\alpha\beta}
\epsilon^{ab}Z\,, \nonumber \\ 
&&\{{\bar Q}_{\dot\alpha a},{\bar Q}_{\dot\beta b}\}= 
2{\sqrt 2}\epsilon_{\dot\alpha\dot\beta}\epsilon_{ab} Z \,.
\label{Susy-C'}
\eea
Since $Z$ commutes with all the generators, we can fix it to be
the eigenvalue for the given representation. Now, let us define:
$$
a_\alpha=\frac{1}{2} \{ Q^1_\alpha +
\epsilon_{\alpha\beta} (Q^2_\beta)^\dagger \}\,, \qquad
b_\alpha=\frac{1}{2} \{ Q_\alpha^1 -
\epsilon_{\alpha\beta} (Q^2_\beta)^\dagger \}\,. 
$$
Then, the algebra (\ref{susy-C}) reduces to 
\beq
\{a_\alpha,a^\dagger_\beta\}=\delta_{\alpha\beta}(M+{\sqrt 2}Z)\,,\qquad
\{ b_\alpha,b^\dagger_\beta\}=\delta_{\alpha\beta} (M-{\sqrt 2}Z)\,,
\label{susy-ab}
\eeq
with all other anticommutators vanishing. Since all physical states
have positive definite norm, it follows that for massless states, the
central charge is trivially realised ({\it i.e.},\,$Z=0$). For massive
states, this leads to a bound on the mass $M\geq {\sqrt 2}|Z|$. When
$M={\sqrt 2}|Z|$, one set of operators in (\ref{susy-ab}) is trivially
realized and the algebra resembles the massless case and the dimension
of representation is greatly reduced. For example, a reduced massive
$N = 2$ multiplet has the same number of states as a massless $N = 2$
multiplet. Thus the representations of the $N=2$ algebra with a
central charge can be classified as either long multiplets (when $M >
{\sqrt 2}|Z|$) or short multiplets (when $M={\sqrt 2}|Z|$).
 
The mass bound $M\geq {\sqrt 2}|Z|$ is reminiscent of the Bogomol'nyi
bound in the Georgi-Glashow model. In fact, it turns out that in the
supersymmetric version of the Georgi-Glashow model (which is based on
the algebra without central charges) the solitonic solutions do give
rise to a central extension term in the supersymmetry algebra, thus
realizing (\ref{susy-C})\cite{Olive-Witten}. The origin of the central
charge is easy to understand: The supersymmetry charges $Q$ and $\bar
Q$ are space integrals of local expressions in the fields (the time
component of the super-currents). In calculating their anticommutators,
one encounters surface terms which are normally neglected. However, in
the presence of electric and magnetic charges, these surface terms are
non-zero and give rise to a central charge. As we will explicitly show
towards the end of this section, it is found that 
\beq
Z = a (q+ig)= ae (n+m\tau)\,,
\label{Ztau}
\eeq
so that $M\geq {\sqrt 2}|Z|$ coincides with the Bogomol'nyi bound
(\ref{Bbound}). From (\ref{susy-ab}) it is clear that the BPS states 
(which saturate the bound) are annihilated by half of the
supersymmetry generators and thus belong to reduced representations of
(\ref{susy-C}). An important consequence of this is that, for BPS
states, the relationship between their charges and masses is dictated
by supersymmetry and does not receive perturbative or non-perturbative
corrections in quantum theory. This is so because a modification of
this relation implies that the states no longer belong to a short
multiplet. On the other hand, quantum correction are not expected to
generate the extra degrees of freedom needed to convert a short
multiplet into a long multiplet. Since there is no other possibility,
we conclude that for short multiplets the relation $M={\sqrt 2}|Z|$ is
not modified either perturbatively or non-perturbatively.    

\subsection{Local Representations of N=1 Supersymmetry}

In this subsection we describe the action of supersymmetry on the
local fields in a quantum field theory. It is well known that the 
Poincar\'{e} group naturally acts on the space-time coordinates. All
other objects transform as components of tensors or spinors defined on
the space-time manifold. Similarly, the supersymmetry transformations
naturally act on an extension of the space-time, called the
``superspace''. The quantum fields then transform as components of a  
``superfield'' defined on the superspace. In the following, we first
describe these notions and then introduce the chiral and vector
superfields.  

\noindent\underline{Superspace :} 
\noindent The superspace is obtained by adding four spinor degrees of
freedom $\theta^\alpha,\bar\theta_{\dot\alpha}$ to the space-time
coordinates $x^\mu$. The spinor index is raised and lowered with the 
$\epsilon$-tensor and $\theta\theta = \theta^\alpha\theta_\alpha = -2
\theta^1\theta^2$. Similarly, $\bar\theta\bar\theta=
\bar\theta_{\dot\alpha}\bar\theta^{\dot\alpha}= 2\bar\theta_{\dot 1} 
\bar\theta_{\dot 2}$. We also have
$$
\theta^{\alpha}\theta^{\beta}=\frac{1}{2} \epsilon^{\alpha \beta}
\theta \theta\,,\qquad
\bar{\theta}_{\dot{\alpha}}\bar{\theta}_{\dot{\beta}}=-\frac{1}{2}
\epsilon_{\dot{\alpha}\dot{\beta}} \bar{\theta}\bar{\theta}\,,\qquad
\theta \sigma^\mu \bar{\theta} \theta \sigma^\nu \bar{\theta}=
\frac{1}{2}\theta \theta \bar{\theta}\bar{\theta} \eta^{\mu\nu}\,.
$$
These formulae are the basis for Fierz rearrangements. 
 
Under the supersymmetry transformations (\ref{susy-noC}) with $N=1$
and transformation parameters $\xi$ and $\bar\xi$, the superspace
coordinates are taken to transform as  
\bea
x^\mu &\rightarrow& x'^\mu=x^\mu + i\theta\sigma^\mu\bar\xi
- i\xi\sigma^\mu\bar\theta \,,\nonumber\\
\theta &\rightarrow&\theta'=\theta + \xi\,, \nonumber\\
\bar\theta &\rightarrow& \bar\theta'=\bar\theta +\bar\xi\,.
\label{superspace}
\eea
Since these transformations are implemented by the operator 
$\xi^\alpha Q_\alpha + \bar{\xi}_{\dot\alpha}\bar{Q}^{\dot\alpha}$, we
can easily obtain the representation of the supercharges acting on the
superspace as 
\beq
Q_\alpha=\frac{\partial}{\partial\theta^\alpha}- 
i\sigma^\mu_{\alpha\dot{\alpha}}\bar{\theta}^{\dot{\alpha}}\,
\partial_\mu\,,\qquad \bar{Q}_{\dot\alpha} = -\frac{\partial}{\partial
\bar{\theta}^{\dot\alpha}}+i\theta^\alpha 
\sigma^\mu_{\alpha\dot{\alpha}}\, \partial_\mu\,.
\label{superspace-Q}
\eeq
These satisfy $\{Q_\alpha ,\bar{Q}_{\dot\alpha}\}= 
2i\sigma^\mu_{\alpha\dot\alpha}\,\partial_\mu$.
Moreover, using the chain rule, it is easy to see that 
$\partial/\partial x^\mu$ is invariant under (\ref{superspace}) but
not $\partial/\partial\theta$ and $\partial/\partial\bar\theta$.   
Therefore, we introduce the super-covariant derivatives
\beq
D_\alpha = \frac{\partial}{\partial \theta^\alpha} +
i\sigma^\mu_{\alpha\dot\alpha}\,\partial_\mu \,,\qquad
\bar{D}_{\dot{\alpha}} = -\frac{\partial}{\partial
\bar{\theta}^{\dot{\alpha}}}-i\sigma^\mu_{\alpha \dot{\alpha}}
\theta^{\alpha} \, \partial_\mu \,.
\label{superspaceD}
\eeq
They satisfy $\{ D_\alpha,\bar{D}_{\dot\alpha}\}=-2i\sigma^\mu_{\alpha
\dot{\alpha}}\,\partial_\mu$ and commute with $Q$ and $\bar Q$.

\noindent\underline{Superfields:} 
\noindent A superfield is a function on the superspace, say,
$F(x,\theta,\bar\theta)$. Since the $\theta$-coordinates are
anti-commuting, the most general $N=1$ superfield can always be
expanded as  
\bea
F(x,\theta,\bar\theta)&=& f(x)+\theta\phi(x) + \bar\theta\bar\chi(x)
+\theta\theta m(x) +\bar\theta\bar\theta n(x) + 
\theta\sigma^\mu\bar\theta v_\mu (x) \nonumber\\
&+&\theta\theta\bar\theta\bar\lambda(x)+\bar\theta\bar\theta\theta\psi(x)
+ \theta\theta\bar\theta\bar\theta d(x)\,. 
\label{superfield}
\eea
Clearly, any function of superfields is, by itself, a superfield. 
Under supersymmetry, the superfield transforms as $\delta F=(\xi
Q+\bar\xi\bar Q)F$, from which, the transformation of the component
fields can be obtained. Note that since $d(x)$ is the component of
highest dimension in the multiplet, its variation under supersymmetry
is always a total derivative of other components. Thus, ignoring
surface terms, the space-time integral of this component is invariant
under supersymmetry. This tells us that a supersymmetric Lagrangian
density may be constructed as the highest dimension component of an
appropriate superfield. To describe physical systems, we do not need
all components of the superfield. The relevant components are selected
by imposing appropriate constraints on the superfield. 

\noindent\underline{Chiral Multiplets:} 
\noindent The $N=1$ scalar multiplet is represented by a superfield
with one constraint:
$$
\bar{D}_{\dot{\alpha}} \Phi = 0 \,.
$$
This is referred to as the chiral superfield. Note that for  
$y^\mu = x^\mu +i\theta\sigma^\mu\bar{\theta}$, we have 
$$
\bar{D}_{\dot\alpha}y^\mu=0,\quad\bar{D}_{\dot\alpha}\theta^\beta=0\,.
$$
Therefore, any function of $(y,\theta)$ is a chiral superfield. It can
be shown that this also is a necessary condition. Hence, any chiral
superfield can be expanded as  
\beq
\Phi(y,\theta)=A(y)+\sqrt{2}\theta\psi(y)+\theta\theta F(y)\,.
\label{csfy}
\eeq
Here, $A$ and $\psi$ are the fermionic and scalar components
respectively and $F$ is an auxiliary field required for the off-shell
closure of the algebra. Similarly, an anti-chiral superfield is
defined by $D_{\alpha}\Phi^\dagger=0$ and can be expanded as 
\beq
\Phi^\dagger(y^\dagger,\bar\theta) =A^\dagger(y^\dagger)+\sqrt{2}
\bar{\theta}\bar{\psi}(y^\dagger)+\bar{\theta}\bar{\theta} 
F^\dagger(y^\dagger)\,,
\label{acsfy}
\eeq
where, $y^{\mu\dagger}=x^\mu-i\theta\sigma^\mu\bar{\theta}$.
The product of chiral superfields is a chiral superfield. In general,
any arbitrary function of chiral superfields is a chiral superfield:
\bea
{\cal W}(\Phi_{i}) &=& {\cal W}(A_{i} + \sqrt{2} \theta \psi_{i} +
\theta \theta F_{i} ) \nonumber \\
&=&{\cal W}(A_{i}) + \frac{\partial {\cal W}}{\partial A_{i}}
\sqrt{2} \theta \psi_{i} + \theta \theta \left (\frac{\partial 
{\cal W}}{\partial A_{i}} F_{i} - \frac{1}{2} \, \frac{\partial^{2}
{\cal W}}{\partial A_i A_j}\, \psi_{i} \psi_{j} \right)\,.
\label{spotential}
\eea
${\cal W}$ is referred to as the superpotential. In terms of the
original variables, $\Phi$ and $\Phi^\dagger$ take the form 
\bea
\Phi(x,\theta,\bar\theta) &=& A(x)+i\theta\sigma^\mu \bar{\theta}
\partial_\mu A-\frac{1}{4}\theta^2 \bar{\theta}^2 \Box A 
\nonumber\\
&&+\sqrt{2}\theta\psi(x)-\frac{i}{\sqrt{2}}\theta\theta
\partial_\mu\psi\sigma^\mu\bar{\theta} +\theta \theta F(x)\,,
\label{csfield}
\\
\Phi^\dagger(x,\theta,\bar\theta)&=&A^\dagger(x)-i\theta\sigma^\mu
\bar{\theta} \partial_\mu A^\dagger - \frac{1}{4} \theta^{2}
\bar{\theta}^{2} \Box A^\dagger \nonumber\\
&&+\sqrt{2} \bar{\theta} \bar{\psi} (x) + \frac{i}{\sqrt{2}} \,
\bar{\theta} \bar{\theta} \, \theta \sigma^\mu \, \partial_\mu
\bar{\psi} + \bar{\theta} \bar{\theta} F^\dagger (x)\,.
\label{asfield}
\eea

\noindent\underline{Vector Multiplet:} 
\noindent This multiplet is represented by a real superfield satifying
$V=V^\dagger$. In components, it takes the form
$$
\begin{array}{lll}
V(x, \theta, \bar{\theta}) =& C + i \theta  \chi - i \bar{\theta}
\bar{\chi} + \frac{i}{2} \theta^{2} (M + i N) - \frac{i}{2}
\bar{\theta}^{2} (M - iN) \\
& - \theta \sigma^\mu \bar{\theta} A_\mu + i \theta^{2} \bar{\theta}
(\bar{\lambda} + \frac{i}{2} \bar{\sigma}^\mu \partial_\mu \chi )\\
& - i \bar{\theta}^{2} \theta (\lambda + \frac{i}{2} \sigma^\mu
\partial_\mu \bar{\chi} ) + \frac{1}{2} \theta^{2} \bar{\theta}^{2}
(D - \frac{1}{2} \Box C)\,.
\end{array}
$$
Many of these components can be gauged away using the abelian gauge
transformation $V \rightarrow V + \Lambda + \Lambda^\dagger$, where  
$\Lambda$ ($\Lambda^\dagger$) are chiral (antichiral) superfields.  
In the so called Wess-Zumino gauge, we set $C=M=N=\chi= 0$, so that
$$
V=-\theta \sigma^\mu\bar{\theta}A_\mu+i\theta^{2}
\bar{\theta}\bar{\lambda} - i \bar{\theta}^{2} \theta \lambda +
\frac{1}{2} \theta^{2} \bar{\theta}^{2} D\,.
$$
In this gauge, $V^2=\frac{1}{2}A_\mu A^\mu \theta^{2}\bar{\theta}^{2}$
and $V^{3}=0$. The Wess-Zumino gauge breaks supersymmetry, but not the
gauge symmetry of the abelian gauge field $A_\mu$. The Abelian field
strength is defined by  
$$
W_{\alpha} = - \frac{1}{4} \bar{D}^{2} D_{\alpha} V\,,\qquad
\bar{W}_{\dot{\alpha}}=-\frac{1}{4}D^{2} \bar{D}_{\dot{\alpha}}V\,.
$$
$W_{\alpha}$ is a chiral superfield. Since it is gauge invariant, it
can be computed in the Wess-Zumino gauge and takes the form
\beq
W_\alpha =-i\lambda_\alpha (y)+\theta_\alpha D-\frac{i}{2} 
(\sigma^\mu\bar{\sigma}^\nu \theta)_\alpha \, F_{\mu\nu}+ 
\theta^2 (\sigma^\mu \partial_\mu \bar{\lambda} )_{\alpha}\,,
\label{W-abelian}
\eeq
where, $F_{\mu\nu}=\partial_\mu A_\nu -\partial_\nu A_\mu$ is the
familiar abelian field strength tensor. 

In the non-Abelian case, $V$ belongs to the adjoint representation of
the gauge group: $V=V_AT^A$, where, $T^{A\dagger}=T^A$. The gauge
transformations are now implemented by 
$$
e^{-2V} \rightarrow e^{-i \Lambda^\dagger} e^{-2V} e^{i \Lambda}
\qquad {\rm where}, \; \Lambda = \Lambda_{A}T^{A}
$$
The non-Abelian gauge field strength is defined by 
$$
W_{\alpha} = \frac{1}{8} \bar{D}^{2} e^{2V} D_{\alpha} e^{-2V}
$$
and transforms as 
$$
W_\alpha\rightarrow W'_\alpha= e^{-i\Lambda}W_\alpha e^{i\Lambda}\,.
$$
In components, it takes the form
\beq
W_\alpha =T^a \left (-i \lambda^{a}_{\alpha} + \theta_{\alpha}
D^{a} - \frac{i}{2}(\sigma^\mu \bar{\sigma}^\nu \theta )_{\alpha}
F^{a}_{\mu\nu}+\theta^2 \sigma^\mu D_\mu\bar{\lambda}^a \right)\,, 
\label{W-nonabelian}
\eeq
where, 
$$
F^a_{\mu\nu}=\partial_\mu A^a_\nu -\partial_\nu A^a_\mu + 
\, f^{abc} A^b_\mu A^c_\nu\,,\qquad
D_\mu\bar{\lambda}^a=\partial_\mu\bar{\lambda}^a +
\,f^{abc} A_\mu^b \bar{\lambda}^c \,.
$$
In the next section, we will construct supersymmetric Lagrangians in
terms of superfields. 

\subsection{Construction of N=1 Lagrangians}

In this section we will construct the $N=1$ Lagrangians for the 
scalar and the vector multiplets. These serve as the building blocks
for the $N=2$ Lagrangian which is our real interest. As stated before,
a supersymmetric Lagrangian can be constructed as the highest
component of a superfield. Thus the problem reduces to that of finding
appropriate superfields. 

\noindent\underline{Lagrangian for the Scalar Multiplet:}
\noindent Let us first consider the product of a chiral and an
anti-chiral superfield $\Phi_i^\dagger\Phi_j$. This is a general
superfield and its highest component can be computed using
(\ref{csfield}) and  (\ref{asfield}) as  
\bea
\Phi^\dagger_i\Phi_j\mid_{\theta^{2}\bar{\theta}^{2}}\,\,\,=
&-&\frac{1}{4}\,A^\dagger_i\Box A_j -\frac{1}{4}\Box A^\dagger_i 
A_j+F^\dagger_i F_j +\frac{1}{2}\,\partial_\mu A^\dagger_i\partial^\mu
A_j \nonumber\\
&-& \frac{i}{2}\,\psi_j\sigma^\mu \partial_\mu \bar{\psi}_{i} +
\frac{i}{2} \partial_\mu \psi_j \sigma^\mu\bar{\psi}_i\,.\nonumber
\eea
Dropping some total derivatives and summing over $i = j$, we get the
free field Lagrangian
$$
{\cal L}=
\Phi^\dagger_i \Phi_i \mid_{\theta^2\bar\theta^2}
=\partial_\mu A^\dagger_i \partial^\mu A_i + F^\dagger_i F_i -i
\bar{\psi}_i \bar{\sigma}^\mu \partial_\mu \psi_i\,.
$$
This is the free Lagrangian for a massless scalar and a massless
fermion with an auxiliary field which can be eliminated by its
equation of motion. Supersymmetric interaction terms can be
constructed in terms of the superpotential (\ref{spotential}) and its
conjugate, which are holomorphic functions of $\Phi$ and
$\Phi^\dagger$, respectively. Moreover, note that the space of the
fields $\Phi$ may have a non-trivial metric $g^{ij}$ in which case the scalar
kinetic term, for example, takes the form $g^{ij}\partial_\mu
A_i^\dagger \partial^\mu A_j$, with appropriate modifications for
other terms. In such cases, the free
field Lagrangian above has to be replaced by a non-linear
$\sigma$-model. Thus, the most general $N=1$ supersymmetric Lagrangian
for the scalar multiplet (including the interaction terms) is given by 
$$
{\cal L}=\int\,d^{4}\theta\,K(\Phi,\Phi^\dagger) +
\int d^{2} \theta {\cal W}(\Phi) + \int d^{2} \bar{\theta} 
\bar{\cal W}(\Phi^\dagger)\,.
$$
Note that the $\theta$-integrals pick up the highest component of the
superfield and in our conventions, $\int d^2 \theta \theta^2=1$ and 
$\int d^2 \bar\theta \bar\theta^2=1$. In terms of the non-holomorphic
function $K(A, A^\dagger)$, the metric on the field space is given by 
$g^{ij}=\partial^2K/\partial A_i\partial A^\dagger_j$. For this
reason, the function $K(\Phi,\Phi^\dagger)$ is referred to as the
K\"{a}hler potential.    

For a renormalizable theory, the forms of $K$ and ${\cal W}$ are not
arbitrary and are constrained by R-symmetry. This symmetry acts on the
chiral superfields as follows
\bea
R\,\Phi (x,\theta) = \Phi'(x,\theta) &=& e^{2 in\alpha}
\Phi (x, e^{-i\alpha}\theta)\,, \nonumber \\
R\,\Phi^\dagger(x,\bar\theta ) = \Phi'^\dagger(x,\theta) &=& e^{-2
in\alpha}\Phi^\dagger(x, e^{i \alpha}\bar{\theta})\,. \nonumber
\eea
Under this, the component fields transform as 
$$
\begin{array}{lll}
A & \rightarrow e^{2i n \alpha} A \,,\\
\psi & \rightarrow e^{2i(n-1/2) \alpha} \psi\,, \\
F & \rightarrow e^{2i(n-1) \alpha} F\,.
\end{array}
$$
We refer to $n$ as the R-character. Since $\theta \rightarrow
e^{+i\alpha}\theta$, or $d^2\theta \rightarrow e^{-2 i\alpha}d^2
\theta$, The R-character of the superfields in each term of ${\cal W}$
must add up to one. Similarly, $K$ should be $R$-neutral. The vector
multiplet is real and it has no natural $R$-symmetry. This symmetry
plays an important role in the study of supersymmetric gauge theories
and we will come back to it in the next section.

\noindent\underline{Lagrangians for the Vector Multiplet:}
\noindent As mentioned in the previous section, the Abelian field
strength $W$, given by (\ref{W-abelian}), is a chiral
superfield. Using the expansion there, one can easily compute that 
$$
W^\alpha W_\alpha\mid_{\theta\theta}=-2i\lambda\sigma^\mu\partial_\mu
\bar{\lambda} + D^2 - \frac{1}{2} F^{\mu\nu} F_{\mu\nu} +\frac{i}{4} \,
\epsilon^{\mu\nu\rho\sigma} F_{\mu\nu}F_{\rho\sigma}\,.
$$
Hence, the usual abelian supersymmetric Lagrangian (which does not
contain the $F\widetilde F$ term) is given by    
$$
{\cal L}=\frac{1}{4g^{2}}\left(\int d^2\theta\, W^\alpha W_\alpha +
\int d^2\bar\theta\, \bar{W}_{\dot\alpha} \bar{W}^{\dot\alpha}\right)\,.
$$
Similarly, in the non-Abelian case, using the normalization 
${\rm Tr}T^aT^b=\delta^{ab}$, we have
\beq
{\rm Tr}(W^\alpha W_\alpha \mid_{\theta\theta})=-2i\lambda^a\sigma^\mu 
D_\mu \bar{\lambda}^a +D^a D^a -\frac{1}{2} F^{a\mu\nu}
F^a_{\mu\nu} +\frac{i}{4}\epsilon^{\mu\nu\rho\sigma} F^a_{\mu\nu} 
F^a_{\rho\sigma}\,,
\label{WW-nonabelian}
\eeq
and, hence, the usual non-Abelian supersymmetric Lagrangian (without
the $F\widetilde F$-term) is given by 
$$
{\cal L}=\frac{1}{4g^2}{\rm Tr}\left(\int d^2\theta\, W^\alpha W_\alpha 
+\int d^2\bar\theta\, \bar{W}_{\dot\alpha}\bar{W}^{\dot\alpha}\right)\,.
$$
However, we are interesed in the supersymmetric analogue of the
Lagrangian (\ref{LXXVIII}) which also contains a $\theta$-term.
From (\ref{WW-nonabelian}), it is obvious that the super Yang-Mills
Lagrangian with a $\theta$-term can be written as 
\bea
{\cal L}&=&\frac{1}{8\pi}{\rm Im}\left(\,\tau\,{\rm Tr}\int d^2\theta\,
W^\alpha W_\alpha \right) \nonumber\\
&=&-\frac{1}{4g^2} F^a_{\mu\nu}F^{a\mu\nu}+ \frac{\theta}{32\pi^2}  
F^a_{\mu\nu}\tilde F^{a\mu\nu}+ \frac{1}{g^2}(\frac{1}{2}D^a D^a 
-i\lambda^a\sigma^\mu D_\mu\bar{\lambda}^a) \,,
\label{tauWW}
\eea
where, $\tau = \theta/2\pi + 4\pi i/g^2$. Note that $\tau$ can be
regarded as a constant chiral superfield.  

\noindent\underline{Interaction Terms and the General $N=1$
  Lagrangian:}
Let the chiral superfields $\Phi_i$ belong to a given
representation of the gauge group in which the generators are
the matrices $T^a_{ij}$. The kinetic energy term
$\Phi_i^\dagger\Phi_i$ is invariant under global gauge transformations
$\Phi^\prime=e^{-i\Lambda} \Phi$. In the local case, to insure that
$\Phi^\prime$ remains a chiral superfield, $\Lambda$ has to be a
chiral superfield. The supersymmetric gauge invariant kinetic energy
term is then given by  $\Phi^\dagger e^{-2V}\Phi$. We are now in a
position to write down the full N=1 supersymmetric Lagrangian as 
\beq
{\cal L}=\frac{1}{8\pi}\,{\rm Im}\left(\tau{\rm Tr}\int d^\theta
\,W^\alpha W_\alpha \right)+\int d^2\theta d^2\bar\theta\,\Phi^\dagger 
e^{-2V}\Phi+ \int d^2\theta\,{\cal W}+\int d^2\bar\theta\,\bar{\cal W}\,.     
\label{fullSF}
\eeq
Note that since each term is separately invariant, the relative
normalisation between the scalar part and the Yang-Mills part is
not fixed by $N=1$ supersymmetry. In the above, we have set the
normalization of the scalar part to one, but later, we will change
this by rescaling the scalar multiplet $\Phi$. In terms of component
fields, the above Lagrangian takes the form  
\bea
{\cal L}\,=&-&\frac{1}{4g^2}F^a_{\mu\nu}F^{a\mu\nu}+\frac{\theta} 
{32\pi^2}F^a_{\mu\nu}\widetilde F^{a\mu\nu}
-\frac{i}{g^2} \,\lambda^a \sigma^\mu D_\mu \bar{\lambda}^a 
+\frac{1}{2g^2}D^a D^a  \nonumber\\
&+& (\partial_\mu A-iA^a_\mu T^a A)^\dagger(\partial^\mu A-iA^{a\mu} 
T^a A) -i\,\bar{\psi} 
\bar{\sigma}^\mu (\partial_\mu\psi -iA^a_\mu T^a \psi)\nonumber\\
&-& D^a A^\dagger T^a A -i\sqrt{2}\,A^\dagger T^a \lambda^a
\psi+i\sqrt{2}\, \bar{\psi} T^a A \bar{\lambda}^a + F^\dagger_i F_i
\nonumber\\ 
&+&\frac{\partial {\cal W}}{\partial A_i} \, F_{i} +
\frac{\partial \bar{\cal W}}{\partial A^\dagger_i} \, F^\dagger_{i} -
\frac{1}{2}\,\frac{\partial^2 {\cal W}}{\partial A_i\partial A_j}\,
\psi_{i}\psi_{j}-\frac{1}{2}\,\frac{\partial^2 \bar{\cal W}} 
{\partial A_i^\dagger\partial
  A_j^\dagger}\,\bar{\psi}_{i}\bar{\psi}_{j}\,. 
\label{fullC}
\eea
Here, ${\cal W}$ denotes the scalar component of the superpotential.
The auxiliary fields $F$ and $D^a$ can be eliminated by using their
equations of motion. The terms involving these fields, thus, give rise 
to the scalar potential  
\beq
V = \sum_i\,\left|\frac{\partial{\cal W}}{\partial A_i}\right|^2
-\frac{1}{2} g^2 (A^\dagger T^a A)^2\,.
\label{Spot}
\eeq

\subsection{The $N=2$ Supersymmetric Lagrangian for Gauge Fields}

The on-shell $N=1$ scalar multiplet $(A, \psi)$ and vector multiplet
$(A_\mu,\lambda)$, put together, have the same field content as the
on-shell $N=2$ vector multiplet $(A,\psi,\lambda, A_\mu)$. The
Lagrangian (\ref{fullC}) contains all these fields but as such is not
$N=2$ supersymmetric. We now assume that $(A,\psi,\lambda, A_\mu)$
form an $N=2$ vector multiplet and discuss the restrictions which this
assumtion imposes on the $N=1$ Lagrangian in (\ref{fullC}). First,
since $A^a_\mu$ and $\lambda^a$ belong to the adjoint representation
of the gauge group, $A_i$ and $\psi_i$ should also belong to the same
representation if they are to be part of the same multiplet. Hence,
$T^a_{ij}=-if^a_{ij}$ and the sets of indices $\{ i \}$ and $\{ a \}$
coincide. Second, since the two supersymmetry generators in the $N=2$
algebra appear on the same footing, the same must be the case with the
fermions $\psi^a$ and $\lambda^a$ in (\ref{fullC}). To satisfy this
condition, we set the superpotential ${\cal W}$ to zero since it
couples only to $\psi^a$. This condition also fixes the arbitrary
relative normalization between the Yang-Mills part and the scalar part
of the Lagrangian since it requires that the kinetic energy terms for
both fermions should have the same normalization.  This is achieved by
scaling $\Phi\rightarrow\Phi/g$ in (\ref{fullC}).  It turns out that
if the Lagrangian (\ref{fullC}) satisfies these conditions, then it
has $N=2$ supersymmetry. The terms containing the auxiliary fields now
take the form
$$
\frac{1}{g^2}{\rm Tr}\left(\frac{1}{2}DD +D\,[A^\dagger\,,\,A] + 
F^\dagger F \right)\,, 
$$
where, we have used the notation $\Phi=\Phi^a T^a$, with $T^a$ in the
fundamental representation. On eliminating $D$ and $F$, we get the
scalar potential    
\beq
V=- \frac{1}{2g^2} {\rm Tr}\left([A^\dagger\,,\,A]^2 \right)\,.
\label{scalarV}
\eeq
The full Lagrangian with $N=2$ supersymmetry can now be written as  
\bea
{\cal L}&=&\frac{1}{8\pi}{\rm Im\, Tr}\left[\,\tau\left(\int d^2\theta\,
W^\alpha W_\alpha +2\,\int d^2\theta d^2\bar\theta\, \Phi^\dagger 
e^{-2V}\Phi \right)\right] 
\nonumber \\ 
&=&\frac{1}{g^2}{\rm Tr}\Big(-\frac{1}{4}F_{\mu\nu}F^{\mu\nu}+ 
g^2\frac{\theta}{32\pi^2}F_{\mu\nu}\widetilde F^{\mu\nu}+
(D_\mu A)^\dagger D^\mu A -\frac{1}{2} [A^\dagger\,,\,A]^2
\nonumber\\
&&{\hphantom {\frac{1}{g^2}{\rm Tr}(\,} }
-i\,\lambda\sigma^\mu D_\mu\bar\lambda -i\,\bar{\psi} 
\bar{\sigma}^\mu D_\mu\psi-i\sqrt{2}\,[\lambda,\psi]\,A^\dagger
-i\sqrt{2}\,[\bar\lambda,\bar\psi]\,A \Big)\,,
\label{N=2}
\eea
where, in the component expansion, the auxiliary fields have been
eliminated. 
Note that (\ref{N=2}) is the supersymmetric generalization of the
Yang-Mills-Higgs theories described in section 1, with the Higgs
potential $V$ given by (\ref{scalarV}), and with a $\theta$-term.    
Therefore, the discussion in section 1 also applies to our $N=2$
supersymmetric theory. In particular, the Higgs vacuum is defined by
$D_\mu A=0, V=0$, and the potential vanishes for non-zero field
configurations provided $A$ commutes with $A^\dagger$. Thus the model
admits monopole and dyonic solutions and 
contains massive gauge bosons. For example, if the gauge group is
$SU(2)$ or $SO(3)$, it is broken down to $U(1)$ and two of the gauge
bosons become massive. 

Now, suppose that we are interested in the behaviour of this model at
energies lower than some cutoff $\Lambda$ which is smaller than the
mass of the lightest massive state in the theory. At such energies, we
will not encounter any on-shell massive states and the physics can be
described by the Wilsonian low-energy effective action. This effective
action is obtained by completely integrating out all massive states as
well as integrating out all massless excitations above the scale
$\Lambda$ \cite{SV1,SV2}. This, in general, is a complicated procedure and
cannot be carried out explicitly. Fortunately, in our model, the
general form of the Wilsonian effective action is severely constrained
by $N=2$ supersymmetry. This is easiest to see when the theory
(\ref{N=2}) is formulated in term of $N=2$ superfields as described
below. 
 
\noindent\underline{The $N=2$ Superspace Formulation:}
The $N=2$ superspace is obtained by adding four more fermionic degrees
of freedom, say, $\tilde\theta$ and $\bar{\tilde\theta}$, 
to the $N=1$ superspace. Thus, a generic $N=2$ superfield can be
written as $F(x,\theta,\bar\theta,\tilde\theta,\bar{\tilde\theta})$. 
We need a superfield which has the same components as the $N=2$ vector
multiplet. This is obtained by imposing the constraints of chirality
and reality on a general $N=2$ superfield \cite{Grimm}. We briefly
describe this following the approach of \cite{bilal}. An $N=2$ chiral
superfield $\Psi$ is defined by the constraints $\bar
D_{\dot\alpha}\Psi=0$ and $\bar{\tilde D}_{\dot\alpha}\Psi=0$, where,
the supercovariant derivative $\bar{\tilde D}_{\dot\alpha}$ is defined
in the same way as $\bar D_{\dot\alpha}$ with $\theta$ replaced by
$\tilde\theta$. The expansion of $\Psi$ can be arranged in powers of
$\tilde\theta$ and can be written as
$$
\Psi=\Psi^{(1)}(\tilde y,\theta)+\sqrt{2}\tilde\theta^\alpha
\Psi^{(2)}_\alpha(\tilde y, \theta) + \tilde\theta^\alpha
\tilde\theta_\alpha \Psi^{(3)}(\tilde y, \theta)\,,
$$
where, $y^\mu =x^\mu +i\theta\sigma^\mu\bar\theta +i\tilde\theta
\sigma^\mu\bar{\tilde\theta}$. This expansion helps us relate the
$N=2$ formalism to the $N=1$ language we have been using so far.
Clearly, the component $\Psi^{(1)}$ has the same form as the $N=1$
chiral superfield $\Phi$. The remaining two components are constrained
by a reality condition. The outcome is that $\Psi^{(2)}_\alpha=
W_\alpha (\tilde y, \theta)$ as given in (\ref{W-nonabelian}), and
$\Psi^{(3)}$ is given by
$$
\Psi^{(3)}(\tilde y,\theta)= \Phi^\dagger(\tilde y-i\theta\sigma 
\bar\theta, \theta,\bar\theta) exp \left[2gV(\tilde y -i\theta\sigma
\bar\theta, \theta,\bar\theta\right]\,\mid_{\bar\theta\bar\theta}\,.
$$
Here, $\Phi(\tilde y -i\theta\sigma\bar\theta,\theta,\bar\theta)$ is
to be understood as the expansion in (\ref{csfield}). Clearly, $\Psi$
has the same field content as the $N=2$ vector multiplet. One can
verify that in terms of the $N=2$ superfield $\Psi$, the $N=2$
Lagrangian (\ref{N=2}) is given by the compact expression
\beq
{\cal L}= \frac{1}{4\pi}{\rm Im\,Tr}\int d^2\theta d^2\tilde\theta\,
\frac{1}{2}\tau\Psi^2\,.
\label{N=2manifest}
\eeq

Using the $N=2$ chiral superfield $\Psi$ (also referred to as the
$N=2$ vector superfield), we can now construct the most general
$N=2$ Lagrangian for the gauge fields: Corresponding to any function
${\cal F}(\Psi)$, we can construct a Lagrangian  
\bea
{\cal L}&=& \frac{1}{4\pi}{\rm Im\,Tr}\int d^2\theta d^2\tilde\theta
\,{\cal F}(\Psi) 
\nonumber\\
&=&\frac{1}{8\pi}{\rm Im}\left(\int d^2\theta\, {\cal F}_{ab}(\Phi) 
W^{a\alpha}W^b_\alpha + 2\int d^2\theta d^2\bar\theta\, (\Phi^\dagger 
e^{2gV})^a {\cal F}_a(\Phi)\right)\,. 
\label{N=2general}
\eea
Here, ${\cal F}_a(\Phi)=\partial{\cal F}/\partial\Phi^a$,
${\cal F}_{ab}(\Phi)=\partial^2{\cal F}/\partial\Phi^a\partial\Phi^b$
and ${\cal F}$ is referred to as the $N=2$ prepotential. From the
above, we can easily read off the K\"{a}hler potential as
${\rm Im}(\Phi^{\dagger a} {\cal F}_a(\Phi)$. This gives rise to a metric 
$g_{ab}={\rm Im}(\partial_a\partial_b{\cal F})$ on the space of fields. 
A metric of this form is called a special K\"{a}hler metric.
If we demand renormalisability, then ${\cal F}$ has to be quadratic in
$\Psi$ as in (\ref{N=2manifest}). However, if we want to write a
low-energy effective action, then renormalisability is not a criterion
and ${\cal F}$ can have a more complicated form. In particular, we can
start from the microscopic theory (\ref{N=2}), corresponding to a
quadratic prepotential, and try to construct the modified ${\cal F}$
for the low-energy Wilsonian effective action. The exact determination
of this function is the subject of the work of Seiberg and Witten.

\subsection{The $N=2$ Supersymmetric Lagrangian for Matter Fields}

Since matter fields and gauge fields transform under different
representations of the gauge group, they cannot be part of the same
multiplet. The $N=2$ matter supermultiplet is called the
hypermultiplet and contains one pair of complex scalars and one pair
of two-component spinors, all transforming under the same
representation of the gauge group. From our discussion of the
representations of the supersymmetry algebra it follows that if a
hypermultiplet is massive, then its mass should appear as a central
extension in the supersymmetry algera. This is so because otherwise
$N=2$ supersymmetry requires a larger number of components than is
contained in a hypermultiplet. In $N=1$ notation, a hypermultiplet
contains a chiral superfield $Q$ and an anti-chiral superfield
$\widetilde Q^\dagger$, both transforming under the same representatin
$N_c$ of the gauge group $SU(N_c)$ We denote the components of $Q$ and
$\widetilde Q$ by $(q,\psi_q,F_q)$ and $(\widetilde q,
\psi_{\widetilde q},F_{\widetilde q})$ respectively. The form of the
Lagrangian for $N_f$ hypermultiplets (labelled by an index $i$),
interacting with a $N=2$ vector multiplet, can be partly inferred from
the $N=1$ theories and is given by
\bea
{\cal L}=&&\int d\theta^4\left( Q^\dagger_i e^{-2V}Q_i +\widetilde Q_i
e^{2V}\widetilde Q^\dagger_i\right)+\int
d\theta^2\left(\sqrt{2}\widetilde Q_i\Phi Q_i+m_i\widetilde Q_i
Q_i\right)+\,h.c.\nonumber\\       &&+\cdots
\label{N=2hyper}
\eea
Here, the dots represent the Lagrangian for the pure $N=2$ vector
multiplet and we have supressed the gauge group indices. The term  
$\sqrt{2}\widetilde Q_i\Phi Q_i$ is related, by $N=2$ supersymmetry, to
the coupling of the hypermultiplet with the $N=1$ vector multiplet
$V$. The presence of this term is not required in an $N=1$ Lagrangian.  
As in normal QCD, when all masses $m_i$ are equal, the theory is
invariant under the global flavour group $SU(N_f)$. 

Eliminating the auxiliary fields $F_q$ and $F_{\widetilde q}$, which
appear in the hypermultiplet, results in a contribution to the scalar
potential given by (compare with (\ref{Spot}))
$$
V=\frac{1}{2}g^2\sum_a D_a D^a\,,
$$
with
\beq
D^a=\sum_{i=1}^{N_f}\left(q^\dagger_i\lambda^a q_i -\widetilde
q_i\lambda^a \widetilde q_i^\dagger\right)\,.
\label{D-term}
\eeq
Here, $\lambda^a$ are the gauge group generators in the fundamental
representaion. This term is referred to as the $D$-term. It must be
mentioned that the $N=2$ algebra has a global $SU(2)_R$ symmetry which
should also be a symmetry of the Lagrangian. However, in our
decomposition of the hypermultiplet in terms of $Q$ and $\widetilde
Q^\dagger$, this symmetry is not manifest since it rotates the scalar
components $q$ and $\widetilde q^\dagger$ as a doublet. It is not
difficult to write an $N=2$ Lagrangian with manifest $SU(2)_R$
symmetry \cite{West}. 
 
\noindent\underline{Minimization of the $D$-term}:
To find the vacuum of the theory, now we also have to minimize the
$D$-term contribution and the hypermultiplet mass term contribution to
the scalar potential \cite{DDS,ADS}. For non-zero quark masses, the
only solution is 
$q=\widetilde q =0$. Thus, only the scalar $A$ can have a non-zero
vacuum expectation value. But when $m_i=0$, then the $D$-term can be
minimized for a set of non-vanishing $q$ and $\widetilde q$ (the
potential has flat directions). However, the $\widetilde Q\Phi Q$ term
in the Lagrangian now requires $A_{vac}=0$. In the following, we
determine these flat directions.

The minimum of the $D$-term corresponds to $D^a=0$. To solve this,
note that $q^{(i)}_\alpha$ (where $\alpha$ denotes a colour index) can
be regarded as a set of $N_c$ vectors in ${\bf C}^{N_f}$. Thus, we can
construct the matrix of scalar products, 
$$
\sum_i q^{(i)}_\alpha q^{\dagger (i)}_\beta=q_\alpha\cdot
q^\dagger_\beta=(qq^\dagger)_{\alpha\beta}\,.
$$
In terms of this,
$$
\sum_i q^\dagger_i\lambda^aq_i={\rm Tr}\,\left(qq^\dagger
\lambda^a\right) \,.
$$
Similarly, we construct the matrix $\widetilde q_\alpha^\dagger \cdot
\widetilde q_\beta = (\widetilde q^\dagger\widetilde q)_{\alpha\beta}$.
Hence, $D^a=0$ can be written as 
$$
D^a={\rm Tr}\left[\left(qq^\dagger-\widetilde q^\dagger
\widetilde q\right)\,\lambda^a\right]=0 \,.
$$
Since $\lambda^a$ are in an irreducible representation of $SU(N_c)$,
the above condition is solved by 
\beq
(qq^\dagger) - (\widetilde q^\dagger\widetilde q)=c^2{\bf 1}_{N_c}\,.
\label{D=0} 
\eeq
Here, we can distinguish two cases: $N_f < N_c$ and $N_f\geq N_c$.

\noindent 1) $N_f < N_c$: In this case, the matrix $(qq^\dagger)$ has
rank $N_f$ and thus it has $N-N_f$ zero eigenvalues. The same applies to
the matrix $(\widetilde q^\dagger\widetilde q)$. By using $SU(N_c)\times
SU(N_f)\times U(1)_R$ rotations, $(qq^\dagger)$ can be diagonalized and
condition (\ref{D=0}) then implies that $(\widetilde
q^\dagger\widetilde q)$ must also be diagonal. Hence, in this basis, we
can solve for $q$ and $\widetilde q$ as
$$
q = \left(
\begin{array}{cccc}
v^{(1)}_1 & 0 &\cdots&0\\
0 & v^{(2)}_2 &&\\
\vdots&&\ddots & \\
&&&v^{(N_f)}_{N_f}\\
0&0&\cdots&0\\
\vdots&\vdots&&\vdots
\end{array}
\right)\,, \quad
\widetilde q = \left(
\begin{array}{cccccc}
\tilde v^{(1)}_1 & 0 & \cdots & & 0 & \cdots \\
 0 & \tilde v^{(2)}_2 & & & 0 & \cdots \\
\vdots  & & \ddots &  &   &  \\
 0 & & &\tilde v^{(N_f)}_{N_f} & 0 & \cdots  \\
\end{array}
\right)\,.
$$
The presence of zero eigenvalues imply that $c = 0$ and therefore,
$v^{(i)}_i =\tilde{v}^{(i)}_i$. In this case, the gauge symmetry is
broken down to $SU(N_c-N_f)$ except for $N_c=N_f-1$, where it is
totally broken. $2N_fN_c-N^2_f$ quark superfields become heavy and the
remaining $N^2_f$ quark superfields remain massless and correspond to
the Goldstone bosons of broken global symmetries.  

\noindent 2) $N_f \leq N$:  In this case $(qq^\dagger)$
has rank $N_c$ and, generically, its eigenvalues are not zero. As a
result, $c\not=0$. Arguing as above, in this case the vacuum values
of $q$ and $\widetilde q$ can be written as  
$$
q = \left(
\begin{array}{cccccc}
v^{(1)}_1 & 0 & \cdots & & 0 & \cdots \\
 0 & v^{(2)}_2 & & & 0 & \cdots \\
\vdots  & & \ddots &  &   &  \\
 0 & & & v^{(N_c)}_{N_c} & 0 & \cdots  \\
\end{array}
\right)\,,\quad
\widetilde q = \left(
\begin{array}{cccc}
\tilde v^{(1)}_1 & 0 &\cdots&0\\
0 &\tilde v^{(2)}_2 &&\\
\vdots&&\ddots & \\
&&&\tilde v^{(N_c)}_{N_c}\\
0&0&\cdots&0\\
\vdots&\vdots&&\vdots
\end{array}
\right)\,.
$$
and
$$
\tilde{v}_{i}^{(i)} = \sqrt{|v_{i}^{(i)}|^2 - c^2} \, .
$$
In this case, the gauge group is completely broken, and depending
on the values of $v_{i}^{(i)}$, the pattern of chiral symmetry
breaking quite complicated with many possibilities.
For $N_f > N$, there are surviving $R$-symmetries. 

\subsection{Central Charges in the $N=2$ Pure Gauge Theory}

We have seen that the $N=2$ supersymmetry algebra with central charge
$Z$ implies the bound $M\ge \sqrt{2}|Z|$ on the particle masses. It
was also stated that this bound is the same as the BPS bound on the
masses which is determined in terms of the electric and magnetic
charges. In this section, we prove the above statement by explicitly
calculating $Z$ following \cite{Olive-Witten}.  

In the supersymmetry algebra, The central charge $Z$ appears in the
commutator of the supercharges $Q^I_\alpha$ which, in turn, are space
integrals of $S^{I0}_\alpha$ (Here, $S^{I\mu}_\alpha$ denotes the
supercurrent). Thus, we first have to compute $S^{I0}$'s in terms of
the basic fields, and then evaluate their commutators. The central
charge is related to a surface term in the space integral of this
commutator which is non-zero if the field configuration corresponds to
electric and magnetic charges.
  
As a warm up exercise, we start with the theory for $N=1$ chiral
superfields. The Lagrangian is given by 
$$
{\cal L}=\int d^4 \theta\,\Phi^\dagger\Phi +\int d^2 \theta \;
{\cal W}(\Phi) + \int d^2 \bar\theta \bar{\cal W}(\Phi^\dagger)\,.
$$
Defining $y^\mu=x^\mu + i \theta \sigma^\mu \bar{\theta}$, the
superfield can be expanded as $\Phi=A(y)+\sqrt{2}\,\theta\psi(y)
+\theta\theta F(y)$. In this basis, $Q_{\alpha}=\partial /
\partial\theta^{\alpha}$, $\bar{Q}_{\dot{\alpha}}=-
\partial/\partial\bar{\theta}^{\dot{\alpha}} + 2i
(\theta \sigma^\mu)_{\dot{\alpha}}\,\partial/\partial y^\mu$, and the
supersymmetry variations, $\delta_{\epsilon}=\epsilon^\alpha Q_\alpha +
\bar{\epsilon}_{\dot{\alpha}} \bar{Q}^{\dot{\alpha}}$, of the fields
are given by 
$$
\begin{array}{ll}
\delta A=\sqrt{2} \epsilon \psi\,,&
\delta\bar A=\sqrt{2} \bar{\epsilon}\bar{\psi}\,, \\
\delta\psi=\sqrt{2}\epsilon F + i\sqrt{2}\sigma^\mu
\bar{\epsilon} \partial_\mu A\,, &
\delta\bar{\psi}=-i\sqrt{2}\epsilon\sigma^\mu
\partial_\mu \bar{A}+\sqrt{2} \bar{F} \bar{\epsilon}\,, \\
\delta F=i\sqrt{2} \bar{\epsilon}\bar{\sigma^\mu}
\partial_\mu \psi\,, & \delta \bar{F}=i\sqrt{2}\epsilon 
\sigma^\mu \partial_\mu \bar{\psi}\,.
\end{array}
$$
Using these transformations, we can compute the variation of the terms
involving the superpotential ${\cal W}$ as  
\bea
\delta ({\cal W}'F - \frac{1}{2} \frac{\partial^2 {\cal W}}{\partial
  A^2}\psi\psi) 
&=& {\cal W}''\delta AF+{\cal W}'\delta F-\frac{1}{2}{\cal W}'''\delta
A\psi\psi+{\cal W}'' \psi \delta \psi \nonumber\\
&=& {\cal W}''\sqrt{2}\epsilon\psi F+i\sqrt{2}{\cal W}'\bar{\epsilon}
\bar{\sigma}^\mu \partial_\mu \psi+\sqrt{2}{\cal W}''\psi(i\sigma^\mu 
\bar{\epsilon}\partial_\mu A + \epsilon F) \nonumber\\
&=&\partial_\mu \left( i \sqrt{2} \frac{\partial {\cal W}}{\partial A}
\bar{\epsilon} \bar{\sigma}^\mu \psi \right )\,.\nonumber
\eea
Note that we have used ${\cal W}'''\delta A\psi\psi={\cal W}'''
\sqrt{2}(\epsilon \psi)\psi\psi=0$. Since ${\cal W}'''$ is totally
symmetric, this statement is also true in the presence of many fields 
$\psi^i$. Thus, we have, 
$$
\delta (\int d^{2} \theta {\cal W} + \int d^{2} \bar{\theta}
\bar{{\cal W}})=i\sqrt{2}\partial_\mu\left(\frac{\partial {\cal W}}
{\partial A_j} \bar{\epsilon}\bar{\sigma}^\mu \psi^{j} -
\frac{\partial \bar{{\cal W}}}{\partial \bar{A}_j} \bar{\psi}_{j}
\bar{\sigma}^\mu \epsilon \right)\,.
$$
Next we calculate the variation of the kinetic terms,
$$
{\cal L}_{D} = \partial_\mu A^\dagger \partial^\mu A +
F^\dagger F-\frac{i}{2}\bar{\psi} \bar{\sigma}^\mu\partial_\mu\psi+
\frac{i}{2}\partial_\mu\bar{\psi}\bar{\sigma}^\mu\psi\,.
$$
A slightly lengthy computation yields:
\bea
\delta {\cal L}_{D}=&-&\frac{i}{\sqrt{2}}\partial_\mu (F\bar{\psi}
\bar{\sigma}^\mu \epsilon ) +
\frac{i}{\sqrt{2}} \partial_\mu (\epsilon\psi\partial^\mu A^\dagger - 
\epsilon \sigma^{\nu\mu}\psi \partial_\nu A^\dagger)\nonumber \\
&+& \frac{i}{\sqrt{2}}\partial_\mu (F^\dagger \bar{\epsilon}
\bar{\sigma}^\mu\psi)+\frac{i}{\sqrt{2}}\partial_\mu (\bar{\epsilon} 
\bar{\psi}\partial^\mu A-\bar{\psi}\bar{\sigma}^{\mu\nu}\bar{\epsilon}
\partial_\nu A )\,.\nonumber
\eea
Adding the variations of the superpotential and the kinetic terms, and
using the definitions 
$$
\sigma^{\mu\nu}=\frac{1}{2} (\sigma^\mu \bar{\sigma}^\nu -
\sigma^\nu \bar{\sigma}^\mu )\,,\quad
\bar{\sigma}^{\mu\nu} = \frac{1}{2} (\bar{\sigma}^\mu \sigma^\nu
- \bar{\sigma}^\nu \sigma^\mu ) \,,
$$
we get the supercurrent as 
\beq
S^\rho_{matter}=\sqrt{2}\epsilon \sigma^\nu\bar{\sigma}^\rho \psi
\partial_\nu A^\dagger + i \sqrt{2}\frac{\partial\bar{{\cal W}}}
{\partial A^\dagger}\bar{\psi}\bar{\sigma}^\rho\epsilon\nonumber 
+ \sqrt{2}\bar{\psi} \bar{\sigma}^\rho\sigma^\nu\bar{\epsilon}
\partial_\nu A - i\sqrt{2}\frac{\partial {\cal W}}{\partial A} 
\bar{\epsilon} \bar{\sigma}^\rho \psi\,. 
\label{N=1scch}
\eeq
Note that for convenience, we have included the supersymmetry
transformation parameters $\epsilon,\bar\epsilon$ in the definition of
the supercurrent. So, expanding in components, we actually have:
\beq
S^\rho =\epsilon^\alpha S^\rho_\alpha + \bar\epsilon_{\dot\alpha}
\bar S^{\rho\dot\alpha}\,.
\label{S-comp}
\eeq
 
Next we consider the inclusion of gauge fields which are described by
a vector multiplet. The supersymmetry variations of the fields in this
multiplet are given by 
\beq
\begin{array}{lll}
\delta A^a_\mu &=& -i\bar{\epsilon}\bar{\sigma}_\mu\lambda^a +
i \bar{\lambda}^a \bar{\sigma}_\mu\epsilon \,,\\
\delta D^{a} &=&\bar{\epsilon}\bar{\sigma}^\mu D_\mu \lambda^a +
D_\mu \lambda^{a} \bar{\sigma}^\mu \epsilon \,,\\
\delta\lambda^a &=& \frac{1}{2} \sigma^{\mu\nu}\epsilon F^a_{\mu\nu} 
+ i\epsilon D^a \,,\\
\delta\bar{\lambda}^a &=&\frac{1}{2}\bar{\epsilon}\bar{\sigma}^{\nu
\mu} F^a_{\mu\nu} - i \bar{\epsilon} D^a \,.
\end{array}
\label{str1-g}
\eeq
Furthermore, in the presence of the gauge interactions, the matter
field transformations need some modification: 
\beq
\begin{array}{lll}
\delta A &=& \sqrt{2} \epsilon \psi \,,\\
\delta \psi &=& \sqrt{2} \epsilon F + i \sqrt{2} \sigma^\mu
\bar{\epsilon} D_\mu A \,,\\
\delta\bar{\psi}&=&\sqrt{2}\bar{\epsilon} F^\dagger -i\sqrt{2}
\epsilon \sigma^\mu D_\mu A^\dagger \,,\\
\delta F &=& i\sqrt{2}\bar{\epsilon} \bar{\sigma}^\mu D_\mu \psi - 
2i T^a A \bar{\epsilon} \bar{\lambda}^a\,.
\end{array}
\label{str1-m}
\eeq
The last term in $\delta F$ is needed to cancel part of the variation
of $\psi$ in the term $A^\dagger T^a\lambda^a\psi$ in the Lagrangian
(\ref{fullC}). The part of the Lagrangian describing the pure vector
superfieled can be written as 
$$
{\cal L}=\frac{1}{g^2}\left(-\frac{1}{4} F^a_{\mu\nu} F^{a\mu\nu}-
\frac{i}{2}\bar\lambda^a\bar\sigma^\mu D_\mu\lambda^a+\frac{i}{2}
D_\mu \bar\lambda^a\bar\sigma^\mu\lambda^a+\frac{1}{2}D^{2} \right) + 
\frac{\theta}{32 \pi^2} F^a_{\mu\nu} \tilde{F}^{a\mu\nu} \,.
$$
If we ignore the $\theta F \tilde{F}$ term, then the supercurrent
obtained from this part of the Lagrangian takes the form
$$
S^\rho_{gauge}=-\frac{i}{2g^2}\left(\bar{\lambda}^a \bar{\sigma}^\rho
\sigma^{\mu\nu}\epsilon F^a_{\mu\nu}+\bar{\epsilon}\bar{\sigma}^{\mu\nu} 
\bar{\sigma}^\rho \lambda^a F^a_{\mu\nu}\right)\,.
$$
For the theory coupled to matter, the supercurrent is not just the sum
of the above, plus the gauge-covariantized version of $S^\rho_{matter}$. 
We also expect some contribution from the terms $D^a A^\dagger T^a A$
and $i A^\dagger T^a \lambda^a\psi+h.c.$. Taking this into account,
the supercurrent for the interacting $N=1$ Lagrangian (\ref{fullC}) is
given by 
\bea
S^\rho = &-&\frac{i}{2g^2}\left(\bar{\lambda}^a \bar{\sigma}^\rho
\sigma^{\mu\nu}\epsilon +\bar{\epsilon}\bar{\sigma}^{\mu\nu} 
\bar{\sigma}^\rho \lambda^a\right) F^a_{\mu\nu} -
\left(\bar{\epsilon} \bar{\sigma}^\rho \lambda^a + \bar{\lambda}^a 
\bar{\sigma}^\rho \epsilon\right) A^\dagger T^a A
\nonumber\\
&+& \sqrt{2}\epsilon\sigma^\mu \bar{\sigma}^\rho \psi D_\mu A^\dagger
+\sqrt{2}\bar{\psi}\bar{\sigma}^\rho\sigma^\mu\bar{\epsilon}D_\mu A
+i\sqrt{2}\frac{\partial \bar{\cal W}}{\partial A^\dagger}\bar{\psi}
\bar{\sigma}^\rho\epsilon-i\sqrt{2}\frac{\partial{\cal W}}{\partial A}
\bar{\epsilon}\bar{\sigma}^\rho \psi\,.
\label{N=1sc}
\eea

Now, let us turn our attention to the pure $N=2$ gauge theory. As
discussed in section 1, this is obtained from the $N=1$ theory
(\ref{fullC}) by setting ${\cal W}=0$ and by scaling the chiral superfield
$\Phi$ to $\Phi/g$. Furthermore, $\Phi$ is now a vector in the adjoint
representation of the gauge group. Thus, relabelling $(\lambda,\psi)$
as $(\lambda_1,\lambda_2)$, we can easily write down one of the $N=2$ 
supercurrents as  
\bea
g^2 S^\rho_{(1)}=&-&\frac{i}{2}\left(\bar{\lambda}^a_1\bar{\sigma}^\rho
\sigma^{\mu\nu}\epsilon + \bar{\epsilon}\bar{\sigma}^{\mu\nu}
\bar{\sigma}^\rho \lambda^a_1\right) F^a_{\mu\nu} -
\left(\bar{\epsilon}\bar{\sigma}^\rho\lambda^a_{1}+\bar{\lambda}^a_{1}
\bar{\sigma}^\rho \epsilon\right) A^\dagger T^a A 
\nonumber \\
&+&\sqrt{2}\,\epsilon\sigma^\mu\bar{\sigma}^\rho\lambda^a_{2}\, D_\mu
A^{a\dagger}+\sqrt{2}\,\bar{\lambda}^a_{2}\bar{\sigma}^\rho \sigma^\mu
\bar{\epsilon}\, D_\mu A^a \,.\nonumber 
\eea

The $N=2$ theory is also invariant under a second set of supersymmetry
transformations (with parameter $\epsilon'$) which is obtained from
(\ref{str1-g}) and (\ref{str1-m}) by the replacement
$\lambda\rightarrow\psi$, $\psi\rightarrow -\lambda$. This corresponds
to a transformation of the ($\lambda,\psi$) doublet by an element of
$SU(2)_R$, which is a symmetry group of the $N=2$ algebra.  The
associated conserved current is then obtained from $S^\rho_{(1)}$ by
the replacement $\lambda_1 \rightarrow\lambda_2$,
$\lambda_2\rightarrow-\lambda_1$:
\bea
g^2 S^\rho_{(2)}=&-&\frac{i}{2}\left(\bar{\lambda}^a_2\bar{\sigma}^\rho
\sigma^{\mu\nu}\epsilon' + \bar{\epsilon'}\bar{\sigma}^{\mu\nu}
\bar{\sigma}^\rho \lambda^a_2 \right) F^a_{\mu\nu} -
\left(\bar{\epsilon'}\bar{\sigma}^\rho\lambda^a_{2}+\bar{\lambda}^a_{2}
\bar{\sigma}^\rho \epsilon'\right) A^\dagger T^a A 
\nonumber \\
&-&\sqrt{2}\,\epsilon'\sigma^\mu \bar{\sigma}^\rho\lambda^a_{1}\,D_\mu
A^{a\dagger}-\sqrt{2}\,\bar{\lambda}^a_{1}\bar{\sigma}^\rho \sigma^\mu
\bar{\epsilon'}\, D_\mu A^a \,.\nonumber 
\eea
Let us first concentrate on $S^\mu_{(1)}$. Using the identities 
\bea
\sigma^a \bar{\sigma}^b \sigma^c &=& \eta^{ab} \sigma^c - \eta^{ac}
\sigma^b + \eta^{bc} \sigma^a + i \epsilon^{abcd} \sigma_d \,, 
\nonumber\\
\bar{\sigma}^a \sigma^b \bar{\sigma}^c &=& \eta^{ab} \bar{\sigma}^c
- \eta^{ac} \bar{\sigma}^b + \eta^{bc} \bar{\sigma}^a - i
\epsilon^{abcd} \bar{\sigma}_d \nonumber\,,
\eea
along with $\chi\sigma^\mu\bar\psi=-\bar\psi\bar\sigma^\mu\chi$, this
supercurrent can be rewritten as 
\bea
g^2S^\mu_{(1)}=&-&\epsilon\sigma_\nu\bar{\lambda}^a_{1}(iF^{a\mu\nu}+ 
\widetilde F^{a\mu\nu})+\sqrt{2}\,\epsilon\sigma^\nu\bar{\sigma}^\mu
\lambda^a_{2}\,D_\nu A^{\dagger a}+\epsilon\sigma^\mu 
\bar{\lambda}^a_{1} A^\dagger T^a A \nonumber \\
&+& (\mbox{\it $\bar\epsilon$ dependent terms})\,.\nonumber
\eea
From the above, we can easily read off the components
$S^\mu_{(1)\alpha}$ (see eq. (\ref{S-comp})) as
$$
g^2 S^\mu_{(1)\alpha}= \sigma_{\nu\alpha\dot\alpha}
\bar{\lambda}^{a\dot\alpha}_{1} (iF^{a\mu\nu}+\widetilde F^{a\mu\nu})
+\sqrt{2}(\sigma^\nu\bar{\sigma}^\mu\lambda^a_{2})_\alpha D_\nu 
A^{\dagger a}+\sigma^\mu_{\alpha\dot{\alpha}}
\bar{\lambda}^{a\dot{\alpha}}_{1} A^\dagger T^a A\,. 
$$
After lowering the spinor index using   
$$
\bar{\lambda}^{\dot{\alpha}}_{1}=\epsilon^{\dot{\alpha}\dot{\beta}}
\bar\lambda_{1\dot\beta}=i(\sigma_y\lambda^\dagger_1)^{\dot\alpha}\,,
\qquad(\mbox{Note that}\;\bar\lambda_{\dot\beta}=\lambda^\dagger_\beta) 
$$
and using vector notation for spatial components of vectors, 
The $\mu=0$ component of the current takes the form 
\bea
g^2S^0_{(1)\alpha}\,\,=&-&i(\vec{\sigma}\sigma_y\lambda^{\dagger a}_1)_\alpha
\cdot (i\vec{F}^a+\vec{\widetilde F})^a \nonumber\\
&+& \sqrt{2} \lambda^a_{2\alpha} D_0 A^{\dagger a} +
\sqrt{2}(\vec{\sigma}\cdot\vec{D}A^{\dagger a}\lambda^a_{2})_\alpha 
+i(\sigma_y\lambda^{\dagger a}_1)_\alpha A^\dagger T^a A\,, \nonumber
\eea
where,
$$
\vec{F}^a=F^{a 0i}\,,\quad \vec{\widetilde F}^a=\widetilde F^{a0i}\,.
$$
As before, the expression for $S^0_{(2)\alpha}$ is obtained from the
above by the replacements $\lambda_1\rightarrow\lambda_2$, $\lambda_2
\rightarrow-\lambda_1$.

To evaluate the central charge, we are interested in the anti-commutator 
$$
\{Q_{(1)\alpha}\,,\,Q_{(2)\beta}\} =
\left\{\int d^3x\,S^0_{(1)\alpha}(\vec{x},0)\,,\,\int d^3y\,S^0_{(2)
\beta}(\vec{y},0)\right\} \,.
$$
As noticed by Olive and Witten \cite{Olive-Witten}, a non-zero
contribution to this comes from certain boundary terms which measure
electric and magnetic charges. To check this result, we have to look,
in the anti-commutator, for terms of the form $\int d^3 x\partial_i 
(A^{\dagger a} F^{aoi} + A^{\dagger a}\widetilde F^{aoi})$. For this,
we need only retain the relevant terms in the supercurrents:  
\bea
g^2S^0_{(1)\alpha}&=&-i(\vec{\sigma}\sigma_y\lambda^{\dagger a}_1)_\alpha
\cdot (i\vec{F}^a+\vec{\widetilde F})^a +\sqrt{2}\,(\vec{\sigma}\cdot
\vec{D}A^{\dagger a}\lambda^a_{2})_\alpha + \cdots \,,\nonumber\\
g^2S^0_{(2)\alpha}&=&-i(\vec{\sigma}\sigma_y\lambda^{\dagger a}_2)_\alpha
\cdot (i\vec{F}^a+\vec{\widetilde F})^a -\sqrt{2}\,(\vec{\sigma}\cdot
\vec{D}A^{\dagger a}\lambda^a_{1})_\alpha + \cdots\,.
\eea
Using this, we get 
\bea
\{Q_{(1)\alpha}\,,\,Q_{(2)\beta}\}&=&\frac{1}{g^4}\int d^3x\int d^3y\;
\big[i\sqrt{2}(\sigma^i\sigma_y)_{\alpha\gamma}\sigma^j_{\beta\lambda}
\{\lambda^{\dagger a}_{1\gamma},\lambda^b_{1\lambda}\}(iF^a_{0i}+
\widetilde F^a_{0i})D_j A^{\dagger b}     \nonumber\\
&&\qquad\qquad\quad -i\sqrt{2}(\sigma^j)_{\alpha\gamma}(\sigma^i
\sigma_y)_{\beta\lambda}\{\lambda^{\dagger a}_{2\gamma},
\lambda^b_{2\lambda}\}(iF^a_{0i}+\widetilde F^a_{0i})D_j 
A^{\dagger b}\big]      \nonumber \\
&=&\frac{1}{g^2}\int d^3x\,i\sqrt{2}\left[(\sigma^i\sigma_y
\sigma^{j\,T})_{\alpha\beta}-(\sigma^i\sigma_y\sigma^{j\,T})_{\beta\alpha}
\right](iF^a_{0i}+\widetilde F^a_{0i})D_j{A^\dagger}^{a}\,. \nonumber
\eea
The term within the square brackets, involving the $\sigma$ matrices,
can be simplified if we use $\sigma^i\sigma_y=-\sigma_y\sigma^{i\,T}$,
so that
$$
(\sigma^i\sigma_y\sigma^{j\,T})_{\alpha\beta}=-(\sigma_y\sigma^{i\,T} 
\sigma^{j\,T})_{\alpha\beta}=-[\sigma_y(\delta^{ij}-i\epsilon^{ijk}
\sigma_k^T)]_{\alpha\beta} \,.
$$
Subtracting from this a similar equation with $\alpha$ and $\beta$
interchanged, we get the term within the square brackets as equal to
$-2(\sigma_y)_{\alpha\beta}\delta^{ij}=2i\epsilon_{\alpha\beta}
\delta^{ij}$. Thus the commutator takes the form
$$
\{Q_{(1)\alpha},Q_{(2)\beta}\}=-\frac{2\sqrt{2}}{g^2}
\epsilon_{\alpha\beta}\int d^3x (iF^{a0i}+\widetilde F^{a0i}) 
D_i A^{\dagger a} \,.
$$
Using the Bianchi identity for the magnetic part, and the equation of
motion for the electric part, one can easily show that this is the
same as 
$$
\{Q_{(1)\alpha},Q_{(2)\beta}\}=-\frac{2\sqrt{2}}{g^2}
\epsilon_{\alpha \beta}\int d^3x\,\partial_i\left[(iF^{a0i}+
\widetilde F^{a0i}) A^{\dagger a}\right] \,.
$$
Similarly,
$$
\{\bar Q_{(1)\dot{\alpha}},\bar Q_{(2)\dot{\beta}}\}=-\frac
{2\sqrt{2}}{g^2} \epsilon_{\dot{\alpha}\dot{\beta}} \int d^3 x\,
\partial_i \left[(-iF^{a0i}+\widetilde F^{a0i}) A^a \right]\,.
$$
Thus, in the commutator of the supercurrents, we have recovered the
total derivatives which are nothing but the electric and magnetic
charge densities:
$$
Q_{ele}=-\frac{1}{ag}\int d^3x \partial_i(F^{a0i} A^a)= gn_e\,,\qquad 
Q_{mag}=-\frac{1}{ag}\int d^3x \partial_i(\widetilde F^{a0i} A^a)=
\frac{4\pi}{g}n_m \,.
$$
Here, $a$ is the value of $A$ in the Higgs vacuum. The charge
quantization condition used above corresponds to integral fundamantal
charges as in $SU(2)$ breaking to $U(1)$ with fileds in the adjoint
representation. Now, comparing with (\ref{Susy-C'}), we can easily
read off the $N=2$ central charge as $Z=-ia(n_e+ (4\pi
i/g^2)n_m)$. Recall that the phase of $Z$ is convention dependent and
we are mainly interested in its magnitude.  
 
To find the effect of the $\theta$-parameter, we can either modify
the calculation by adding the contribution from $\theta F\widetilde F$,
or simply use the Witten effect described in section (1.10). As we
learnt there, the effect of the $\theta$ parameter is to shift the
electric charge to $Q_{ele}=gn_e+(\theta g^2/8\pi^2)Q_{mag}$. Using
this (and ignoring an overall factor of $-i$), we get the central
charge as  
$$
Z= a\,(n_e + \tau_{cl} n_m)\,,\quad\rm{where},\quad 
\tau_{cl} = \frac{\theta}{2\pi} + \frac{4\pi i}{g^2}\,.
$$
Using supersymmetry algebra this implies a mass bound 
$$
M \geq \sqrt{2}\, |Z|= {\sqrt 2}\,|\,a\,(n_e + \tau_{cl} n_m )\,|\,,
$$
which is the BPS bound. So far, we have considered the microscopic
$N=2$ action (\ref{N=2}). At the level of the effective action
(\ref{N=2general}) specified by a prepotential ${\cal F}$, the central
charge takes the form 
$$
Z = an_e + a_D n_m \,,
$$
where $a_D=\partial{\cal F}/\partial a$. This formula can be
motivated as follows: As we will see in the next section, the theory
(\ref{N=2general}) has a dual description in which the magnetic
monopoles, and not the electric charges, appear as fundamental
objects. In this description, $n_e$ and $n_m$ are interchanged and
also $a$ is replaced by $a_D$. Thus, in the dual theory, one can
easily infer the contribution of the monopole to the BPS bound to be
$a_Dn_m$. The full duality group, as we will see in the next section,
is $SL(2,Z)$ under which $a$ and $a_D$ transform as a
doublet. Combining these facts leads to the central extension formula
given above. Similar constructions apply to arbitrary groups $SU(N)$
with adjoint matter.

\subsection{Central Charge in N=2 Gauge Theory with Matter}

When matter in the fundamental representation is added to the $N=2$
pure gauge theory described above, there is an important change: the
central charge also receives contributions from the masses of the
matter fields. In $N=2$ supersymmetry, matter fields are part of
hypermultiplets. A hypermultiplet can be most easily described in the
$N=1$ notation as consisting of a chiral superfield $Q$ and an
anti-chiral superfield $\widetilde Q^\dagger$. $Q$ ($\widetilde Q$)
contains components $q,\psi_q$ ($\widetilde q,\psi_{\widetilde q}$)
and transforms in the $N$ ($\bar{N}$) representation of the gauge
group $SU(N)$. Under the $SU(2)_{R}$ symmetry of the $N=2$ algebra,
$q,\widetilde q^\dagger$ form a doublet while $\psi_q$ and
$\psi^\dagger_{\tilde q}$ are singlets. Since the fields in a
hypermultiplet have spin $\leq 1/2$, they belong to a short
representation of the $N=2$ algebra and, therefore, must satify the
relation $M=\sqrt{2} Z$. The fact that $Z$, as derived in the previous
section, does not satisfy this relation in the presence of bare
hypermultiplet masses, indicates that the central charge must receive
further contributions from the hypermultiplets. We will calculate this
contribution below. 

Beside the standard kinetic terms and gauge couplings for the chiral
fields $Q$ and $\widetilde Q$, the $N=2$ Lagrangian in the presence of
matter also contains the $N=1$ superpotential given by 
$$
\sum_{i=1}^{N_f}\sqrt{2}\widetilde{Q}_i\Phi Q_i+\sum_{i=1}^{N_f} m_i 
\widetilde{Q}_i Q_i + h.c.
$$
Here, the index $i$ runs over the $N_f$ quark flavours and $\Phi$ is
the chiral superfield of the vector multiplet in the adjoint
representation. The first term is related to the gauge coupling of
the matter fields by $N=2$ and the second term is an $N=2$ invariant
mass term. When all masses are equal, the theory also has a $SU(N_f)$
flavour symmetry which is broken down to smaller subgroups when the
masses are not equal. For all masses unequal, it breaks down to
$U(1)^{N_f}$.  

The contribution of $Q$ and $\widetilde Q$ to the supercurrents can be
easily calculated using (\ref{N=1scch}), with the superpotential
${\cal W}$ as given above. Thus, as the contribution to
$S^\mu_{(1)\alpha}$, we get (supressing both colour and flavour
indices):
\bea
\epsilon^\alpha S^\mu_{(1)\alpha}=\cdots&+&\sqrt{2}\,\epsilon\sigma^\nu  
\bar{\sigma}^\mu\psi_q D_\nu q^\dagger+i\sqrt{2}\,m\,
\widetilde{q}^\dagger \bar{\psi}_q\bar{\sigma}^\mu \epsilon\nonumber\\
&+& \sqrt{2}\,\epsilon\sigma^\nu\bar{\sigma}^\mu\psi_{\widetilde{q}}
D_\nu {\widetilde{q}^\dagger} + i \sqrt{2}\, m\, q^\dagger
\bar{\psi}_{\widetilde{q}} \bar{\sigma}^\mu \epsilon +\cdots\,.\nonumber
\eea
To obtain the corresponding terms in the second supercurrent, we have
to make the replacements $q\rightarrow\widetilde q^\dagger$, 
$\widetilde q^\dagger\rightarrow -q$,
\bea
\epsilon^\alpha S^\mu_{(2)\alpha}=\cdots&+&\sqrt{2}\,\epsilon\sigma^\nu  
\bar{\sigma}^\mu\psi_q D_\nu \widetilde q - i\sqrt{2}\,m\,
q \bar{\psi}_q\bar{\sigma}^\mu \epsilon\nonumber\\
&-& \sqrt{2}\,\epsilon\sigma^\nu\bar{\sigma}^\mu\psi_{\widetilde{q}}
D_\nu q + i \sqrt{2}\, m\, \widetilde q \bar{\psi}_{\widetilde{q}} 
\bar{\sigma}^\mu \epsilon +\cdots\nonumber\,.
\eea
Using the standard canonical commutation relations and following a
procedure similar to the previous section, we can calculate the
contribution of these extra terms to the anti-commutator 
of the supercharges. The additional term is of the form
$2i\epsilon_{\alpha\beta}\,\sum_im_i S_i$, with
$$
S_i=\int d^3x\left(D_0q_i^\dagger q_i + q_iD_0q_i^\dagger
-\frac{i}{2}\psi^\dagger_{qi}\psi_{qi} + \frac{i}{2}\psi_{qi}
\psi^\dagger_{qi} -(q\rightarrow \widetilde q, \psi_q\rightarrow
\psi_{\widetilde q})\right)\,,
$$   
where, the index $i$ is not summed over. Clearly, this is a conserved
charge associated with a global $U(1)$ symmetry under which
$Q_i$ and $\widetilde Q_i$ carry charges $+1$ and $-1$,respectively. 
These are the $U(1)$ factors of the broken flavour group. Taking this
extra term into account, the formula for the central charge takes the
form 
\beq
Z = n_e a + n_m a_D + \sum_{i} \;\frac{1}{\sqrt{2}} \; m_i S_i \,.
\label{bps-m}
\eeq
This formula is crucial in the Seiberg-Witten analysis of $N = 2$
$SU(2)$ gauge theory with quark hypermultiplets.

\section{The Seiberg-Witten Analysis of $N=2$ Supersymmetric
Yang-Mills Theory}

In this section, we analyze the $N = 2$ pure $SU(2)$ theory following
Seiberg and Witten \cite{SW-I}.  $N=2$ has powerful Ward identities
which, together with some physical input, lead to interesting
conclusions like monopole condensation and - after $N = 2$ is softly
broken to $N = 1$ - to confinement due to condensation of the massless
monopoles. The chain of reasoning is long and elaborate.  For the
theories under consideration, when the $\theta$-angle is set to zero,
the BPS bound is given by
\bea
M &\geq& \sqrt{2}\,\vert Z \vert\,,\nonumber \\
Z &=& a\,(n_e + {i\over \alpha} n_m)\,,\quad{\rm with}\quad 
\alpha = {g^2\over 4\pi}\,. \nonumber
\eea
The form of $Z$ is invariant under $n_e \leftrightarrow n_m$
accompanied by $\alpha\leftrightarrow 1/\alpha$ and $a \leftrightarrow
a/\alpha$. This may be regarded as some evidence for the
Montonen-Olive conjecture of electromagnetic duality, although we have
seen that $Z$ gets an interesting renormalization in $N=2$
theories. While the conjecture in its original form makes sense in
$N=4$ theories, for $N=2$, we can still have an $SL(2, Z)$
transformation acting on the parameter $\tau = \theta/2\pi +
i4\pi/g^2$ which embodies a weaker form of Montonen-Olive
conjecture. By combining these transformations with global symmetries
and the requirement of positivity of kinetic energy, Seiberg and
Witten were able to formulate a procedure for determining the exact
form of the low-energy theory as will be described below.

\subsection{Parametrization of the Moduli Space}

The classical action for the $N=2$ supersymmetric Yang-Mills theory
(\ref{N=2}) containts the scalar potential
$$
V={1\over 2g^2}{\rm Tr}\left([\phi^\dagger,\phi]^2\right)\,.
$$
Therefore, the Higgs vacuum is defined by $[\phi,\phi^\dagger]=0$,
which implies that $\phi$ takes values in the Cartan subalgebra of the
gauge group: $\phi= \phi_i H^i$. Thus, generically, the gauge group
$G$ is broken to the subgroup $H$ which is generated by elements from
the Cartan subalgebra. Elements in $G/H$ do not leave the Higgs vacuum
invariant, but being gauge transformations, they relate physically
equivalent vacua. On the other hand, once a given basis for the Cartan
subalgebra is chosen, then different vacuum values of $\phi$, within
this subalgebra, correspond to different physical theories. Thus these
degrees of freedom in $\phi$ ({\it i.e.}, the $\phi_i$'s) parametrize
the space of physically inequivalent vacua, or the moduli space of the
theory.  The dimension of this moduli space is equal to the rank $r$
of the gauge group $G$. However, this parametrization of the moduli
space is not the desired one as there is still some residual gauge
invariance: The coset $G/H$ contains elements which, while not leaving
the vacuum invariant, do not take $\phi$ out of the Cartan
subalgebra. These transformations are precisely the Weyl 
reflections. Therefore, the correct parametrization of the moduli
space is given not by $\phi$, but by Weyl invariant functions
constructed out of it.

The Weyl invariants are obtained from the characteristic equation, 
$$
\det (\lambda - \phi ) = 0\,.
$$
Since Weyl reflections act on $\phi$ by conjugation, $\det (\lambda
-\phi)$ is invariant. Hence, if we expand this quantity as a
polynomial in powers of $\lambda$, then the coefficients are Weyl
invariant quantities. In the following, we express these coefficients
as simple functions of $\phi$. Let $a_1,\cdots , a_N$ denote the roots
of the characteristic equation, or the eigenvalues of $\phi$. If the
gauge group is $SU(N)$ (or $SO(N)$), we have ${\rm Tr}\phi=\sum
a_i=0$. For generic values of $a_i$, the gauge group is broken to
$U(1)^r$. When some of the eigenvalues coincide, the unbroken group
jumps from $U(1)^r$ to something at least as big as $U(1)^{r-1}\times
SU(2)$. In terms of $a_i$, the characteristic polynomial takes the
form: 
$$
\lambda^N + \lambda^{N-2} \sum_{i<j} a_ia_j - \lambda^{N-3}
\sum_{i<j<k} a_ia_ja_k + \cdots + (-1)^N \prod_{i=1}^N a_i = 0\,.
$$
For $SU(2)$, $\phi={1\over 2}a\sigma_3$ and it can be easily checked
that the desired Weyl invariant is $u={\rm Tr}(\phi^2)={1\over2}
a^2$. For $SU(3)$, the coefficients in the characteristic polynomial
are  $a_1a_2+a_1a_3+a_2a_3$ and $a_1 a_2 a_3$, where $a_1, a_2, a_3$
are the eigenvalues of $\phi$. Using $({\rm Tr}\phi)^2=0$, and $({\rm 
Tr}\phi)^3=0$, one can easily write the Weyl invariants as    
$$
\begin{array}{ll}
u=\frac{1}{2}{\rm Tr}(\phi^2)=-(a_1a_2+a_1a_3+a_2a_3)\,,  \\
v=-\frac{1}{3}{\rm Tr}(\phi^3)=a_1 a_2 a_3\,.
\end{array}
$$
In general, for $SU(N)$ similar formulae can be worked out.  The
coefficients of the characteristic polynomial are the ``Chern" classes
of $\phi$, 
$$
\det(\lambda-\phi)=\lambda^N -\lambda^{N-1} c_{1}(\phi)+
\lambda^{N-2}c_{2}(\phi)+\cdots +(-1)^{j} \lambda^{N-j} c_{j} (\phi )+
\cdots + (-1)^N c_N(\phi)= 0\,.
$$
The coefficients $c_j(\phi)$ can be easily determined from the following
formal expansion,
\bea
\det(\lambda -\phi) &=& \lambda^N \det (1-\phi/\lambda) =
\lambda^N e^{{\rm Tr\,ln}(1-\phi/\lambda)} 
=\lambda^N {\rm exp}\left(-\sum_{n=1}^\infty \frac{{\rm Tr}\,(\phi^n)}
{n\lambda^n}\right) \nonumber \\
&=& \lambda^N - \frac{1}{2}{\rm Tr}(\phi^2)\lambda^{N-2} -\frac{1}{3}
{\rm Tr}(\phi^3) \lambda^{N-3}+\cdots\,. \nonumber
\eea 
Note that the series expansion for ${\rm ln} (1-\phi/\lambda)$ makes
sense only for $\lambda >> \phi$. Therefore, in the above expansion
only terms with positive powers of $\lambda$ are relevant and they
provide all the Weyl invariant quantities we need for a gauge
invariant parametrization of the moduli space.

In the above, we have treated $\phi$ classically. In quantum field
theory, we parametrize the moduli space by the vacuum expectation
values of the corresponding classical Weyl invariants. For example,
the moduli space of the $SU(2)\,\,N=2$ supersymmetric Yang-Mills
theory is parametrized by $u=<{\rm Tr}(\phi^2)>$, which at the
classical limit, reduces to $a^2/2$.
 
\subsection{R-Symmetry and its Breaking}

The $N$-extended supersymmetry algebra (\ref{susy-noC}) is invariant
under global $U(N)$ rotations of the $N$ supercharges. Therefore, a
supersymmetric theory should also exhibit such a global symmetry,
usually referred to as R-symmetry. The action of this global symmetry
on the supercharges can be easily translated into a transformation of
the superspace variables. For example, for $N=1$, the $U(1)$
R-symmetry acts on the supercharge as $Q\rightarrow
e^{-i\alpha}Q$. From this we can obtain its action on the superspace
coordinates as $\theta\rightarrow e^{i\alpha}\theta$ and
$\bar\theta\rightarrow e^{-i\alpha}\bar\theta$ (see
(\ref{superspace-Q})).  For $N=2$, we have a $U(1)$ which acts as
above on the $N=2$ superspace coordinates $\theta^I, \bar\theta_I$,
along with an $SU(2)$ R-symmetry which rotates the index $I$ of the
supercharges. In order to keep the supersymmetric Lagrangian invariant
under these transformations, we have to assign appropriate
transformation properties to various superfields. From this, we can
easily obtain the behaviour of the component fields under
R-symmetry. Below, we will describe this in a little more detail.

\noindent\underline{Action on the $N=2$ Vector Multiplet:}
As discussed in the previous section, the $N=2$ vector multiplet
contains a vector field $A_\mu$, two Weyl spinors $\lambda, \psi$ and a
scalar $\phi$, all transforming in the adjoint representation of the
gauge group $G$. These components can be arranged as
$$
\matrix{ &A_\mu& \cr \lambda &&\psi \cr &\phi &}\,.
$$
The $SU(2)_R$ transformation acts on the rows in the above diagram and
rotates the fermions $(\lambda,\psi)$ into each other while keeping
$A_\mu$ and $\phi$ invariant. In the
$N=1$ formalism, this multiplet decomposes into a vector superfield
$V(A_\mu, \lambda)$, and a chiral superfield $\Phi(\phi,\psi)$.    
Therefore, the only part of $SU(2)_R$ which remains manifest in the
$N=1$ language is a $U(1)_J$ subgroup which does not mix $\lambda$ and
$\psi$. This subgroup is generated by $\sigma_3$ and acts as 
$(\lambda,\psi)\rightarrow(e^{i\alpha}\lambda,e^{-i\alpha}\psi)$. 
The action of the $U(1)_{\cal R}$ on the $N=1$ superfields was
discussed in the previous section. Below, we summarize these
transformations:
\beq
\begin{array}{lll}
U(1)_{\cal R} &: \Phi\rightarrow e^{2i\alpha}\Phi(e^{-i\alpha}\theta),
&V\rightarrow V(e^{-i\alpha}\theta)\,,
\\ \\
U(1)_J &: \Phi \rightarrow \Phi (e^{-i\alpha}\theta), &
V\rightarrow V(e^{-i\alpha}\theta)\,.
\end{array}
\label{Rb1}
\eeq

\noindent\underline{Action on the $N=2$ Hypermultiplet:}
In $N=2$, the matter fields appear in hypermultiplets, each containing
two complex scalars $(q,\widetilde q^\dagger)$ and two Weyl fermions
$(\psi_q,\widetilde\psi^\dagger_q)$. All these components transform in
the same (usually the fundamental) representation of the gauge group.
The components of a hypermultiplet can be arranged as 
$$
\matrix{ &\psi_q& \cr q &&\widetilde q^\dagger \cr
  &\widetilde\psi_q^\dagger&}\,. 
$$
Again, $SU(2)_R$ acts on the rows and thus rotates $q,\widetilde
q^\dagger$. In the $N=1$ language, the hypermultiplet is decomposed in
terms of chiral multiplets $Q(q,\psi_q)$, and $\widetilde Q(\widetilde
q,\widetilde\psi_q)$, which carry dual gauge quantum numbers. In this
decomposition, only the $U(1)_J$ subgroup of $SU(2)_R$ is
manifest. Moreover, in the $N=1$ decomposition, a hypermultiplet
interacting with a vector multiplet gives rise to a superpotential
term
$$
{\cal W} = \sqrt{2} \widetilde Q \Phi Q.
$$
Since ${\cal W}$ should carry two units of $U(1)_{\cal R}$ charge,
$Q$ and $\widetilde Q$ are neutral. We summarize these transformations
below: 
\beq
\begin{array}{lll}
U(1)_{\cal R}: &Q \rightarrow Q(e^{-i\alpha}\theta)\,,
& \widetilde Q \rightarrow \widetilde Q (e^{-i\alpha}\theta)\,, \\ \\
U(1)_J : &Q\rightarrow e^{i\alpha} Q(e^{-i\alpha}\theta)\,,
&\widetilde Q\rightarrow e^{i\alpha}\widetilde Q(e^{-i\alpha}\theta)\,. 
\end{array}
\label{Rb2}
\eeq
For later convenience, we list below the transformations of all
component fields under $U(1)_{\cal R}$ and $U(1)_J$:
$$
\begin{array}{lccc} 
U(1)_{\cal R}:\qquad & \phi &\rightarrow & e^{2i\alpha}\phi\,, \\
&(\psi,\lambda)&\rightarrow & e^{i\alpha}(\psi,\lambda)\,, \\
&(\psi_q,\widetilde\psi_q)&\rightarrow &e^{-i\alpha} 
(\psi_q,\widetilde\psi_q)\,, \\
&(A_\mu, q, \widetilde q)&\rightarrow & (A_\mu, q, \widetilde q)\,.
\\ \\
U(1)_J: &(\lambda, q)&\rightarrow & e^{i\alpha}(\lambda, q)\,,\\
&(\psi,\widetilde q^\dagger)&\rightarrow & e^{-i\alpha}
(\psi,\widetilde q^\dagger)\,,\\
&(\phi, A_\mu, \psi_q,\widetilde\psi_q^\dagger)&\rightarrow &
(\phi, A_\mu, \psi_q,\widetilde\psi_q^\dagger)\,.
\end{array}
$$
Note that we can combine the two-component spinors $\lambda$ and
$\bar\psi$ into a four-component Dirac spinor $\psi_D$ (see
(\ref{sp-comp})). The spinor $\psi_D$ transforms as $e^{i\alpha}
\psi_D$ under $U(1)_J$ and as $e^{i\alpha\gamma_5}\psi_D$ under
$U(1)_{\cal R}$. Similarly, $U(1)_{\cal R}$ acts as a chiral $U(1)$ on
the Dirac spinor constructed out of $\psi_q$ and $\widetilde
\psi_q^\dagger$\,, though now with the opposite charge. Thus
$U(1)_{\cal R}$ is a chiral symmetry and is, therefore, broken by  
a chiral anomaly as will be discussed below. 

\noindent\underline{Breaking of R-Symmetries:} 
Classically, our theory has the full global $SU(2)_R\times U(1)_{\cal
R}$ as a symmetry group. However, quantum mechanically, $U(1)_{\cal
R}$ is broken to a discrete subgroup due to anomalies. This subgroup
can be easily determined from elementary instanton considerations (the
more direct method will be described in the next subsection). To
compute the anomaly for the gauge group $SU(N_c)$, note that by the
index theorem, in the presence of an instanton, there is one zero-mode
for each left moving fermion in the fundamental or antifundamental
representaion and $2N_c$ zero-modes for each left-handed fermion in
the adjoint representation. A correlation function in this theory
involves integrations over the fermionic collective coordinates
corresponding to these zero-modes. For a correlator to be non-zero, it
should contain enough fermion insertions to soak the zero-modes. 
Hence, in the presence of $N_f$ flavours, the first non-vanishing
correlator is 
\beq
G=\langle\lambda(x_1)\cdots\lambda(x_{2N_c})\psi(y_1)\cdots\psi(y_{2N_c})
\psi_q(z_1)\cdots\psi_q(z_{N_f})\widetilde\psi_q(u_1)\cdots
\widetilde\psi_q(u_{N_f})\rangle \,.
\label{G}
\eeq
Under the $U(1)_{\cal R}$, $G$ transforms as
$$
G \rightarrow e^{i\alpha(4N_c-2N_f)}\,G\,.
$$
Hence $U(1)_{\cal R}$ is broken to the discrete group 
${\bf Z}_{4N_c-2N_f}$. In the following, we focus on the pure
Yang-Mills theory so that $N_f=0$. In this case, $U(1)_{\cal
  R}\rightarrow {\bf Z}_{4N_c}$ and is represented by $e^{2\pi
  i\alpha}$, where  $\alpha=n/4N_c\,, n=1,\cdots 4N_c$. Thus the
global symmetry group is $SU(2)_R\times {\bf Z}_{4N_c}$. However, note
that the centre of $SU(2)_R$, which acts as $(\lambda,\psi)\rightarrow
e^{i\pi}(\lambda,\psi)$, is also contained in ${\bf Z}_{4N_c}$
(corresponding to $n=2N_c$). Hence, the true symmetry group is  
$$
\left(SU(2)_R\times {\bf Z}_{4N_c}\right)/{\bf Z}_2\,.
$$

This surviving R-symmetry is broken further by the Higgs vacuum. 
The field $\phi^2$ has charge 4 under ${\bf Z}_{4N_c}$ and transforms
to $e^{2\pi i n/N_c}\phi^2$, which is invariant only for
$n=N_c,2N_c,3N_c,4N_c$. Therefore, if the vacuum is characterized by
non-zero $\phi^2$, then ${\bf Z}_{4N_c}$ is broken down to ${\bf
  Z}_4$. This is the situation for the $SU(2)$ gauge theory ($N_c=2$). 
In this case, all elements which do not keep $\phi^2$ invariant, act
as a ${\bf Z_2}: \phi^2\rightarrow -\phi^2$. Therefore, the final
R-symmetry group for the $SU(2)$ gauge theory is $(SU(2)_R\times 
{\bf Z_4})/{\bf Z_2}$. 

For the gauge group $SU(3)$, we have ${\bf Z_{4N_c}}={\bf Z_{12}}$.
The vacuum is parametrized by $\phi^2$ and $\phi^3$, which, under
${\bf Z}_{4N_c}$, transform to $e^{2\pi i n/N_c}\phi^2$ and $e^{3\pi i
n/N_c}\phi^3$ respectively. Invariance of $\phi^2$ requires
$n=N_c,2N_c,3N_c,4N_c$, breaking ${\bf Z_{12}}$ to ${\bf
Z_{4}}$. Invariance of $\phi^3$ picks $n=2N_c, 4N_c$. Thus, for the
$SU(3)$ gauge theory, ${\bf Z}_{12}$ is broken to ${\bf Z}_2$. The
broken ${\bf Z}_6$ subgroup acts non-tivially on the moduli space. For
$SU(4)$ and higher gauge groups, no subgroup of ${\bf Z}_{4N_c}$
survives.

\subsection{Low-Energy Effective Action for $N=2$ Gauge Theory}

Let us consider the $N=2$ supersymmetric Yang Mills theory based on a
group $G$ of rank $r$ which is spontaneously broken by a non-zero
$\langle\phi\rangle$. Far from the points where two eigenvalues of
$\langle\phi\rangle$ coincide, the only massless fields are the vector
supermultiplets associated with the unbroken subgroup $U(1)^{r}$ of
$G$. At sufficiently low energies, non of the massive states will
appear as physical states and an effective description of the theory
can be given in terms of the massless fields alone. In principle, this
low-energy theory can be obtained by integrating out all the massive
modes as well as massless modes above a low-energy cutoff. In
practice, this procedure is not easy to implement. Seiberg and Witten
formulated an indirect procedure for determining the exact low-energy
theory as we will in the remaining part of this section.

As discussed in the previous section, the effective action (the part
with at most two derivatives) is fully determined by a prepotential
${\cal F}$ which is only a function of $r$ massless vector
supermultiplets.  In the $N=1$ language, the corresponding Lagrangian
takes the form (with $A_i$ denoting chiral superfields):
$$
\frac{1}{4\pi} \, {\rm Im} \, \left [ \int \, d^{4}\theta \,
\frac{\partial {\cal F}}{\partial A^{i}} \bar{A}^{i} +
\int \, d^{2} \theta \, \frac{1}{2}
\frac{\partial^{2}{\cal F}}{\partial A_{i} \partial A_{j}}
W^{\alpha i} W^j_\alpha
\right ]\,.
$$
In terms of ${\cal F}(A)$, the  K\"{a}hler potential is given by
$K={\rm Im}\left(\bar{A}_{i}\partial{\cal F}(A)/\partial A_{i}\right)$. 
If $a_i$ denotes the scalar component of the chiral superfield $A_i$,
then the metric on the space of fields, and therefore, the one on the
space of Higgs vacuua, is given by
$$
ds^2 = g_{i\bar{j}}\,da^i d\bar{a}^j =
{\rm Im}\,\frac{d^2{\cal F}}{\partial a_i \partial a_j}
\, da^i d\bar{a}^j \,.
$$
Note that, by virtue of $N=2$ supersymmetry, this metric is the same
as the generalized gauge coupling which appears in front of the
$F_{\mu\nu}^iF^{\mu\nu i}$ term in the above Lagrangian. This
relationship is not modified in the quantum theory.   

In the following, we will be mainly interested in the $SU(2)$ gauge
group spontaneously broken to $U(1)$. In this case the effective
low-energy Lagrangian takes the form
\beq
{\cal L}_{eff}=
\frac{1}{4\pi} \, {\rm Im} \, \left [ \int \, d^{4}\theta \,
\frac{\partial {\cal F}(A)}{\partial A} \bar{A} +
\int \, d^{2} \theta \, \frac{1}{2}
\frac{\partial^{2}{\cal F}}{\partial A^{2}} W^{\alpha} W_{\alpha}
\right ]\,, 
\label{EA}
\eeq
and the metric on the field space is given by
\beq
ds^2= {\rm Im}(\tau)\, da d\bar a\,,\quad {\rm where},\quad
\tau (a) = \frac{\partial^{2} {\cal F}}{\partial a^{2}}.
\label{metric}
\eeq
Our aim in this subsection is to determine the perturbative form of
${\cal F}$ following \cite{Seiberg88}. The determination of the exact
form of the prepotential, including non-perturbative corrections, is
the subject of the later subsections. 

To determine the one-loop perturbative correction to ${\cal F}$, we
start from the exact microscopic theory (\ref{N=2}) with gauge group
$SU(2)$. This theory is asymptotically free and therefore at high
energies one can perform reliable perturbative calculations. As
indicated in the previous subsection, the $U(1)_{\cal R}$ symmetry of
this theory is broken by the standard chiral anomaly. Thus we have 
$$
\partial_\mu J^\mu_5= -\frac{N_c}{8\pi^2}F_{\mu\nu}\widetilde
F^{\mu\nu} \,.
$$
This implies that, to one-loop, under a $U(1)_{\cal R}$
transformation, the effective Lagrangian changes by 
\beq
\delta {\cal L}_{eff}= - \frac{\alpha N_c}{8\pi^2} F\widetilde F\,.
\label{anom}
\eeq
If the theory also contains $N_f$ fermions in the fundamental
representation, then in the above expression, $N_c$ has to be replaced
by $N_c-N_f/2$. Note that since $(32\pi^2)^{-1}\int F\widetilde F$ is
an integer, the anomaly breaks $U(1)_{\cal R}$ to ${\bf Z}_{4N_c}$ (or
to ${\bf Z}_{4N_c-2N_f}$ in the presence of matter). The same result
was obtained in the previous subsection from a consideration of
fermion zero modes in an instanton background.  Moreover, the anomaly
can be regarded as causing a shift in the $\theta$-angle. In the
following, we use this freedom to set $\theta$ to zero by an
appropriate chiral rotation of the fermions.
 
The one-loop form of ${\cal F}$ can be determined from the requirement
that under a $U(1)_{\cal R}$ transformation, the low-energy effective
action (\ref{EA}) change as in (\ref{anom}). The variation $\delta
{\cal L}_{eff}$ could come only from terms in ${\cal L}_{eff}$ which
are quadratic in $F_{\mu\nu}$. Writing only the relevent terms from
the Lagrangian (\ref{EA}), this means
$$ 
\frac{1}{16\pi}{\rm Im}\left[{\cal F}''(e^{2i\alpha}A)(-FF + iF
\widetilde F)\right] = \frac{1}{16\pi}{\rm Im}\left[{\cal F}''(A)
(-FF+iF\widetilde F)\right]-\frac{\alpha N_c}{8\pi^2}F\widetilde F\,.
$$ 
The form of the prepotential is therefore restricted by 
$$
{\cal F}''(e^{2i\alpha}A)={\cal F}''(A) - \frac{2\alpha N_c}{\pi}\,,
$$
or, for infinitesimal $\alpha$, by
$$
\frac{\partial^3 {\cal F}}{\partial A^3}=\frac{N_c}{\pi}\frac{i}{A}\,.
$$
The above equation can be easily integrated to give the one-loop
expression for the prepotential as 
\beq
{\cal F}_{1-loop}(A)=\frac{i}{2\pi} A^2 {\rm ln} \frac{A^2}{\Lambda^2}\,.
\label{oneloop}
\eeq
Here, $\Lambda$ is a fixed dynamically generated scale like
$\Lambda_{QCD}$.  It is known that, due to $N=2$ supersymmetry, the
above one-loop expression for the prepotential does not receive higher
order perturbative corrections. This is related to the fact that in
this theory the perturbative $\beta$-function is only a one-loop
effect. Thus, in this theory, the one-loop nature of the perturbative
$\beta$-function is consistent with the well known one-loop nature of
the anomaly. This is not the case in $N=1$ theories where the anomaly
is, of course, still a one-loop effect but not the $\beta$-function
(see, for example, \cite{SV1, SV2}).

Although, the prepotential, as given in (\ref{oneloop}), is exact in
perturbation theory, it does receive non-perturbative corrections due
to instanton effects as argued in \cite{Seiberg88}. The general form
of these corrections can be fixed as follows: First, it is clear that
a correction to ${\cal F}$ coming from a configuration of instanton
number $k$ should be proportional to the $k$-instanton factor
$exp(-8\pi^2k/g^2)$ (since the prepotential is a holomorphic function,
it cannot receive corrections from anti-instanton configurations). 
Using the one-loop $\beta$-function of the theory given by
$\beta(g)=-g^3/4\pi^2$, the $k$-instanton factor can be written as 
$$
e^{-8\pi^2k/g^2} = \left(\frac{\Lambda}{a}\right)^{4k}\,.
$$
Furthermore, following the approach of Seiberg in \cite{Seiberg-n,
Seiberg-poh}, we note that the broken $U(1)_{\cal R}$ symmetry is
restored if we assign a charge of $2$ to $\Lambda$. With this
modification, the prepotential should transform under $U(1)_{\cal R}$
as a field of charge $4$, without a non-homogeneous term. This implies
that the $k$-instanton correction should also be proportional to
$A^2$. Putting these together, the prepotential including generic
non-perturbative corrections can be written as
$$
{\cal F}=\frac{i}{2\pi} A^2 {\rm ln} \frac{A^2}{\Lambda^2}
+\sum_{k=1}^\infty\,{\cal F}_k\left(\frac{\Lambda}{A}\right)^{4k}A^2\,.
$$
The coefficients ${\cal F}_k$ are not field dependent due to the fact
that in a supersymmetric theory, instantons contribute to the path
integral only through zero-modes\cite{DD}. The coefficient ${\cal
F}_1$ was calculated in \cite{Seiberg88} and found to be non-zero. The
determination of the exact form of ${\cal F}$ is the subject of the
work of Seiberg and Witten. 

The one-loop expression for ${\cal F}$ can also be obtained from
integrating the expression for the $\beta$-function, choosing $A$ as
one of the integration limits: In the classical low-energy theory,
which is obtained by simply dropping the massive fields from the
microscopic Lagrangian (and not integrating over them), the
prepotential is given by ${\cal F}(A)=\frac{1}{2} \tau_{cl}A^2$,
where, $\tau_{cl}=\frac{\theta}{2\pi}+i\frac{4\pi}{g^2}$. In the
quantum theory, for large $a$ (which is the {\it vev} of the scalar
somponent of $A$), asymptotic freedom takes over and the theory is
weakly coupled. Therefore, in this limit, a good approximation to the
quantum behaviour of the theory can be obtained by replacing $g$ by
the corresponding running couplings $g(a)$ (we have set $\theta$ to
zero by a chiral redefinition of the fermion fields).  In the limit of
large $a$, $g(a)$ and hence ${\cal F}(a)$ can be obtained in
perturbation theory by integrating the $\beta$-function.

For other gauge groups we note that for every root $\alpha$, ($\alpha
> 0$), we have massive fields $W^{\alpha}, W^{- \alpha}$. For
simply-laced groups, $\alpha^{2} = 2$, and
$$
{\cal F} = \frac{i}{4\pi} \, \sum_{\alpha > 0} \, (\vec{\alpha} \cdot
\vec{a})^{2} \, {\rm ln}\, \frac{(\vec{\alpha}\cdot
\vec{a})^{2}}{\Lambda^{2}}\,.
$$
The coefficient of the logarithm follows from the one-loop
$\beta$-function and it also leads to the unbroken global
$U(1)_{R}$-symmetry.

The one-loop expression for ${\cal F}$, coupled with the fact that a
well defined theory should have a positive kinetic energy term, leads
to very interesting consequences.  For large $|a|$, using
(\ref{oneloop}), we can calculate $\tau(a)=\frac{i}{\pi}({\rm
ln}\,\frac{a^2}{\Lambda^2}+ 3)$. This is a multi-valued function,
though the metric on the field space given by ${\rm Im}\tau$ is single
valued. However, since ${\rm Im}\tau(a)$ is a harmonic function, it
cannot have a global minimum. Thus, if it is globally defined, it
cannot be positive everywhere. Therefore, the positivity of the
kinetic energy requires that $\tau(a)$ is defined only locally. This
means that in the region of the complex plane where $\tau(a)$ becomes
negative, one needs a different description of the theory. In the next
subsection we describe how these equivalent descriptions could be
obtained.

\subsection{Duality}
To find the duality transformations which relate different
descriptions of the same theory, we consider the gauge field terms in
the bosonic part of the action. Working in the Minkowski space with
conventions $(F_{\mu\nu})^2=-(\widetilde F_{\mu \nu})^2$ and
$\widetilde{\widetilde F}=-F$, these
terms can be written as 
$$
\frac{1}{32 \pi} \, {\rm Im}\, \int \, \tau (a) (F+i\widetilde F)^2 =
\frac{1}{16 \pi} \, {\rm Im} \, \int \, \tau (a) (F^{2}+i\widetilde
FF) \,.
$$
Now we regard $F$ as an independent field and implement the Bianchi 
identity $dF = 0$ by introducing a Lagrange multiplier vector field
$V_{D}$. To fix the Lagrange multiplier term, $U(1)\subset SU(2)$ is
normalized such that all $SO(3)$ fields have integer charges. Then, as
discussed in section 1, all matter fields in the fundamental
representation of $SU(2)$ will have half-integer charges. With this 
convention, a magnetic monopole satisfies $\epsilon^{0\mu\nu\rho}
\partial_{\mu}F_{\nu\rho}=8\pi\delta^{(3)}(x)$. The Lagrange
multiplier term can now be constructed by coupling $V_D$ to a
monopole:
$$
\frac{1}{8\pi}\,\int\,V_{D \mu}\epsilon^{\mu\nu\rho\sigma}
\partial_{\nu}F_{\rho\sigma}=\frac{1}{8\pi}\,\int\,\widetilde F_{D}F = 
\frac{1}{16\pi}\,{\rm Re}\,\int\,(\widetilde
F_{D}-iF_{D})(F+i\widetilde F)\,,
$$
where, $F_{D\mu\nu}=\partial_\mu V_{D\nu}-\partial_\nu V_{D\mu}$. 
Adding this to the gauge field action and integrating over $F$,
gives the dual theory
$$
\frac{1}{32\pi}\,{\rm Im}\int \left(-\frac{1}{\tau}\right)
(F_D +i\widetilde F_D)^2=\frac{1}{16\pi}\,{\rm Im}\int\left(-\frac{1}{\tau}
\right) (F_D^2 +i\widetilde F_D F_D)\,.
$$
The dualization can also be performed in an $N=1$ supersymmetric
language. In this case, the relevant term in the action is 
$$
\frac{1}{8 \pi} \, {\rm Im} \, \int \, d^{2}\theta
\tau (A) W^{2}\,.
$$
The Bianchi identity is replaced by ${\rm Im}{\cal D}W = 0$. This can
be implemented by introducing a vector superfield $V_{D}$ and the
corresponding Lagrange multiplier term becomes
$$
\frac{1}{4\pi}\,{\rm Im}\,\int d^4x d^4\theta\, V_{D}{\cal D}W 
=\frac{1}{4 \pi}\,{\rm Re}\,\int d^4x d^4\theta\, i{\cal D}V_D W
=-\frac{1}{4\pi}\,{\rm Im}\,\int d^4x d^2\theta\, W_{D} W\,.
$$
Adding this to the action and integrating out $W$, gives the dual
action 
$$
\frac{1}{8\pi}\,{\rm Im}\,\int d^2\theta \left(-\frac{1}{\tau(A)}
W^2_D \right)\,. 
$$
Thus, the effect of the duality transformation is to replace a gauge
field which couples to electric charges by a dual gauge field which
couples to magnetic charges, and at the same time, transform the
gauge coupling as  
\beq
\tau\rightarrow\tau_D = -\,\frac{1}{\tau}\,.
\label{tautauD}
\eeq
We recognize this as the electric-magnetic duality of section 1.
Also, note that the action is invariant under the replacement
$\tau\rightarrow\tau+1$. This transformation, along with the one in
(\ref{tautauD}), generates the $SL(2,{\rm Z})$ group which, therefore,
is the full duality group of our theory. This group acts on $\tau$ as 
\beq
\tau\rightarrow{\frac{a\tau + b}{c\tau + d}}\,,
\label{tauSLtwoZ}
\eeq
where, $ad-bc=1$ and $a,b,c,d \in {\rm Z}$. 

$N=2$ supersymmetry relates the dual description of the gauge fields
to a dual description for the scalar fields. To see this, let us
introduce $h(A)=\partial{\cal F}/\partial A$. In terms of this,
$\tau(A)=\partial h(A)/\partial A$ and the scalar kinetic energy term
becomes ${\rm Im}\int d^4\theta\,h(A)\bar A$. For the dual theory
corresponding to $(\ref{tautauD})$, let us introduce the corresponding
dual variables $A_D,{\cal F}_D,h_D(A_D)$ and $\tau_D$. Then, equation
(\ref{tautauD}) implies that $A_D=h=\partial{\cal F}/\partial A$ and 
$h_D = -A$. Under this transformation, the scalar kinetic energy term
retains its form,
$$
{\rm Im}\int d^4\theta\,h(A)\bar A=
{\rm Im}\int d^4\theta\,h_D(A_D)\bar A_D\,. 
$$
In the following, we use the notations $A_D$ and $h(A)$
interchangeably. 

We now consider the effect of the full $SL(2,Z)$ group on $A$ and
${\cal F}$, or equivalently on $A$ and $A_D=\partial{\cal F}/\partial
A$. The transformation (\ref{tauSLtwoZ}) implies that
\beq
\left(\begin{array}{c}A_D\\ A\end{array}\right)\rightarrow
\left(\begin{array}{cc} a & b \\ c & d \end{array}\right)
\,\left(\begin{array}{c}A_D\\ A \end{array}\right) \,.
\label{ASLtwoZ}
\eeq 
Note that, in general, we could also add a constant column matrix $C$
to the right-hand side of the above equation. However, for non-zero
$C$, the BPS mass formula for the theory in the absence of matter is
not invariant under the above transformation. Thus in this case we
should set $C=0$. However, in the presence of matter fields, the BPS
mass formula is modified and a non-zero $C$ is allowed. This case will
be discussed in the last section. On the space of the scalar fields,
the transformations $\tau\rightarrow -1/\tau$ and $\tau\rightarrow
\tau+1$ are implemented by the matrices
\beq  
S=\left(\begin{array}{cc}0&1\\-1&0\end{array}\right),\quad
T=\left(\begin{array}{cc}1&1\\0&1\end{array}\right).
\label{ST}
\eeq
These matrices generate the $SL(2,{\rm Z})$ group.

The transformation of ${\cal F}$ can be easily obtained from
(\ref{ASLtwoZ}), or equivanlently from 
\bea
A'_D &=& a A_D + b A\,, \nonumber\\
A'   &=& c A_D + d A\,. \nonumber
\eea
The first equation can be integrated with respect to $A'$ by using the
second equation and the result is 
$$
{\cal F}'=\frac{1}{2}\beta\delta A^2 +\frac{1}{2}\alpha\gamma A^2_D
+\beta\gamma A A_D + {\cal F}\,.
$$

Let us now come back to the metric on the moduli space as given by 
(\ref{metric}). In terms of the variable $a_D$, this takes the form
$$
ds^2={\rm Im}\,da_D d\bar{a}=-\frac{i}{2}(da_D d\bar{a}-da d\bar{a}_D)\,,
$$
which is invariant under the $SL(2,{\rm Z})$ transformations. To make
the description more precise, note that the moduli space ${\cal M}$ is
a complex one dimensional manifold and let $u$ be a holomorphic
coordinate on this manifold. Finally, we will identify $u$ as 
$\langle{\rm Tr}\phi^2\rangle$. Let $a$ and $a_D$ be the coordinates
on a space $X \cong {\rm C}^2$ on which we can choose a symplectic form
$\omega={\rm Im}\,da_D\wedge d\bar{a}$. The functions
$(a_{D}(u),a(u))$ give a map $f$ from ${\cal M}$ to $X$. In other
words, they determine a section of $X$ regarded as an $SL(2,{\rm Z})$
bundle over ${\cal M}$. In terms of the coordinate $u$, the metric on
the moduli space takes the form 
$$
ds^2 ={\rm Im}\,\frac{da_D}{du}\,\frac{d \bar{a}}{d\bar{u}}\,
dud\bar{u}=-\frac{i}{2}\left(\frac{da_{D}}{du}\,\frac{d\bar{a}}{d\bar{u}}
-\frac{d\bar{a}_D}{d\bar{u}}\,\frac{da}{du}\right)dud\bar{u}\,.
$$
This is the pull-back of the K\"{a}hler metric associated with the
symplectic form $\omega$ and is, therefore, manifestly $SL(2,{\rm Z})$
invariant. Choosing $u=a$, we get back to the original formula. Note
that for arbitrary $(a_D(u),a(u))$, the metric is not
positive. However, the solution we will obtain later determines these
functions such that the metric is always positive.  

\noindent\underline{Higher Dimensions:}
{}For a group of rank $r$, the metric takes the form
$$
ds^2 ={\rm Im}\,\frac{\partial^2{\cal F}}{\partial a_i\partial a_j}\, 
da_i d\bar{a}_j\,.
$$
Introduce a space $X \simeq {\bf C}^{2r}$ with coordinates $a^{i},
a_{Dj}=\partial {\cal F}/\partial a^{j}$. Endow $X$ with the symplectic
form $\omega=\frac{i}{2}\,\sum_{i}(da^i\wedge d\bar{a}_{Di}-da_{Di} 
\wedge d\bar{a}^i)$ and the holomorphic 2-form $\omega_h =\sum_{i} \, 
da^{i} \wedge da_{Di}$. Choose $u^s\, (s=1,...,r)$ as coordinates on
the moduli space ${\cal M}$. Then, $a^i(u),a_{Dj}(u)$ give a map $f$
from ${\cal M}$ to $X$ such that $f^{*}(\omega_{h}) = 0$ (this
condition ensures that $a_{Dj}=\partial {\cal F}/\partial a_{j}$). 
Thus, the metric takes the form 
$$
ds^{2} = {\rm Im} \, \frac{\partial a_{Di}}{\partial u^{r}} \,
\frac{\partial \bar{a}^{i}}{\partial \bar{u}^{s}} \,
du^{r} d\bar{u}^{s}\,.
$$
This is the metric associated with $f^*(\omega)$ and is therefore
invariant under $Sp(2r,{\rm R})$: If we write $v^T =(a_D, a )$, then
the metric is invariant under $v\rightarrow Mv + c$, where, $M \in
Sp(2r,{\bf R})$. In the absence of matter, $c=0$ and moreover, only
the $Sp(2r,{\bf Z})$ part survives.

If the moduli space ${\cal M}$ has a non-trivial structure, then, on
being taken around a close loop, the vector $v^T=(a_D, a)$ will get
transformed by an element of the monodromy group. As  will be shown
later, the monodromy group is a subgroup of $SL(2,{\rm Z})$, and its
determination is essential for solving the problem. 

\subsection{Dyon Masses}

As described in sections 1 and 2, for the microspcopic $SU(2)$ theory,
the BPS bound is given by $M \geq \sqrt{2} |Z|$, where,
$$
Z=a(n_e + \tau_{cl}\, n_m)\,, 
$$
and $\tau_{cl}=\frac{\theta}{2\pi}+i\,\frac{4\pi}{g^2}$. All states
that saturate this bound fall in a short multiplet of the $N=2$
algebra. In a more general $N=2$ theory, like the low-energy effective
action we have been studying, this formula is slightly modified. 
Suppose, the theory contains matter fields in hypermultiplets. When
$a\ne 0$, these fields, which in the $N=1$ formalism are described by
chiral fields $M, \widetilde M$, become massive. If a hypermultiplet
carries charge $n_e$, then its coupling to the chiral field $A$ is
uniquely fixed by $N=2$ supersymmetry as 
$$
\sqrt{2}\,n_e \, A M \widetilde M\,.
$$
From this we can easily see that for such a state, $Z=a\,n_e$. On the
other hand, if we consider a magnetic monopole carrying $n_m$ units of
magnetic charge, then a manipulation very similar to the one in
subsection 1.6 leads to the BPS bound $\sqrt{2}|n_m\,a_D|$ or,
equivalently, $Z=n_m\,a_D$. In general, one can compute this formula
for dyons and obtain 
\beq
Z = a n_e + a_D n_m\,.
\label{ZaaD}
\eeq
If the gauge group is of rank $r$, then we have
$$
Z = a^i n_{e,i} + h_i(a) n^i_m =a\cdot n_e +a_D \cdot n_m\,.
$$
Since masses are physically observable, the mass formula should be
invariant under the monodromies on the moduli space. Therefore, if
$v=(a_D, a)^T$ transforms to $M v$ (with $M\in Sp(2r,{\rm Z}) )$, then
the vector $w= (n_m, n_e)$ should transform to $w M^{-1}$. Note that
this once again requires $M$ to be an integral matrix. 

Under the action of the monodromy, the coupling matrix $\tau_{ij}=
\partial a_Di/\partial a_j$ transforms to $(A\tau+B)(c\tau+D)^{-1}$,
where, $M=\left(\begin{array}{cc}A&B\\C&D\end{array}\right)$. This is
very similar to the transformation of the period matrix of a genus $r$
Riemann surface under the action of the monodromy group on the moduli
space of genus $r$ surfaces. This makes it reasonably appealing to
identify our $r$-dimensional moduli space of vacua ${\cal M}$ with the
moduli space of genus $r$ Riemann surfaces. The variables $a_i,a_{Dj}$
are then related to the periods of these surfaces and can be
calculated. Moreover, ${\rm Im}\,\tau$ is always positive definite,
which resolves the problem of geting a negative kinetic term. To check
this hypothesis for our $SU(2)$ theory, we first have to determine the
monodromy structure on the moduli space ${\cal M}$. This is the
subject of the next subsection.

\subsection{Monodromies on the Moduli Space of the $SU(2)$ Theory:}

We have discussed the possibility of the existence of monodromies on
the moduli space and pointed out that understanding the monodromy
structure may help us solve the theory, {\it i.e.,} to determine the
exact non-perturbative prepotential for the effective low-energy
theory. This is equivalent to determining the functions $a_D(u)$ and
$a(u)$. In this subsection, we set out to identify the singularities
on the moduli space ${\cal M}$ and calculate the associalted
monodromies.  

\noindent\underline{Monodromy at large $u$:} 
At large $|a|$, the theory is asymptotically free and
$u=\frac{1}{2}a^2$. To a good approximation, the prepotential is given
by the one-loop formula ${\cal F}(a)=(i/2\pi)a^2\,{\rm ln}\,
(a^2/\Lambda^2)$, from which we obtain 
$$
a_D=\frac{\partial{\cal F}}{\partial a}=
\frac{2ia}{\pi}\,{\rm ln}\,(\frac{a}{\Lambda})+\frac{ia}{\pi}\,.
$$
If we make a close loop on the $u$-plane around $u=0$, we get 
${\rm ln}\,u \rightarrow {\rm ln}\,u+2\pi i$, and ${\rm ln}\,a
\rightarrow {\rm ln}\,a+i\pi$, hence, 
$$
\begin{array}{rcl}
a_D & \rightarrow & - a_D + 2a \,,\\
a & \rightarrow & - a \,.
\end{array}
$$
This is implemented by the monodromy matrix
\bea
M_{\infty}&=& PT^{-2}=\left(\begin{array}{cc}
-1 & 2 \\ 0 & -1 \end{array} \right )\,,
\label{mondone}
\eea
acting on $(a_D, a)^T$.  Here, $P$ is the negative of the identity
element in $SL(2,{\rm Z})$ and $T$ is as defined in (\ref{ST}). Under
the action of this monodromy, the magnetic and electric quantum
numbers of BPS states change as $(n_m,n_e)\rightarrow
(-n_m,-n_e-2n_m)$ so that the mass formula is unchanged.

\noindent\underline{Monodromies at finite $u$:}
The monodromy at $u=\infty$ implies that there exist other monodromies
somewhere else on the $u$-plane. If these monodromies commute with
$M_\infty$, then $a^2$ is a good global coordinate. However, this
cannot be the case as then the positivity of the kinetic energy is
violated. To get a non-abelian monodromy group, we need at least two
singularities on the $u$-plane, and at finite $u$, with non-trivial
monodromies around them. These singularities will be related by the
broken discrete symmetry $u\rightarrow -u$. We make this minimal
assumption on the number of extra singularities. Then a loop enclosing
both these singularities should reproduce the monodromy $M_\infty$

What is the origin of these singularities on the moduli space? To
obtain the Wilsonian low-energy effective action, we have integrated
out all massive states in the theory. The massive particle loops then
give rise to a non-trivial metric on the moduli space. The values of
the masses, however, depend on the modulus $u$, and it may so happen
that for certain values of $u$, some of the masses become zero. Then,
at these points, we end up integrating out massless states and thus
create singularities on the moduli space. The nature of a singularity,
{\it i.e.,} the monodromy associated with it, depends on the
properties of the particle which becomes massless at the singularity.
Since, for finite $u$ (the non-perturbative regime), the mass spectrum
as a function of $u$ is not known, the singularities cannot be found
in a straightforward way. The way to proceed is to assume that some 
generic states become massless at certain values of $u$ (say $u=1$ and
$u=-1$) and find the corresponding monodromies (say $M_1$ and
$M_{-1}$). The massless states are then specified by the condition  
$M_1 M_{-1}=M_\infty$.

Naively, one may expect that the massive gauge boson multiplets
contribute to the singularity when they become massless. Classically,
this happens at the point $u=0$, which, in quantum theory, may
get shifted to a non-zero $u$. However, as argued in \cite{SW-I}, a
spin-1 multiplet becoming massless does not lead to a consistent
picture. We will not repeat this argument here.  

The only other massive states in the theory with spin $\le 1/2$ are
monopoles and dyons, which, due to the spin condition, belong to short
$N=2$ multiplets and, therefore, are BPS states. These states are
described by hypermultiplets. Seiberg and Witten suggested that the
singularities arise when some of these non-perturbative states become
massless.  The problem now is to calculate the associated
monodromies. Note that the hypermultiplets for monopoles and dyons
cannot be coupled to the fundamental fields of our theory in a local
way. However, in the subsection on duality it was seen that it is
possible to go to a dual description of the theory in which some dual
gauge fields couple to the monopoles or dyons in exactly the same way
that the usual gauge fields couple to a particle of unit electric
charge. Thus we only have to calculate the monodromy when a massive
electrically charged hypermultiplet becomes massless and then, using
the duality transformation, find the monodromy for a generic monopole
or dyon. Let us consider a dual description of the theory in which a
certain monopole or dyon appears as an elementary state, and let us
label this description by a letter, say $q$. Near the point where this
state is massless, all massive fields can be integrated out and the
theory is essentially a $U(1)$ theory coupled to a hypermultiplet. If
we denote the {\it vev} of the scalar field in this description of the
theory by $a(q)$, then the mass of a BPS state of unit electric charge
goes to zero when $a(q)=0$ at some $u=u_q$. Thus, near this point,
$a(q)$ is a good local coordinate and can be expanded as $a(q)\approx
c_q\,(u-u_q)$. Moreover, near this point, the one-loop $U(1)$
$\beta$-function implies (see the next subsection):
$$
\tau (a(q))=-\frac{i}{\pi}\,{\rm ln}\,\frac{a(q)}{\Lambda}\,,
$$
from which we obtain
$$
a_D(q)=-\frac{i}{\pi}\,a(q)\,{\rm ln}\,\frac{a(q)}{\Lambda}+
\frac{i}{\pi} \,.
$$
Moving on a closed loop around $u_q$ so that $(u-u_q)\rightarrow
e^{2\pi i} (u-u_q)$, we get the monodromy
\beq
\begin{array}{rcl}
a_D(q)& \rightarrow & a_D(q) + 2 a(q)\,,\\
a(q)  & \rightarrow & a(q)\,.
\end{array}
\label{mondq}
\eeq

Let us now calculate the monodromy when a $(n_m,n_e)$ dyon becomes
massless (the dyon is stable or marginally stable if $n_m$ and $n_e$
are coprime). The first step is to find a dual description of the
theory in which this dyon appears as an elementary stat of charge
$(0,1)$. Under a generic $SL(2,{\rm Z})$ transformation we get a 
$\left(n_m(q),n_e(q)\right)$ dyon with
\beq
\left(\begin{array}{c}a_D(q)\\a(q)\end{array}\right)=
\left(\begin{array}{c}\alpha a_D + \beta a\\
                      \gamma a_D + \delta a \end{array}\right)\,,
\qquad
\left(\begin{array}{c}n_m(q)\\n_e(q)\end{array}\right)=
\left(\begin{array}{c}n_m\delta - n_e\gamma \\
                    -n_m\beta + n_e\alpha \end{array}\right)\,,
\label{asns}
\eeq
so that $Z=n_m a_D + n_e a$ is invariant. In the above, $\alpha\delta
- \beta\gamma =1$. Now we choose the parameters such that $n_m(q) =0,
n_e(q)=1$. With this choice, $(a_D(q), a(q))$ become the variables in
terms of which the dyon couples to the $SL(2,{\rm Z})$ transformed
gauge field in the same way that the unit electric charge couples to
usual gauge fields. In particular, when the dyon becomes massless at
some point on the moduli space, the associated monodromy, in this
description, is given by (\ref{mondq}). Inverting the first equation
in (\ref{asns}), we get $a_D=-\beta a(q) + n_e a_D(q)$ and $a=\alpha
a(q) - n_m a_D (q)$. Thus we can easily find the action of the
monodromy on the original variables as 
\beq
\left(\begin{array}{c}a_D\\a\end{array}\right)\rightarrow 
\left(\begin{array}{cc}1+2n_e n_m & 2n_e^2\\
-2n_m^2 & 1-2n_e n_m \end{array}\right)\,
\left(\begin{array}{c}a_D\\a\end{array}\right)\,.
\label{mondn}
\eeq
Denoting the monodromy matrix as $M(n_m,n_e)$, we note that ${\rm
  Tr}\,M(n_m,n_e)=2$. Thus the monodromy always belongs to the parabolic
subgroup of $SL(2,{\rm Z})$. 

Now we calculate the monodromies at $u=\pm 1$. Let us assume that a
$(m,n)$ dyon becomes massless at $u=1$ and a $(m',n')$ dyon becomes
massless at $u=-1$. The the associated monodromies should satisfy
\beq
M_1(m,n)\,M_{-1}(m',n') = M (\infty)\,.
\label{mondeq}
\eeq
Using (\ref{mondone}) and (\ref{mondn}), this can be written as 
$$
\left ( \begin{array}{cccc} 1 + 2mn & 2n^{2} \\ -2m^{2} & 1-2mn
\end{array} \right ) =
\left( \begin{array}{cccc} -1 & 2 \\ 0 & -1 \end{array} \right )
\left ( \begin{array}{cccc} 1-2m'n' & -2n'^{2} \\
2m'^{2} & 1 + 2m'n' \end{array} \right )\,,
$$
leading to the equations
$$
\begin{array}{rcl}
1 + mn &=& m'n' + 2m'^{2} \,,\\
m^{2} &=& m'^{2} \,,\\
n^{2} &=& n'^{2} + 1 + 2m'n'\,, \\
1 - mn &=& -m'n'\,.
\end{array}
$$
These imply that $m=\pm 1$ and $m'=\pm 1$. For each combination of
$(m,m')$, we can easily determine $n'$ in terms of $n$ and get the
following possible sets of solutions
$$
\begin{array}{cccccccc}
(m,n): &(1,n)&,&(-1, n)&,&(-1, n)&,&(1, n)\,,\\
(m',n'): &(1, n-1)&,&(1, -n-1)&,&(-1, n+1)&,&(-1, -n+1)\,. 
\end{array}
$$
There do not seem to be any solutions where $M_{\infty}$ is factorized
into a product of more than two such parabolic $M(m,n)$. Moreover,
note that the solution allows only dyons of unit magnetic charge to
contribute to the monodromy. This is  consistent with the result that,
semiclassically, only these dyons are stable. 

Let us consider another consistency check: In general, under the
action of the monodromy, the quantum numbers of dyons will
change. However, we expect that the particular dyon which becomes
massless and is the source of the singularity should remain invariant
under the monodromy (as it is the properties of this dyon which
determine the monodromy matrix). This can be easily checked. The
eigenvalue equation
\beq
(q_m, q_{e}) \left( \begin{array}{cc}
1+2mn & 2n^{2} \\ -2m^{2} & 1-2mn \end{array} \right ) =
(q_m, q_{e})\,,
\label{mon-inv}
\eeq
leads to $nq_m-mq_e=0$, provided $m$ and $n$ do not vanish
simultaneously. Clearly, $q_m=m, q_n=n$ is a solution of this
equation. If we restrict ourselves to stable dyons, then this is the
only solution. Thus, knowing the monodromy, we can find the dyon which
gives rise to it.   

The simplest solution to the equation (\ref{mondeq}) corresponds to 
$m=m'=1, n=0, n'=-1$. The monodromy matrices are then given by 
\beq
M_1=\left(\begin{array}{cc}1&0\\-2&1\end{array}\right),\qquad
M_{-1}=\left(\begin{array}{cc}-1&2\\-2&3\end{array}\right)\,.
\label{mondothers}
\eeq
A comparison with (\ref{mon-inv}) implies that $M_1$ arises due to a
monopole becoming massless at $u=1$ and $M_{-1}$ arises when a
$(1,-1)$ dyon becomes massless at $u=-1$. At the point where the
monopoles become massless, we have $a_D=0$ and where the $(1,-1)$ dyon
becomes massless, we have $a-a_D=0$ as is evident from $Z=n_m a_D +
n_e a$. Note that the monodromy at infinity, $M_\infty$, shifts the
electric charge by $2$ units. Hence, at the points where condensation
takes place, the electric charge is really defined modulo $2$. We can
conjugate the representation of the fundamental group by
$M^{n}_{\infty}$.

\subsection{$U(1)\; \beta$-function}

To construct the monodromies in the previous subsection, we have used
the $\beta$-function for a $U(1)$ theory interacting with a
hypermultiplet. Let us look at this in some more detail. If we have
Weyl fermions with charge $Q_{f}$ and complex scalars with charge
$Q_{s}$, then their contribution to the $U(1) \beta$-functions is
$$
\beta(g)\equiv\mu \frac{d}{d \mu}g = \frac{g^{3}}{16\pi^{2}} \,
\left( \sum_{f} \; \frac{2}{3} Q^{2}_{f} +
\sum_{s} \; \frac{1}{3} Q^{2}_{s} \right )\,.
$$
If we denote the coefficient of $g^3$ by $b$ and define $\alpha= 
g^2/4\pi$, then the above equation can be rewritten as
$$
\mu \frac{d}{d\mu} \left( \frac{1}{\alpha} \right) = - 8\pi b\,.
$$ 
Consider hypermultiplet which is a reduced multiplet of $N = 2$ with
spin $\leq 1/2$. In $N=1$ language, this is described by chiral
superfields $M,\widetilde M$ and contains two Weyl fermions and two
complex scalars, all with the same charge $Q$. Hence we get
$$
b = \frac{1}{16\pi^{2}} \; Q^{2} \left(2 \cdot \frac{2}{3} +
2 \cdot \frac{1}{3} \right) = \frac{1}{8\pi^{2}}\; Q^{2}\,.
$$
Now remember that, using the anomaly, we have set the
$\theta$-parameter to zero by a chiral rotation of the fermions. As a
result we have $\tau=i/\alpha$, so that, 
$$
\mu \frac{d\tau}{d\mu} = - \frac{i}{\pi} Q^{2}\,.
$$
Identifying $\mu$ with the natural scale of the theory which is $a$
and setting $Q=1$, we obtain
$$
\tau \simeq - \frac{i}{\pi}\,{\rm ln}\,\frac{a}{\Lambda}\,.
$$
This is the expression which was used in the determination of the
monodromies at finite $u$. If we are interested in the contribution
from a monopole multiplet, then we can perform the above calculation
in terms of the dual variables. The answer then becomes $\tau_D\simeq
-(i/\pi)\, {\rm ln}\,(a_D/\Lambda)$.   

\subsection{Monopole Condensation and Confinement}

In this subsection we describe how confinement in $N=1$ gauge theory
can be understood in terms our macroscopic $N=2$ theory. The $N = 2$
vector superfield ${\cal A}$ can be decomposed, in the $N=1$ formalism,
into a vector superfield $W_{\alpha}$ and a chiral superfield $\Phi$. To
break $N = 2$ down to $N = 1$, one can add a superpotential $W=m\;
{\rm Tr}\Phi^2$ to the action. This gives a bare mass to the $\Phi$
multiplet and, therefore, the low-energy theory is a pure Abelian
gauge theroy. This low-energy theory has ${\bf Z}_{4}$ chiral
symmetry and is believed to have a mass gap with confinement 
and spontaneous breaking of ${\bf Z}_4$ to ${\bf Z}_2$. Seiberg and
Witten gave a macroscopic description of this phenomenon based on the
$N=2$ picture as will be described below.

In $N=2$ theory, the massless spectrum in the semiclassical limit
contains only the Abelian vector multiplet ${\cal A}$.  Let us analyze
the effect of turning on a mass $m$ for the scalar multiplet
$\Phi$. In the low-energy theory, ${\rm Tr}\,\Phi^2$ is represented by
a chiral superfield $U$. Its scalar component is $u =\langle{\rm
Tr}\phi^2\rangle$, which is a holomorphic function on the moduli
space. For small $m$, we can simply add $W_{eff} = mU$ to the 
low-energy Lagrangian. This presumably removes the vacuum degeneracy
and gives a mass to the scalar multiplet. To make the Abelian gauge
field also massive (so that the theory has a mass gap), we need either
(i) extra light gauge fields giving rise to a strongly coupled
non-Abelian theory, or (ii) light charged fields giving rise to a
Higgs mechanism.  Thus, in either case, somewhere on ${\cal M}$ extra
massless states must appear. As indicated before, one cannot get extra
light gauge fields. Therefore, we consider option (ii) with the light
charged fields being monopoles or dyons. Near the point with massless
monopoles, we go to a dual description of the theory and use the $N=1$
chiral superfields $M, \widetilde M$ to describe the monopole
hypermultiplet. In this dual description, the full $N=1$
superpotential then becomes
$$
\hat{W} = \sqrt{2} A_{D} M \widetilde M + m \; U(A_{D})\,.
$$
The low-energy vacuum structure is easy to analyze. Vacua correspond
to solutions of
$$
d \hat{W} = 0\,,
$$
satisfying $|M| = |\widetilde M|$, so that the $D$-term vanishes. For $m
=0$, $M=\widetilde M =0$ and $a_{D}$ is arbitrary. Thus we recover the
$N=2$ moduli space. If $m \not= 0$, then 
\bea
&{\displaystyle
\sqrt{2} M \widetilde M + m \; \frac{du}{dA_{D}} = 0} &\,,\nonumber \\
& a_{D}M = a_{D}\widetilde M = 0 &\,.\nonumber 
\eea
Assuming that $du/da_{D} \not= 0$, we get $M, \widetilde M\not= 0$, so
that $a_D=0$ and $M=\widetilde M=(-m u'(0)/\sqrt{2})^{1/2}$. Since $M$
is charged, its vacuum expectation value generates a mass for the
gauge field through the Higgs mechanism. Thus, by a simple analysis,
we reproduce the expected mass gap of the microscopic theory. To
understand charge confinement, note that, since the hypermultiplet
$M\widetilde M$ describes monopoles, we have a magnetic Higgs
mechanism. Thus, $M\ne 0$ means that massless magnetic monopoles
condense in vacuum. This gives rise to the confinement of electric
charges by the dual ({\it i.e.,} magnetic) Meissner effect.  

\subsection{The Solution of the Model}

In this subsection, we will identify the moduli space of the $N=2$
theories with the moduli space of genus 1 Riemann surfaces and then
use this identification to calculate $a_D(u)$ and $a(u)$. 

Let us first summarize what we have learnt about the structure of the
moduli space: The moduli space ${\cal M}$ is the $u$-plane with
singularities at $1, -1, \infty$ and a ${\bf Z}_{2}$ symmetry acting
as $u \rightarrow -u$ (Fig. 3). Over this punctured plane there is a
flat $SL(2,{\bf Z})$ bundle $V$, which has $(a_D, a)^T$ as a
section. This bundle has monodromies  
\beq
M_{1} = \left( \begin{array}{cc}
1 & 0 \\ -2 & 1 \end{array} \right )\,, \quad
M_{-1} = \left( \begin{array}{cc}
-1 & 2 \\ -2 & 3 \end{array} \right )\,, \quad
M_{\infty} = \left( \begin{array}{cc}
-1 & 2 \\ 0 & -1 \end{array} \right )\,,
\label{mondall}
\eeq
around the singularities. 
\vglue1cm
\epsfxsize=6cm
\centerline{\epsffile{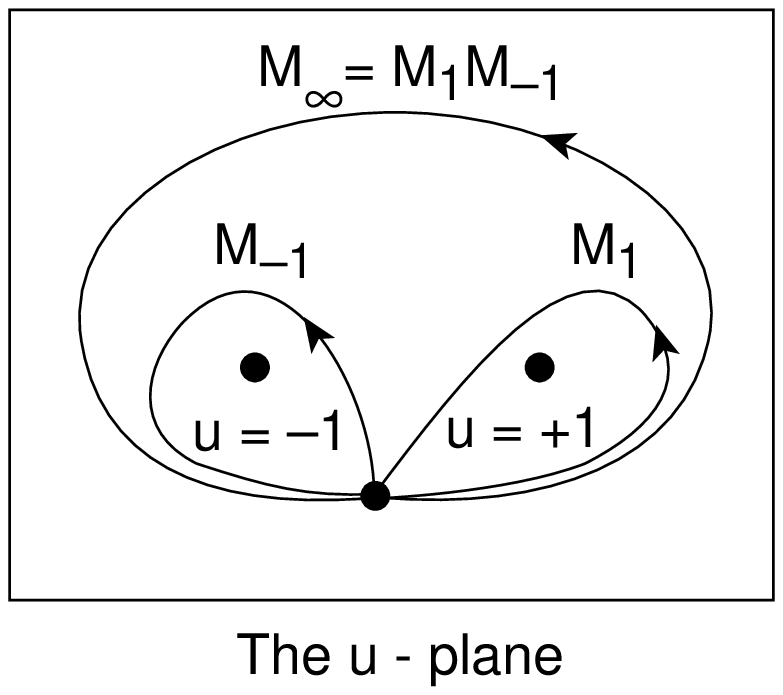}}
\begin{center}Figure 3 \end{center}

To be more precise, the quantities $(a_D(u), a(u))$ form a holomorphic
section of the bundle $V\otimes {\bf C}$, and have the following
asymptotic behaviour
\bea
u\approx \infty\,:&\qquad& \left\{\, \begin{array}{l}
a \cong \sqrt{2u}\,  \\
a_D \approx i\;\frac{\sqrt{2u}}{\pi}\;{\rm ln}\; u
\end{array} \right. \,, \nonumber\\
u=1\,:&\qquad& \left\{\, \begin{array}{l}
a_D \approx \,c_0 (u-1) \\
a \approx a_0 + \frac{i}{\pi} \; a_{D}{\rm ln}\; a_{D}
\end{array} \right. \,.\nonumber
\eea
where, $a_0$ and $c_0$ are constants. For $u=-1$ we get a behaviour
similar to $u=1$ but with $a_D$ replaced by $a-a_D$. The metric on the
moduli space is $ds^2={\rm Im}(\tau)\,|da|^2$ with 
\beq
\tau =\frac{da_D/du}{da / du}\,.
\label{tau-metric}
\eeq
To insure positivity of kinetic energy, ${\rm Im}(\tau)$ should be
positive definite. The monodromies generate a subgroup $\Gamma(2)$ of
$SL(2,{\rm Z})$ and, in fact, the $u$-plane with its singularities is
the quotient of the upper half plane $H$ by $\Gamma(2)$. This quotient
has three cusps corresponding to the three singularites.

The space $H/\Gamma(2)$, which is the $u$-plane, also parametrizes the
family of curves $E_u$ described by the equation 
\beq
y^2 = (x-1)(x+1)(x-u)\,, 
\label{cubic}
\eeq
where $x$ is a complex variable. First, note that this equation is
invariant under the transformations  
$$
w:\{u\rightarrow -u\,,x\rightarrow -x\,,y\rightarrow \pm iy \}\,,
$$
that generate a ${\bf Z}_{4}$ symmetry, out of which only a ${\bf
Z}_{2}$ subgroup acts on $u$. This is the same as the symmetry
structure on ${\cal M}$. Now, let us describe the curve which the
above equation represents. The curve basically is the $x$-space the
topology of which is determined by the requirement that $y$ is a
single valued function. Sine the equation is quadratic in $y$, if we
move along a close loop on the $x$-space around anyone of the three
zeros of $y$, then we get $y\rightarrow -y$. The same is also true for
a loop which contains all the zeros, or equivalently, a loop around
the point $x=\infty$ (this is because there is an odd number of
zeros).  Therefore, if $y$ is to be a single valued function, then the
$x$-space should be a double cover of the complex plane ${\bf C}$ with
the point at infinity added to it. Furthermore, this space should have
four branch points at $x=-1,1,u,\infty$ which are joined pairwise by
two cuts (Fig. 4). To fix attention, consider one branch cut from $-1$
to $1$, and another from $u$ to $\infty$. The two  sheets are joined
along these cuts so that on crossing a cut, we move from one sheet to
the other. It is on this space that $y$ is single valued.

\epsfxsize=8cm
\centerline{\epsffile{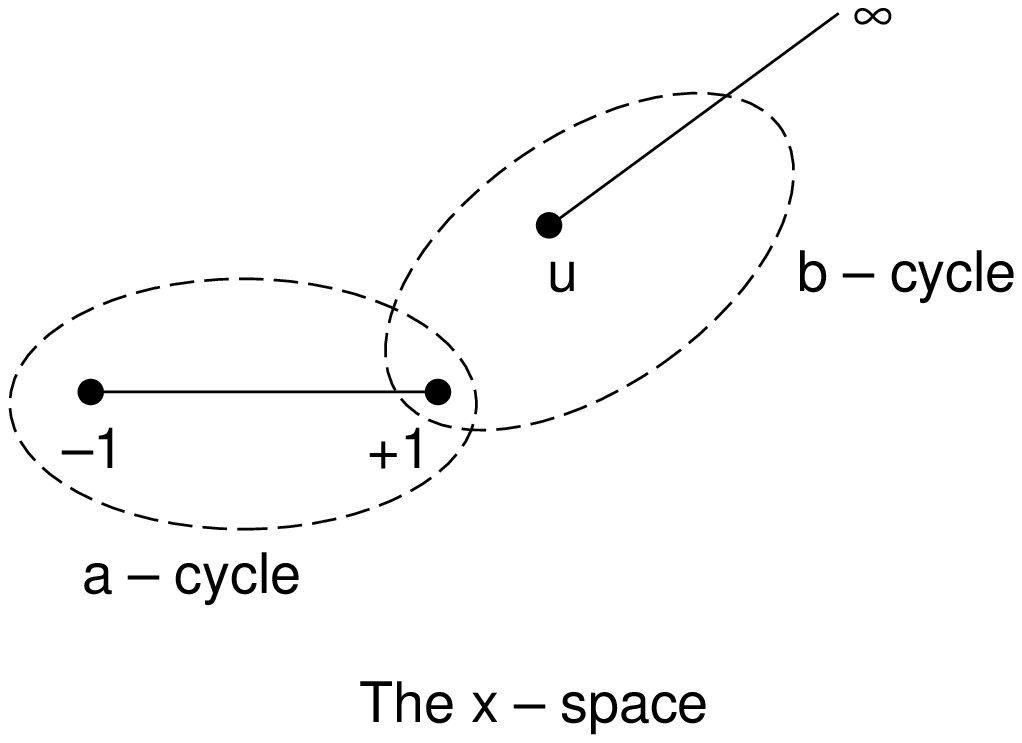}}
\begin{center}Figure 4\end{center}

The $x$-space so obtained is nothing but a genus one Riemann
surface. This can be easily visualized as follows: On a torus, draw a
circle $c_1$ along the $a$-cycle and translate this along the torus to
get a circle $c_2$ (Fig. 5(a)). Now, squash the circles $c_1$ and
$c_2$ into line segments $l_1$ and $l_2$ (Fig. 5(b)). This divides the
torus into two halves joined along these segments. If we now cut open
both the two halves, the surface we get is the same as the $x$-space
described above with $l_1$ and $l_2$ as the two branch cuts and with
the point at infinity mapped to a point at finite $x$ (Fig. 5(c)).

\epsfxsize=12cm
\centerline{\epsffile{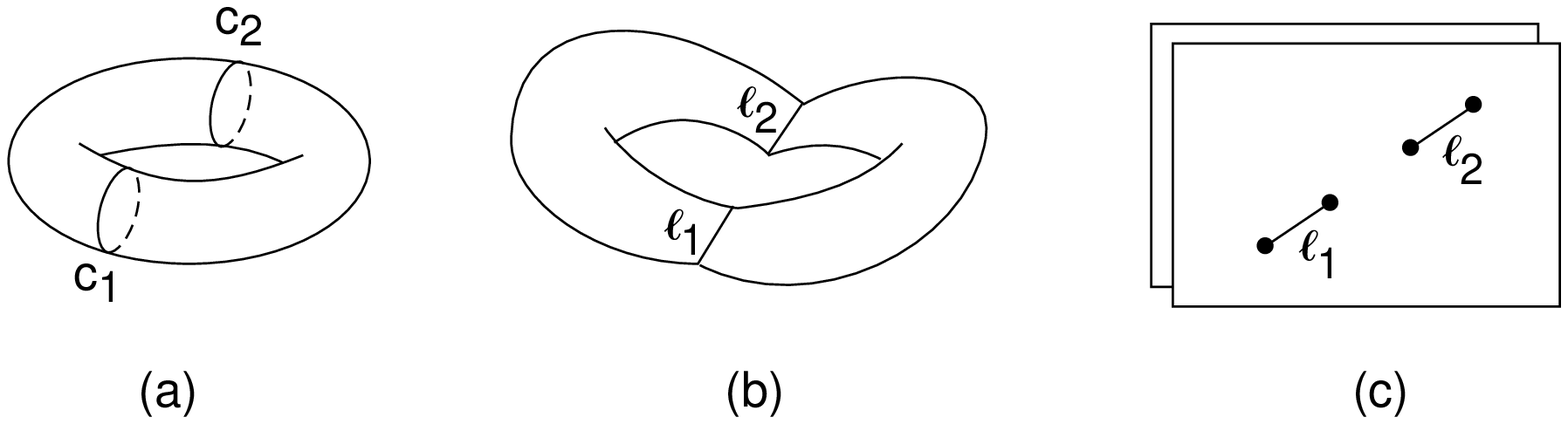}}
\begin{center}Figure 5 \end{center}

From this, it is clear that a loop on the $x$-plane that goes around
one of the two cuts corresponds to the $a$-cycle of the torus. On the
other hand, a loop which intersects both cuts (in our case, goes
around $1$ and $u$), corresponds to the $b$-cycle on the torus.
Note that when two of the branch points on the $x$-space coincide, a
cycle on the curve $E_u$ shrinks to zero size and the curve becomes
singular. For example, in our case, if $u=\infty$, the $a$-cycle 
shrinks to zero size and when $u=1$, this happens to the $b$-cycle.
Thus the singularities on the $u$-plane are at the points where a
curve in the family $E_u$ develops a vanishing cycle.

To identify $a_D$ and $a$, on the genus one Riemann surface $E_u$, we
pick up two independent one cycles $\gamma_1$ and $\gamma_2$,
normalized such that their intersection number is one: 
$$
\gamma_1 \cdot \gamma_2 = 1\,.
$$
These cycles, which continuously vary with $u$, form a local basis for
the first homology group $V_u=H^1(E_u,{\rm C})$ of $E_u$. A cycle can
be paired with elements $\lambda$ from the first cohomology group 
$$
\gamma \rightarrow  \oint_{\gamma} \lambda\,.
$$
$\lambda$ can be thought of as a meromorphic $(1,0)$-form on $E_u$
with vanishing residue, modulo exact forms. The vanishing of the
residue insures that the pairing is invariant under continuous
deformations of $\gamma$ even across poles of $\lambda$. By virtue of
this pairing, we can also regrad $\lambda$ as an element of $V_u$. For
the one-forms on $E_u$, we can choose a basis   
$$
\lambda_{1} = \frac{dx}{y}\,, \qquad \lambda_{2} = \frac{xdx}{y}\,.
$$
Upto scalar multiplication, $\lambda_1$ is the unique holomorphic
differential on $E_u$ and if we define
$$
b_i=\oint_{\gamma_i}\lambda_1\,,\qquad{\rm for}\quad i = 1,2\,,
$$
then the torus is characterized by a parameter
$$
\tau_u = b_1/b_2\,,\qquad {\rm with} \quad {\rm Im}(\tau_u) > 0\,. 
$$
Let us consider an arbitrary section
$$
\lambda = a_1(u) \lambda_1 + a_2 (u) \lambda_2\,,
$$
of $V_u$ and, for the moment, make the identification
$$
a_D=\oint_{\gamma_1}\lambda\,,\qquad a=\oint_{\gamma_2}\lambda\,.
$$
If $\lambda$ is a form with vanishing residue then, on circling a
singularity, $a_D$ and $a$ transform in the right way, simply
according to how $\gamma_1$ and $\gamma_2$ transform under a subgroup
of $SL(2,Z)$ (this is further explained in subsection 4.6 below). On
the other hand, if $\lambda$ has a pole with a non-vanishing residue,
then it is possible that the integration path may move across this
pole and, as a result, $a_D$ and $a$ may no longer transform under a
pure $SL(2,Z)$. This second possibility is of course not consistent
with the symmetries of the BPS mass formula in the absence of fermions
with non-zero bare masses, as we already discussed in a previous
subsection. Therefore, $\lambda$ should not have a pole with a
non-vanishing residue (In the presence of fermions with non-zero bare
masses, the situation is different and will be discussed at the end of
the next section). The above identification of $a_D$ and $a$ implies
that  
$$
\frac{da_D}{du}=\oint_{\gamma_1}\frac{d\lambda}{du}\,,\qquad
\frac{da}{du}=\oint_{\gamma_2}\frac{d\lambda}{du}\,.
$$
To fix the arbitrariness in $\lambda$, we use the condition 
${\rm Im}\tau > 0$ for the metric on ${\cal M}$ as defined
in (\ref{tau-metric}). First, suppose that 
$$
\frac{d\lambda}{du} = f(u) \lambda_{1} = f(u) \frac{dx}{y}\,.
$$
Then,
$$
\frac{da_{D}}{du} = f(u)\,b_1\,,\qquad \frac{da}{du} = f(u)\,b_2\,,
$$
so that
$$
\tau = \frac{b_1}{b_2} = \tau_u\,. 
$$
Since ${\rm Im}\,\tau_u > 0$, we get ${\rm Im}\,\tau >0$. As argued by
Seiberg and Witten, the converse is also true, so $d\lambda/du$ does
not depend on $\lambda_2$. The function $f(u)$ is fixed by the
asymptotic behaviour of the theory near the singularities on the $u$
plane and is given by $f(u)=-\sqrt{2}/4\pi$. With this, we can obtain
$\lambda$ as
$$
\lambda=\frac{\sqrt{2}}{2\pi}\;\frac{dx\sqrt{x-u}}{\sqrt{x^2-1}}=  
\frac{\sqrt{2}}{2\pi}\;\frac{dx y}{x^2-1}=\frac{\sqrt{2}}{2\pi}\;
\frac{dx}{y} (x-u) \,.
$$
To calculate $a$ and $a_D$, we have to choose a specific basis of 
one-cycles on $E_u$. We identify $\gamma_2$ with the $a$-cycle on the
torus, or equivalently, with a curve which loops around the points
$-1$ and $1$ on the $x$ plane. We can deform this curve so that it lies
entirely along the cut from $-1$ to $1$ and back. Thus, $a(u)$ is
given by 
\beq
a(u)=\frac{\sqrt{2}}{\pi}\,\int_{-1}^1\,\frac{dx\sqrt{x-u}}{\sqrt{x^2-1}}\,.
\label{au}
\eeq
For $\gamma_1$, we choose the curve which loops around the points $1$
and $u$ and get
\beq
a_D(u)=\frac{\sqrt{2}}{\pi}\,\int_{1}^u\,\frac{dx\sqrt{x-u}}{\sqrt{x^2-1}}\,.
\label{adu}
\eeq
It can be checked that with this choice of the one-cycles, $a$ and
$a_D$ have the desired  behaviour near the singularities \cite{SW-I}. 

\noindent\underline{Explicit Formulae for $a(u)$ and $a_D(u)$}:
One can easily find explicit formulae for the $a$ and $a_D$ in terms
complete elliptic integrals $E(k)$ and $K(k)$ by using the integral
representation of hypergeometric functions $F(\alpha,\beta,\gamma;z)$ 
\cite{GR}. The hypergeometric functions are given by
\bea
F(\alpha,\beta,\gamma;z)&=& \frac{\Gamma(\gamma)}{\Gamma(\beta)
\Gamma(\gamma-\beta)}\int_0^1\,dx\,x^{\beta-1} (1-x)^{\gamma-\beta-1}
(1-zx)^{-\alpha}\nonumber\\ 
&=& \frac{\Gamma(\gamma)}{\Gamma(\alpha)\Gamma(\beta)}\sum_{n\geq 0}
\frac{\Gamma(\alpha +n)\Gamma(\beta +n)}{\Gamma(\gamma +n)}\,
\frac{z^n}{n!}\,.\nonumber
\eea
Comparing this with the integral representation of $a$ in (\ref{au}),
we can easily see that 
$$
a(u)=\sqrt{2(1+u)}\, F\left(-1/2, 1/2, 1; 2/(1+u)\right)\,.
$$
As for $a_D$, let us first make the substitution $x=(u-1)t+1$ in 
(\ref{adu}). This gives
$$
a_D(u)=\frac{i}{\pi} (u-1)\int_0^1\,dt\,t^{-\frac{1}{2}}
(1-t)^{\frac{1}{2}} (1-\frac{1-u}{2} t)^{-\frac{1}{2}}\,. 
$$
Comparing this with the expression for the hypergeometirc functions,
we get
$$
a_D(u)=\frac{i}{2} (u-1) F\left(1/2, 1/2, 2; (1-u)/2\right)\,.
$$

In terms of the hypergeometric functions, the complete elliptic
integrals are given by 
\bea
K(k) &=& \frac{\pi}{2}\, F\left(1/2, 1/2, 1; k^2\right)\,,\nonumber\\
E(k) &=& \frac{\pi}{2}\, F\left(-1/2, 1/2, 1; k^2\right)\,.\nonumber
\eea
Further, we define $k'^2=1-k^2$ and set 
$$
E'(k)\equiv E(k')\,,\qquad K'(k) \equiv K(k')\,.
$$ 
Now, using the identity
$$
c(1-z)F(a,b,c;z) - c F(a,b-1,c;z) + (c-a)z F(a,b,c+1;z)=0\,,
$$ 
with $c=1, a=b=1/2$ and $z=(1-u)/2$, in the expression for $a_D$, we
get 
\bea
a(u)   &=& \frac{4}{\pi k} E(k)\,,\nonumber\\
a_D(u) &=& \frac{4}{i\pi} E(i\frac{k'}{k}) + 
\frac{4i}{\pi k^2} K(i\frac{k'}{k})\,.\nonumber
\eea
Here, $k^2=2/(1+u)$ and $(1-u)/2=-k'^2/k^2$. These expressions can be
further simplified if we note that  
$$
F(a, b, c; z)=(1-z)^{-a} F(a, c-b, c; z/(z-1))=
(1-z)^{-b} F(b, c-a, c; z/(z-1))\,.
$$
This implies that $K(ik'/k)=kK(k')=kK'(k)$ and $E(ik'/k)=k^{-1}E(k')
=k^{-1}E'(k)$, so that
\bea
a(u)  &=& \frac{4}{\pi k}\, E(k)\,,\nonumber\\
a_D(u)&=& \frac{4}{\pi i}\,\frac{E'-K'}{k}\,.\nonumber
\eea
The generalized coupling $\tau$ can be calculated by using
\bea
\frac{dE}{dk}&=&\frac{E-K}{k}\,,\quad 
\frac{dK}{dk}=\frac{1}{kk'^2}\,(E-k'^2 K)\,,\nonumber\\
\frac{dE'}{dk}&=&-\frac{k}{k'^2}\,(E'-K')\,,\quad 
\frac{dK'}{dk}=-\frac{1}{kk'^2}\,(E'-k^2 K')\,,\nonumber
\eea
and is given by
$$
\tau=\frac{\partial a_D}{\partial a}=\frac{da_D/dk}{da/dk}
=\frac{iK'}{K}\,.
$$

\section{The Seiberg-Witten Analysis of $N=2$ Gauge Theory With Matter}

In this section, we consider the $N=2$ supersymmetric gauge theory,
with gauge group $SU(2)$, coupled to $N_f$ matter multiplets. The
analysis involves many technical issues and the properties of the
theory depend on the value of $N_f$. As in usual QCD, matter fields
(fermions) and gauge bosons contribute to the $\beta$-function with
opposite signs. Thus, to insure asymptotic freedom, we require $N_f
\leq 4$. The theory with $N_f =4$ is finite to all orders in
perturbation theory and, when the quarks are massless, it is very
likely conformally invariant non-perturbatively. Seiberg and Witten
provided evidence that this theory is an example of an $SL(2,Z)$
invariant theory. The only other $SL(2,Z)$ invariant theory known in
four dimensions is the $N=4$ supersymmetric gauge theory which also
has a vanishing $\beta$-function. However, in the following, we will
mainly concentrate on $N_f\leq 3$. For the $N_f=4$ case, the reader is
referred to original work of Seiberg and Witten \cite{SW2}.

In the presence of matter, it is convenient to choose a charge
normalization different from the previous sections. In $N=2$ pure
gauge theory all fields transform in the adjoint representation and
the charges of particles and monopoles are integers. Hence, in the
formula $Z=a n_e + a_D n_m$, $n_e$ and $n_m$ are integers. With this
normalization, when quarks are present, $n_e$ could be half-integral.
In this section we choose a slightly different convention: To ensure
that $n_e$ is always integral, we multiply it by 2 and divide $a$ by 2
to keep the mass formula unchanged. Since $a_D$ is kept unchanged, it
is now given by $2 a_D=\partial{\cal F}/\partial a$. In terms of the
gauge invariant quantity $u={\rm Tr}\phi^2$, as $|u|\rightarrow\infty$,
the asymptotic behaviour of $a$ and $a_D$ with this normalization
becomes 
$$
a \cong \frac{1}{2} \sqrt{2u}\,, \quad
a_D \simeq i \frac{4}{\pi} a \log a\,.
$$
Furthermore, the effective coupling is also rescaled to 
$\tau=\partial a_D/\partial a = \frac{\theta}{\pi} + \frac{8 \pi
i}{g^2}$. This rescaling of the charge affects the form of the curve
which determines the solution of the model, keeping the physics
unchanged.  

In this section we are going to study the theory with gauge group
$SU(2)$. In this case, the most interesting dynamics appear in the
Coulomb branch of the moduli space, where $SU(2)$ breaks down to 
$U(1)$. It is, therefore, instructive to first study $N = 2$
supersymmetric QED.

\subsection{$N = 2$ Supersymmetric QED}

The $N=2$ supersymmetric QED contains the photon vector multiplet and
$k$ quark hypermultiplets. The photon multiplet consists of the $N=1$
chiral superfields $(W_{\alpha}, A)$ and each quark hypermultiplet
consists of two $N=1$ chiral superfields $M^i$ and $\widetilde M_i$,
with $U(1)$ charges $+1$ and $-1$, respectively. The only
renormalizable $N = 2$ compatible superpotential is: 
$$
{\cal W}=\sqrt{2} A M^i\widetilde M_i + \sum_i m_i M^i\widetilde
M_i\,. 
$$
When $m_i = 0$, the theory has a global symmetry group $SU(k)\times
SU(2)_R \times U(1)_{\cal R}$. $SU(k)$ is a flavour symmetry group
acting on the $k$ hypermultiplets, while $SU(2)_R$ rotates the two
supersymmetries and $U(1)_{\cal R}$ is the usual $R$-symmetry which is
afflicted by an anomaly. As in the case without matter, the strategy
is to first determine the classical moduli space and then to
see whether the quantum moduli space is qualitatively different.
The symmetries of the theory are very useful in determining the
structure of these moduli spaces.

In general, the classical moduli space has a Higgs branch and a
Coulomb branch. The Higgs branch is defined by some of the $M^i$'s
acquiring vacuum expectation values. In this case, both in QED and in
QCD with gauge group $SU(2)$, the gauge group is completely broken.
For $k=1$, there is no Higgs phase.  When $k \geq 2$, and $m_i = 0$,
then, using the global symmetries together with the the vanishing
condition for the $D$-term, it is possible to rotate $M$ and
$\widetilde M$ to the form $M=(B,0,0,...)$ and $\widetilde M = (0,
B,0,...)$. From this one can read off the different patterns of
symmetry breaking as a function of the number of flavours $k$. A
theory on the Higgs branch does not contain monopoles or dyons and
hence, the dynamical possibilities are not as rich as on the Coulomb
branch. The classical moduli space is a hyperK\"ahler manifold and the
symmetries of the theory lead to a unique hyperK\"ahler metric on
it. As a result of this uniqueness, the metric does not receive
quantum mechanical corrections. Therefore, on the Higgs branch, the
classical and the quantum moduli spaces are the same.

In the case at hand, if all $m_i = 0$, then the Coulomb branch is
defined by $<A> \not= 0$ which, in turn, implies that
$<M^i>=<\widetilde M_i>=0$. The $U(1)$ gauge group remains unbroken
while all the $M$'s become massive. At a generic point on the moduli
space, the effective low-energy threory (which involves only the
massless modes) is a pure $N=2$ gauge theory and the K\"ahler
potential is of the special geometry type : $K={\rm Im} a_D (a)
\bar{a}$. This is related by $N=2$ supersymmetry to the the gauge
kinetic term,  
$$
\int d^2 \theta \; \frac{\partial a_D}{\partial a} \;
W^{\alpha}W_{\alpha}\,.
$$
The K\"ahler potential (and thus the metric) receive quantum
corrections, but the one-loop approximation to $K$ is exact and also
there are no non-perturbative corrections since this theory does not
contain instantons. Using the one-loop $\beta$-function for QED with
$k$ hypermultiplets, we obtain 
\beq
a_D = - \frac{ik}{2\pi} \; a \ln (a/\Lambda)\,.
\label{beta-hyper}
\eeq
To reproduce this formula one should be careful about the extra
factors of $2$ in the supersymmetric way of defining $g$ and the
normalization convention mentioned earlier. The metric ${\rm Im}
(\tau)$ obtained from this is zero at $|a| = \Lambda / e$ and thus the
effective coupling constant is singular. Because of this Landau pole
singularity, the theory does not make sense in the ultra-violet
region unless embedded in a larger theory which is asymptotically
free. If fermion mass terms are added, the singularities on the  
Coulomb branch can move.  Since, when $a = - \frac{1}{\sqrt{2}} m_i$
one electron becomes massless, we have
$$
a_D = - \frac{i}{2\pi} \sum_i (a+m_i / \sqrt{2}) \;\ln
\frac{(a+m_i / \sqrt{2})}{\Lambda}\,.
$$
To each bare mass is now associated a singularity on the moduli space
where the particle becomes massless.  Depending on the possible
equality of two or more masses, one can have Higgs and Coulomb
branches touching, leading to an intricate structure.

Here, we also see a manifestation of the modified form of the central
charge of the $N = 2$ algebra discussed in section 2. For pure $N=2$,
we have $Z = n_e a + n_m a_D$. However, now the electron masses are
not just $\sqrt{2} |a|$, rather the $i$-th multiplet has mass 
$|\sqrt{2} a + m_i |$. As shown in section 2, this is consistent with
the fact that the $U(1)$ charges $S_i$ of the massive hypermultiplets
appears in the formula for the central charge $Z$ as
$$
Z = n_e a + n_m a_D + \sum_i S_i m_i / \sqrt{2}\,.
$$

\subsection{$N=2$ Supersymmetric QCD with Matter}
The $N=2$ supersymmetric QCD coupled to $N_f$ matter hypermultiplets 
contains the $N=1$ superpotential
$$
W=\sum_{i=1}^{N_f}(\sqrt{2}\widetilde{Q}_i\Phi
Q^i+m_i\tilde{Q}_iQ^i)\,. 
$$
In general, when $m_i=0$, this theory has a global symmetry group
$SU(N_f)\times SU(2)_R\times U(1)_{\cal R}$. In the special case, when
the gauge group is $SU(2)$, the flavour group is enlarged to
$O(2N_f)$. This is due to the fact that for $SU(2)$ the fundamental
representation and its conjugate are isomorphic and, therefore, $Q^i$
and $\widetilde Q_i$ can be combined into a $2N_f$-dimensional vector
transforming under $O(2N_f)$. Thus, for the special case of the
$SU(2)$ gauge group, the theory also has a parity symmetry group ${\bf
Z}_2\subset O(2N_f)$ acting as
\beq
\rho : Q_1 \leftrightarrow \widetilde{Q}_1\,,
\label{rho}
\eeq
with all other fields remaining unchanged. This parity plays an
important role in the analysis of the theory.  The global symmetry
group of the theory is actually a quotient of $O(2N_f) \times SU(2)_R
\times U(1)_{\cal R}$. The quotient is to be taken because a $Z_2
\subset U(1)_{\cal R}$ is the same as $(-1)^F$ contained in the
Lorentz group. This, combined with the center of $SU(2)_R$, is the
same as the ${\bf Z}_2$ in the center of $O(2N_f)$.

For $N_f = 0,1$, the theory has only a Coulomb branch with $(<\phi>
\not= 0)$. On this branch $SU(2)$ is broken to $U(1)$ and all quarks
are massive. Moreover, $U(1)_{\cal R}$ is spontaneously broken because
$\Phi$ has $U(1)_{\cal R}$ charge $2$. For $N_f \geq 2$ we can either
have a Coulomb branch or Higgs  branches. On a Higgs branch, the
gauge symmetry is fully broken and the dynamics is not very rich. We
will not analyse this branch in detail, but will only mention that, as
in the QED case, the metric on the moduli space is uniquely
determined by the symmetries and does not receive quantum correction. 

\noindent\underline{Some Properties of the Quantum Theory}: 
The perturbative $\beta$-function of our theory (which, due to
supersymmetry, is only a one-loop effect) is given by
\beq
\beta (g) = - \frac{4-N_f}{16\pi^2} g^3.
\label{betanf}
\eeq
Therefore, to insure asymptotic freedom, we consider only
$N_f=0,1,2,3,4$. 

In the previous section, from the counting of fermion zero-modes in an
instanton background, we found that $U(1)_{\cal R}$ is broken to a
discrete subgroup ${\bf Z}_{4N_c-2N_f}$. This was obtained by
requiring the invariance of the correlation function $G$ given by
(\ref{G}) under $U(1)_{\cal R}$. For $N_c=2$, we have an extra
discrete symmetry group (\ref{rho}) which permutes the fermion
zero-modes $\psi_{q_1}$ and $\widetilde\psi_{q_1}$. This group is
anomalous as it changes $G$ to $-G$. Therefore, now we can also allow
$U(1)_{\cal R}$ transformations which do not keep $G$ invariant, but
change it by a sign. This sign can be compensated for by an anomalous
${\bf Z}_2$ transformation. Thus, for $N_c=2$, $U(1)_{\cal R}$ is
broken to the discrete subgroup ${\bf Z}_{2(4N_c-2N_f)}={\bf
Z}_{4(4-N_f)}$. This can be combined with the anomalous ${\bf Z}_2$ to
get a discrete ${\bf Z}_{4(4-N_f)}$ anomaly-free subgroup with the
action (see (\ref{Rb1}) and (\ref{Rb2})):
$$
\begin{array}{l}
\left.\begin{array}{l}
W_{\alpha} \rightarrow \omega W_{\alpha} (\omega^{-1} \theta )\\
\Phi \rightarrow \omega^2 \Phi (\omega^{-1} \theta )\\
Q^1  \rightarrow \widetilde Q_1 (\omega^{-1} \theta )\\
\widetilde Q_1  \rightarrow Q^1 (\omega^{-1} \theta ) \\
\end{array}\right. \\
\left.\begin{array}{l}
Q^i  \rightarrow Q^i (\omega^{-1} \theta )\\
\widetilde Q_i  \rightarrow \widetilde Q_i (\omega^{-1} \theta )\\
\end{array}\right\} \qquad i\not= 1\,,
\end{array}
$$
where, $\omega={\exp}(2i\pi/4(4-N_f))$. For $N_f = 0$ we do not have the
quarks to cancel the anomaly and only the square of the above
transformations is anomaly free. This case was discussed in the
previous section. Furthermore, it can be seen that a subgroup ${\bf
Z}_2 \subset {\bf Z}_{4(4-N_f)}$ is the same as $(-1)^F$ in the
Lorentz group. Combining the above transformations with the $U(1)_J$
subgroup of $SU(2)_R$ (see (\ref{Rb1}) and (\ref{Rb2})), we find a
${\bf Z}_{4(4-N_f)}$ symmetry which commutes $N=1$ supersymmetry
\beq
\begin{array}{l}
\left.\begin{array}{l}
\Phi \rightarrow \omega^2 \Phi(\theta) \\
Q^1 \rightarrow \omega^{-1} \widetilde{Q}_1(\theta) \\
\widetilde{Q}_1 \rightarrow \omega^{-1} Q^1(\theta) \\
\end{array} \right.\\
\left.\begin{array}{l}
Q^i \rightarrow \omega^{-1} Q^i(\theta) \\
\widetilde{Q}_1 \rightarrow \omega^{-1} \widetilde{Q}_i(\theta) 
\end{array} \right\}
\qquad i \not= 1\,.
\end{array}
\label{symm-A}
\eeq
Under this transformation, the gauge invariant order parameter $u={\rm
Tr}\phi^2$ has charge $4$ and transforms as $u \rightarrow \exp 2 \pi
i /(4-N_f) u$. This further breaks ${\bf Z}_{4(4-N_f)}$ down to ${\bf
Z}_4$. The remaining ${\bf Z}_{4-N_f}$ acts non-trivially on the
$u$-plane. Note that for $N_f=0$, the subgroup which does not keep $u$
invariant is only a ${\bf Z}_2$.

As in the $N_f = 0$ case, the large $u$-behaviour of $a_D(u)$ is
determined by the one-loop $\beta$-function (\ref{betanf}):
\bea
a &\cong& \frac{1}{2} \sqrt{2u} + \cdots \,,\nonumber\\
a_D &\simeq& i \;\frac{4-N_f}{2\pi}\,a(u)\ln\frac{u}{\Lambda^2_{N_f}}
+ \cdots \,.\label{betanf-mon}
\eea
Here, the dots represent non-perturbative instanton corrections. The
generic form of these corrections can be obtained by arguments similar
to the ones used for the $N_f=0$ case in the previous section. First,
a correction coming from a $k$-instanton configuration is proportional
to the $k$-instanton factor, which, using the $\beta$-function
(\ref{betanf}), can be written as 
\beq
e^{-8\pi^2k/g^2} =\left(\frac{\Lambda_{N_f}}{a}\right)^{k(4-N_f)}\,. 
\label{inst-nf}
\eeq
Following Seiberg \cite{Seiberg-n,Seiberg-poh}, we can restore the
broken part of the $U(1)_{\cal R} \times {\bf Z}_2 (\rho )$ symmetry
by assigning appropriate charges to $u$ and $\Lambda_{N_f}$.  Thus we
assign charge $4$ and even $\rho$-parity to $u$ and charge $2 (4-N_f)$
and odd $\rho$-parity to $\Lambda_{N_f}^{4-N_f}$. Note that with this
assignment, the one-instanton factor $\exp(-8\pi^2/g^2)$ will have an
odd $\rho$-parity which compensates for the odd parity under ${\bf
Z}_2(\rho)$ of the correlation function $G$ in (\ref{G}), thus keeping
it invariant . For the special case of $N_f=4$, $U(1)_{\cal R}$ is
non-anomalous and $u$ has charge $4$. $Z_2(\rho)$ in $O(8)$ is still
anomalous, but again, it can be treated as unbroken by assigning odd
parity to the instanton factor $\exp(-8\pi^2/g^2)$. Invariance under
$U(1)_{\cal R}$ with the above charge assignments implies that each
correction term should contain a factor of $\sqrt u$. Moreover, since
the metric on the $u$-plane is invariant under the $\rho$-parity,
configurations with odd instanton numbers cannot contribute to $a$ and
$a_D$. Putting these facts together, we can write the generic form of
the corrected $a$ and $a_D$ as
\bea
a &=& \frac{1}{2} \sqrt{2u} \left ( 1 + \sum_{n=1}^\infty a_n (N_f)
\left(\frac{\Lambda^2_{N_f}}{u}\right)^{n(4-N_f)} \right )\,,
\nonumber\\
a_D &=& i \frac{4-N_f}{2 \pi} a(u) \ln \frac{u}{\Lambda^2_{N_f}} +
\sqrt{u} \sum_{n=0}^\infty  a_{Dn} (N_f) \left(\frac{\Lambda^2_{N_f}}
{u}\right)^{n(4-N_f)} \,.\nonumber
\eea
The difficult part is to compute the coefficients $a_n$ and $a_{Dn}$.

Since the gauge group $SU(2)$ breaks to $U(1)$, the theory will
have massive charged states and we want to know how the unbroken
global symmetry acts on them. The unbroken part of the global
symmetry, which is the part that keeps $u$ invariant, is obtained by
raising the transformations (\ref{symm-A}) to the power $4-N_f$.
This unbroken transformation changes the sign of $\phi$ and therefore
acts as charge conjugation on the charged fields. For $N_f=1,3$;
$4-N_f$ is odd and the unbroken transformation contains odd powers of
$\rho$. It, therefore, acts as the parity in $O(2N_f)$. Hence, in this
case, the parity (\ref{rho}) is realized on the spectrum but it
reverses the signs of electric and magnetic charges. States of given
charge belong to $SO(2)$ (for $N_f=1$) or $SO(6)$ (for $N_f=3$)
multiplets. For $N_f=2,4$; $4-N_f$ is even and the unbroken
transformation contains only even powers of $\rho$. All the odd
powers of $\rho$, which amount to the parity transformation, are part
of the broken transformations. Thus, in this case, the parity symmetry
is broken and the states are only in $SO(4)$ (for $N_f=2$) or $SO(8)$
(for $N_f=4$) representations. 

\subsection{BPS Saturated States}

On the Higgs branch $SU(2)$, is completely broken and there are no
electric or magnetic charges. Thus the central charge of the $N=2$
algebra only contains contributions from the $U(1)$ charges of the
hypermultiplets. We will not discuss this in any detail. 

On the Coulomb branch, the simplest BPS saturated states are the
quarks with zero bare mass and which acquire masses $M = \sqrt{2} |a|$
after the spontaneous breaking of the gauge symmetry. These form a set
of BPS states which transform as a vector of $SO(2N_f)$. Besides
these, there are BPS states which transform as a spinor of $SO(2N_f)$
arising as follows: Since the gauge symmetry is broken to $U(1)$, the
theory contains monopoles. In the presence of a monopole each $SU(2)$
doublet of fermions has one zero-mode. Since there are $N_f$
hypermultiplets, there are $2N_f$ fermion doublets and, therefore,
$2N_f$ fermion zero-modes. After quantization, these zero-modes give
rise to a $2N_f$-dimensional Dirac algebra which provides a spinor
representation of $SO(2N_f)$. Thus the fermion zero-modes turn the
monopole into a spinor of $SO(2N_f)$. This is very similar to the
quantization of the Ramond sector of the Superstring theory. The
presence of spinors indicate that, at the quantum level, the symmetry
group is a universal cover of $SO(2N_f)$, or $Spin(2N_f)$.

One of the monopole's collective coordinates is a charge rotation.
Upon quantization this leads to a spectrum of electrically charged
states for the monopole. A $2\pi$ rotation by the electric charge
operator, however, is not the identity.  Using the Witten effect, such
a rotation gives a topologically non-trivial gauge transformation with
eigenvalue $e^{i \theta} (-1)^H$ for a monopole with $n_m = 1$. Here,
$(-1)^H$ is the centre of $SU(2)$ which is odd for states in the
hypermultiplet and even for states in the vector multipler. In this
section we normalize the charge operator so that elementary quarks
have charges $\pm 1$ and massive gauge bosons have charges $\pm 2$. 
With this normalization, the above mentioned gauge rotation can be
written as an operator statement 
$$
e^{i \pi Q} = e^{i \theta n_m} (-1)^H.
$$
where, $Q = n_e + n_m \theta/\pi$. Since, $n_e \in {\bf Z}$, this
relation implies that states of even $n_e$ have $(-1)^H = 1$ and
states of odd $n_e$ have $(-1)^H = -1$. The factor $(-1)^H$ can be
identified with the chirality operator for the spinor representation
of $SO(2N_f)$.  For $N_f = 1,3$, the $SO(2N_f)$ parity transformation 
acts on the spectrum and it guarantees that a dyon transforming as a
positive chirality spinor of $SO(2N_f)$ is degenerate with a particle
of opposite electric and magnetic charges and opposite $SO(2N_f)$
chirality.  No such relation exists for $N_f = 2,4$.

For $N_f > 0$ we will see that the spectrum contains states with $n_m
> 1$. A convenient way of labelling states is in terms of their
behaviour under the centre of $Spin(2N_f)$.
\begin{itemize}
\item

$N_f = 2$: \, $Spin(4) = SU(2) \times SU(2)$ with centre ${\bf Z}_2
\times {\bf Z}_2$. The representations of the center can be labeled
by $(\epsilon , \epsilon ')$, where $\epsilon = 0$ for vector-like
irreducible representation and $\epsilon = 1$ for spinor-like
irreducible representations.  An elementary quark transforms as
$(\epsilon , \epsilon ') = (1,1)$. Multiple quark states then
transform as $(n_e \; mod \; 2, n_e \; mod \; 2)$. Since monopoles
behave like spinors, we have
$$
(\epsilon , \epsilon ') = ((n_e + n_m) \; mod \; 2, n_e \; mod \; 2)
\qquad {\rm for}\,\, {N_f = 2}\,.
$$
\item
$N_f = 3$: \, $Spin(6) = SU(4)$ with centre ${\bf Z}_{4}$.  The
elementary quarks are in the $6$ of $Spin (6)$, thus they have charge
$2$ with respect to the centre.  Since monopoles behave like spinors,
we conclude that ${\bf Z}_{4}$ acts as:
$$
\exp \; \frac{i \pi}{4} (n_m + 2n_e)
\qquad{\rm for} \,\,{N_f = 3}\,.
$$
\item
$N_f=4$:\, $Spin(8)$ has centre ${\bf Z}_2\times {\bf Z}_2$. Hence,
using similar arguments as above, we get,
$(\epsilon,\epsilon')$$=$$(n_m\; mod\; 2, n_e\; mod \; 2) $. The four
representations of the center are labeled by the representations of of
$Spin(8)$ which realize them: $(0,1)\equiv v$, $(1,0)\equiv s$,
$(1,1)\equiv c$ and $(0,0)\equiv o$.  Here, $v$ refers to the vector
representation, $s$ and $c$ to two spinor representatios and $o$ to
the trivial representation. $Spin(8)$ has a triality group of outer
automorphisms which is isomorphic to the permutation group ${\bf S}_3$
acting on $v$, $s$ and $c$.
\end{itemize}

If an $N = 2$ invariant mass is turned on, say $m_{N_f} \not= 0$,
then $SO(2N_f)$ explicitly breaks to $SO(2N_f-2) \times SO(2)$, and
the global Abelian charge associated with $SO(2)$ appears in the
central charge 
$$
Z = n_e a + n_m a_D +S_{N_f} \frac{m_{N_f}}{\sqrt{2}},\qquad
M = \sqrt{2} |Z|\,.
$$
Hence for $a = \pm m_{N_f} / \sqrt{2}$ one of the elementary quarks
becomes massless. This will be crucial later in determining the global
structure of the quantum moduli space on the Coulomb branch.

\subsection{Duality}

As in the $N_f=0$ case, there is an $SL(2,Z)$ duality group which acts
on the fields and the couplings. In the presence of matter, some new
issues appear which will be discussed below.

As before, we can compute the monodromy matrices for $(a_D,a)$ around
the singular points on the moduli space.  The simplest case
corresponds to having a single quark with non-zero bare mass, and
investigating what happens as $a \approx a_0 \equiv m_{N_f} /
\sqrt{2}$.  From the QED analysis we know that there is a logarithmic
singularity at $a_0$ where this quark becomes massless. Near the
singularity we have 
$$
\begin{array}{rcl}
a& \approx & a_0\,, \\
a_D & \approx & - \frac{i}{2\pi} (a-a_0) \ln (a-a_0) +
c\,. 
\end{array}
$$
The monodromy can now be easily computed to be
$$
\begin{array}{rcl}
a &\rightarrow & a \,,\\
a_D &\rightarrow &\displaystyle{ a_D + a - \frac{m_{N_f}}{\sqrt{2}}}\,.
\end{array}
$$
Unlike the $N_f = 0$ case, now we have an inhomogeneous transformation
as the column vector $(a_D, a)^T$ picks up a shift under the monodromy
(besides the usual $SL(2,Z)$ transformation). The possibility of such
a shift was also noticed for the pure gauge theory case in the
previous section. There, however, the shift was not part of the
monodromy group and, furthermore, it was not compatible with the
symmetries of the BPS mass formula. In the presence of matter, this
shift is allowed since the BPS mass formula is modified. Moreover, now
the shift naturally appears as a part of the monodromy group. To write
a monodromy matrix, we construct a column vector $(m/ \sqrt{2} , a_D ,
a )^T$. The monodromy can now be written as 
$$
\left ( \begin{array}{c} m/ \sqrt{2} \\ a_D \\ a \end{array} \right)
\rightarrow {\cal M} 
\left ( \begin{array}{c} m/ \sqrt{2} \\ a_D \\ a \end{array}\right)\,, 
$$
with the monodromy matrix ${\cal M}$ given by 
$$
{\cal M } = \left ( \begin{array}{ccc}
1 & 0 & 0 \\ -1 & 1 & 1 \\ 0 & 0 & 1
\end{array} \right )\,,\qquad
{\cal M}^{-1} = \left(\begin{array}{ccc}
1 & 0 & 0 \\ 1 & 1 & -1 \\ 0 & 1 & 1
\end{array}\right) \,.
$$
If we arrange the charges as a row vector $W = (S, n_m, n_e)$, then
invariance of $Z$ (or of $M$) implies that, under the monodromy,
$W \rightarrow W{\cal M}^{-1}$. In general the matrix ${\cal M}$ can
be of the form
$$
{\cal M} = \left ( \begin{array}{ccc}
1 & 0 & 0 \\ r & k & l \\ q & n & p
\end{array} \right )\,,
\qquad
{\cal M}^{-1} = \left ( \begin{array}{ccc}
1 & 0 & 0 \\ lq-pr & p & -l \\
nr-kq & -n & k
\end{array} \right )\,,
$$
with $\det {\cal M} = 1$. 

Note that under the monodromy, the electric and magnetic charges can
mix with each other but will not pick up contributions proportional to
$S$ which is a global symmetry charge.  On the other hand, $S$ can
pick up contributions proportional to $n_e$ and $n_m$ which are
related to the local gauge symmetry.  For the example considered
above,
$$
(S, n_m, n_e ) \rightarrow (S + n_m, n_m, -n_m + n_e )\,.
$$
This leads once again to issues of marginal stability: For large
values of $m_{N_f}$, the singularity is in the weak-coupling region of
large $u$ where a semi-classical treatment of monopole is reliable.
In the semi-classical quantization, the global $U(1)$ charge is
carried only by fermionic zero-modes. Since there is only a finite
number of these modes, one cannot construct states with arbitrarily
large values of $S$. This means that, although, by acting with ${\cal
M}^{-1}$ we can increase the value of $S$ at will, what might have
started as a one-particle state comes back as a multiparticle state.  
This is possible if at some point, when going around the singularity,
the single-particle state becomes unstable and decays into a
multi-particle state. Thus, although the formalism is $SL(2, Z)$
covariant, the spectrum is not. 

Seiberg and Witten provide many compelling arguments to suggest that
the theory with $N_f = 4$, like the theory with $N = 4$, is
$SL(2,Z)$ invariant, though the details are quite
different. In the $N_f=4$ theory, the global symmetry group is
$Spin(8)$ which is the universal cover of $SO(8)$. The states $(n_m,
n_e)=(0,1)$ are the elementary hypermultiplets in the vector
representation $v$ of $Spin(8)$.  The states $(n_m, n_e)=(1,0)$ are in
a spinor representaion $s$, and $(n_m, n_e)=(1,1)$ are in a spinor
representation $c$ of $Spin(8)$. While $SL(2,Z)$ alone cannot keep
this spectrum unchanged, a combination of $SL(2,Z)$ and the $Spin(8)$
triality group (which permutes the representations $v$, $s$ and $c$)
could do so. This is provided one is willing to assume that 
there are monopole bound states for every pair of relatively prime
integers $(p,q)$, analoguous to the situation discussed by Sen for the
$N=4$ theory \cite{sen}.  For each such pair, there should exist eight
states transforming in a representation $8_v$, $8_s$ or $8_c$ of
$Spin(8)$, depending on the $mod \; 2$ reduction of $(p,q)$.
$SL(2,Z)$ mixed with triality will then keep the spectrum
invariant. The solution Seiberg and Witten provided for this model
gives strong support to this possibility.  The global symmetry group
is then a semi-direct product of $Spin(8)$ and $SL(2,Z)$.

\subsection{A First Look at Singularities}

As in the pure $N = 2$, we first try to locate the singularities on
the moduli space before computing their monodromies.  We recall that,
with the standard normalization of the gauge coupling constant $g$,
the one-loop $\beta$-function which is given by (\ref{betanf}), can be
integrated to give
$$
\frac{1}{\alpha_{N_f}(\mu)} = \frac{4-N_f}{2 \pi} \ln 
\frac{\mu}{\Lambda_{N_f}}\,,
$$
where, $\alpha=4\pi/g^2$.  Now, if some (say $N_f-N_f'$) of the quarks
have masses $m_i=m$, and we are looking at the theory at some scale
$\mu < m$, then the low energy theory contains only $N_f'$
hypermultiplets as the degrees of freedom. The coupling $\alpha(N_f')$
is then given by an expression similar to the above one with $N_f$ and
$\Lambda_{N_f}$ replaced by $N_f'$ and $\Lambda_{N_f'}$. The scales of
the theories can be related by the matching condition $\alpha_{N_f}(m)
=\alpha_{N_f'}(m)$, which implies
$$
\Lambda^{4-N_f'}_{N_f'} = m^{N_f - N_f'}\;\Lambda_{N_f}^{4-N_f}\,.
$$
For instance,
$$
\begin{array}{lcl}
N_f = 3 \;, N'_f = 0\, &\Rightarrow& \Lambda^{4}_{0} = m^3 \Lambda_3\,,\\
N_f = 1 \;, N'_f = 0\, &\Rightarrow& \Lambda^{4}_{0} = m \Lambda_{1}^{3}\,.
\end{array}
$$

To determine the singularity structure on the moduli space, we first
consider theories with $N_f \leq 3$, and with hypermultiplet bare
masses very large as compared to $\Lambda$. The singularities which
arise from hypermultiplets becoming massless are now in the
semi-classical (large $u$) region of the moduli space and can be easily
identified. In the small $u$ region, one is effectively left with an
$N_f=0$ theory with two singularities corresponding to massless
monopoles and dyons. Then we slowly decrease the bare masses to zero
and follow the movement of the sigularities on the moduli space.

\noindent\underline{$N_f = 3$}:
Let us start with equal masses $m_i = m >> \Lambda$. In this case, the
global symmetry $Spin(6)\approx SU(4)$ of the massless theory is
broken to $SU(3) \times U(1)$. Classically, there is a singularity at
$a = m/ \sqrt{2}$ where the three quarks become massless. These
electrically charged massless fields form a \underline{3}
representation of $SU(3)$. For $m >> \Lambda$ the singularity is in
the semiclassical region, $u \sim 2a^2 >> \Lambda^2$. For $u<< m^2$
the three massive quarks can be integrated out giving a $N_f=0$ theory
with scale $\Lambda^{4}_{0} = m^3 \Lambda_3$. Hence, for small $u$ the
moduli space is that of a pure $N = 2$ theory with scale $\Lambda_{0}$
which has two singularities with $(n_m , n_e) = (1,0)$ and
$(1,1)$. These are the points where monopoles and dyons become
massless. Clearly the massless states at these singularities are
$SU(3)$ invariant. 

As $m$ is decreased, the singularity at large $u$ moves, and although
the charges of the associated states can change (through monodromy
matrices), their non-Abelian charges cannot change. Hence, for any
$m$, the massless fields at the various singularities transform as
\underline{3}, \underline{1} and \underline{1} of $SU(3)$. In the $m =
0$ limit, the original global symmetry is restored and the massless
states must combine into representations of $SU(4)$. The only
possibility is to have two singularities combining into a
\underline{4} of $SU(4)$ and the other sigularity moves somewhere else
remaining a singlet. Thus, we conclude that the massless $N_f = 3$
theory has two singularities with massless particles in the
\underline{4} and \underline{1} of $SU(4)$. From our study of how
different states transform under the centre, the smallest charges for
the massless particles at the singularities are $(n_m , n_e) = (1,0)$
for the \underline{4}, and $(n_m, n_e) = (2,1)$ for the \underline{1}.
Semiclassically the first state exists, but it is not known whether
the monopole-monopole bound state implied by the second also exists or
not.

\noindent\underline{$N_f = 2$}:
In this case, with two equal masses $m_i=m >>\Lambda$, the $Spin(4)$
global symmetry of the massless theory is broken to $SO(2)
\times SO(2)$. There is a singularity at $a = m/ \sqrt{2}$ and the
massless states there transform under one or the other of the two
$SO(2)$'s. In the region $u<<m^2$, we can again integrate out the
quarks, leading to a $N_f=0$ theory with $\Lambda^4_0=m^2\Lambda^2_2$,
and two singularities with $(n_m, n_e) = (1,0), (1,1)$. The massless
fields at these singularities are singlets under $SO(2)\times SO(2)$.
In all, we have four massless states associated with three
singularities. As $m \rightarrow 0$, we recover the full global
symmetry group of the massless theory with two flavours which is 
$Spin (4) \approx SU(2) \times SU(2)$. The singularities have to
combine in such a way that the massless states form a representation
of this unbroken group. Since each $SO(2)$ is contained in a different
$SU(2)$, the only way to combine the massless states into
representations of $SU(2)\times SU(2)$ is as (\underline{2},
\underline{1}) and (\underline{1}, \underline{2}). Hence, for $m = 0$,
there are two singularities, and the simplest charge assignments are
$(n_m , n_e ) = (1,0)$ in one spinor of $SO(4)$, and $(n_m , n_e ) =
(1,1)$ in the other spinor. Recall that the transformation under the
centre is $(\epsilon,\epsilon')=((n_m+n_e) mod\; 2, n_e\; mod\;2)$. 

\noindent\underline{$N_f=1$}:
Now the massive theory has the same $SO(2)$ symmetry as the massless
theory, however, the same arguments as before imply the presence of
three singularities for large $m$. This number does not change as
$m\rightarrow 0$ due to the ${\bf Z}_3$ symmetry acting on the
$u$-plane. We recall that this symmetry is the subgroup ${\bf
Z}_{4-N_f}={\bf Z}_{3}$ of ${\bf Z}_{4(4-N_f)}$ in (\ref{symm-A})
which is broken by a non-zero $u$. From 
$$
a_D = i \frac{4-N_f}{2\pi} a(u) \ln \frac{u}{\Lambda^2_{N_f}}+...\,,
$$
we see that as $a$ transforms homogeneously under (\ref{symm-A}),
$a_D$ is shifted by $a$, {\it i.e.}, $a\rightarrow \omega^2 a$, $a_D
\rightarrow \omega^2 (a_D + a)$ with $\omega = e^{i\pi/6}$. Therefore,
if one of the singularities in the $m=0$ limit corresponds to massless
states with $(n_m , n_e ) = (1,0)$, then the ${\bf Z}_3$ symmetry
implies the existance of two other singularities characterized by
$(n_m , n_e )= (1,1)$ and $(1,2)$. Hence, even for $m=0$, the moduli
space of the $N_f=1$ theory has three singularities corresponding to
massless states $(n_m , n_e )= (1,0), (1,1)$ and $(1,2)$.

\subsection{Monodromies and the Determination of the Metric}

As in the case of $N=2$ theory without matter, a solution to the
theory can be found by regarding the $u$-plane as the moduli space of
a family $E_u$ of elliptic curves parametrized by $u$. The quantities
$a_D(u)$ and $a(u)$ can then be related to the periods of these
curves. A curve in $E_u$ becomes singular when one of the cycles on it
shrinks to zero size. This corresponds to a singularity on the
$u$-plane with a non-trivial monodromy around it. If we know the
singularities and the associated monodromies on the $u$-plane (which
are associated with the appearence of massless particles in the
spectrum), then we can work backwords and determine the families of the
elliptic curves from which the periods $a_D$ and $a$ could be
computed. In this subsection, we sketch the physical arguments
used by Seiberg and Witten in \cite{SW2} to find the curves for
$N_f=1,2,3$ theories exploiting the general features of the curve for
the $N_f=0$ theory (This reference also contains a more rigorous
treatment of this problem which we will not reproduce here).

For $N_f = 0$, the solution was given by a family of elliptic curves
described by the equation
$$
y^2 = (x - \Lambda^2 )(x + \Lambda ^2 )(x-u) \,,
$$
with the monodromy in $\Gamma(2)$. In the present section we have
changed our conventions so that $(n_m , n_e )$ are both integers even
in the presence of matter fields. This was achieved by multiplying
$n_e$ by $2$ and dividing $a$ by $2$ so that $Z = a_D n_m + a n_e$ is
unchanged. This change of convention can be implemented as a
transformation  
$$
\left ( \begin{array}{cc} a_D \\ a \end{array} \right ) \rightarrow
\left ( \begin{array}{cc} a_D \\ a' \end{array} \right ) =
\left ( \begin{array}{cc} 1 & 0 \\ 0 & 1/2 \end{array} \right )
\left ( \begin{array}{cc} a_D \\ a \end{array} \right ),
$$
which also changes the monodromy matrix as
$$
\left ( \begin{array}{cc} m & n \\ p & q \end{array} \right )
\rightarrow
\left ( \begin{array}{cc} 1 & 0 \\ 0 & 1/2 \end{array} \right )
\left ( \begin{array}{cc} m & n \\ p & q \end{array} \right )
\left ( \begin{array}{cc} 1 & 0 \\ 0 & 2 \end{array} \right ) =
\left ( \begin{array}{cc} m & 2n \\ p/2 & q \end{array} \right )\,.
$$
Since $p$ and $n$ are integers modulo 2, the new monodromy matrix
contains the entry $2n$ which is an integer modulo 4, while all other
entries are integers. These martices form the subgroup $\Gamma_{0}(4)$
of $SL(2,Z)$. In the new convention, the monodromies in (\ref{mondall})
take the form
\beq
M_{\infty} = \left( \begin{array}{cc}
-1 & 4 \\ 0 & -1 \end{array} \right )\,, \quad
M_{1} = \left( \begin{array}{cc}
1 & 0 \\ -1 & 1 \end{array} \right )\,, \quad
M_{-1} = \left( \begin{array}{cc}
-1 & 4 \\ -1 & 3 \end{array} \right )\,,
\eeq
and the solution is given by a family of curves described by  
\beq
y^2 = x^3 - ux^2 + \frac{1}{4} \Lambda^4 x \,.
\label{f2}
\eeq
As discussed in the previous section, any genus-one curve can be
represented by a cubic
$$
y^2 = F(x) = (x-e_1)(x-e_2) (x-e_3)\,.
$$
This describes a space $x$ as a double cover of the complex plane
branched over $e_1$, $e_2$, $e_3$ and $\infty$. The curve becomes
singular when two branch points coincide ({\it e.g.}, $e_1=e_2\not=
e_3$ or $e_3 \rightarrow \infty$). In this case the singularity is
called stable.  If more than two branch points coincide, then the
singularity is not stable, but a $u$-dependent reparametrization of
$x$ and $y$ can always convert this into a stable singularity. For the
curve (\ref{f2}) the branch points are at 
$$
x=0,\, \frac{1}{2} (u \pm \sqrt{u^2 - \Lambda^4}),\, \infty\,.
$$
For $u = \pm \Lambda^2$ we have two stable singularities, but the $u
\rightarrow \infty$ singularity is not stable.  

To understand the properties of this curve better, let us first
consider a generic situation: For a stable singularity at, for
instance, $u = 0$, the family of curves near $u=0$ can be written in
the form 
$$
y^2 = (x-1)(x^2 - u^n)\,, 
$$
for some integer $n$. The monodromy around $u=0$ is then conjugate to
$T^n$ where the matrix $T$ is defined in (\ref{ST}). This can be
understood as follows: Consider the holomorphic Abelian differential
$\omega = dx/y$ on a curve $y^2=(x-1)(x^2-\lambda)$, where $\lambda =
u^n$. The periods can be written as  
$$
\omega_1 = \int_{\Gamma_1}\;\frac{dx}{y}\,, \qquad
\omega_2 = \int_{\Gamma_2}\;\frac{dx}{y}\,,
$$
where $\Gamma_1$ is a path from $u=-\lambda^{1/2}$ to
$u=\lambda^{1/2}$, and $\Gamma_2$ is a path from $u=\lambda^{1/2}$ to  
$u=1$ (Fig. 6(a)). As $\lambda \rightarrow e^{2 \pi i} \lambda$,
$\lambda^{1/2} \rightarrow -\lambda^{1/2}$ and the cut $\Gamma_1$ 
moves as in Fig. 6(b). This simply exchanges the branches of the
integrand and therefore $\omega_1 \rightarrow \omega_1$.

\noindent As for $\omega_2$, the path $\Gamma_2$ is transformed as in
Fig. 7.  

\noindent Thus, $\omega_2 \rightarrow \omega_2 + \omega_1$. Since
$\lambda = u^n$, when $u \rightarrow e^{2\pi i}u$, $\lambda$ makes $n$
turns and we obtain the $n$-th power of the monodromy. Thus, in terms
of $\lambda$, the monodromy is conjugate to $T$ while in terms of $u$,
it is conjugate to $T^n$ as we wanted to show.

\epsfxsize=12cm
\centerline{\epsffile{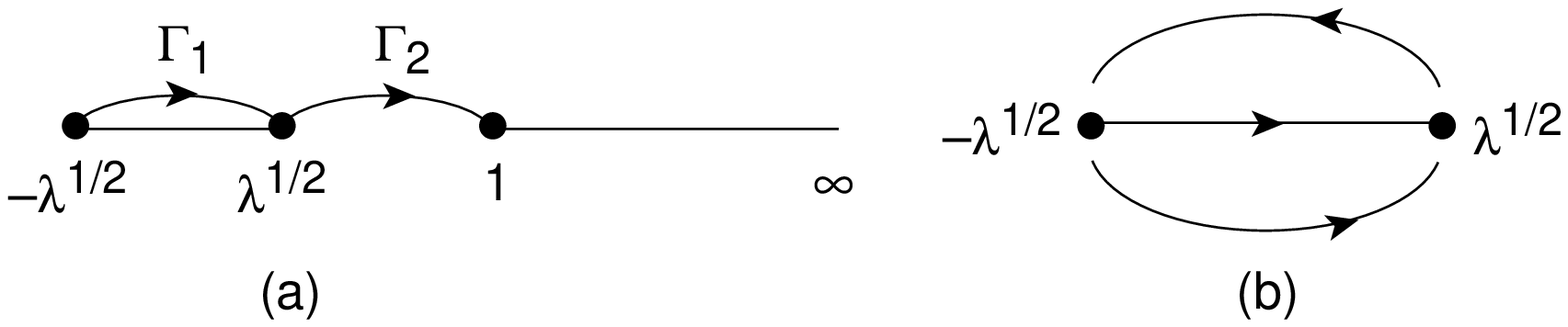}}

\begin{center}Figure 6\end{center}

\epsfxsize=7cm
\centerline{\epsffile{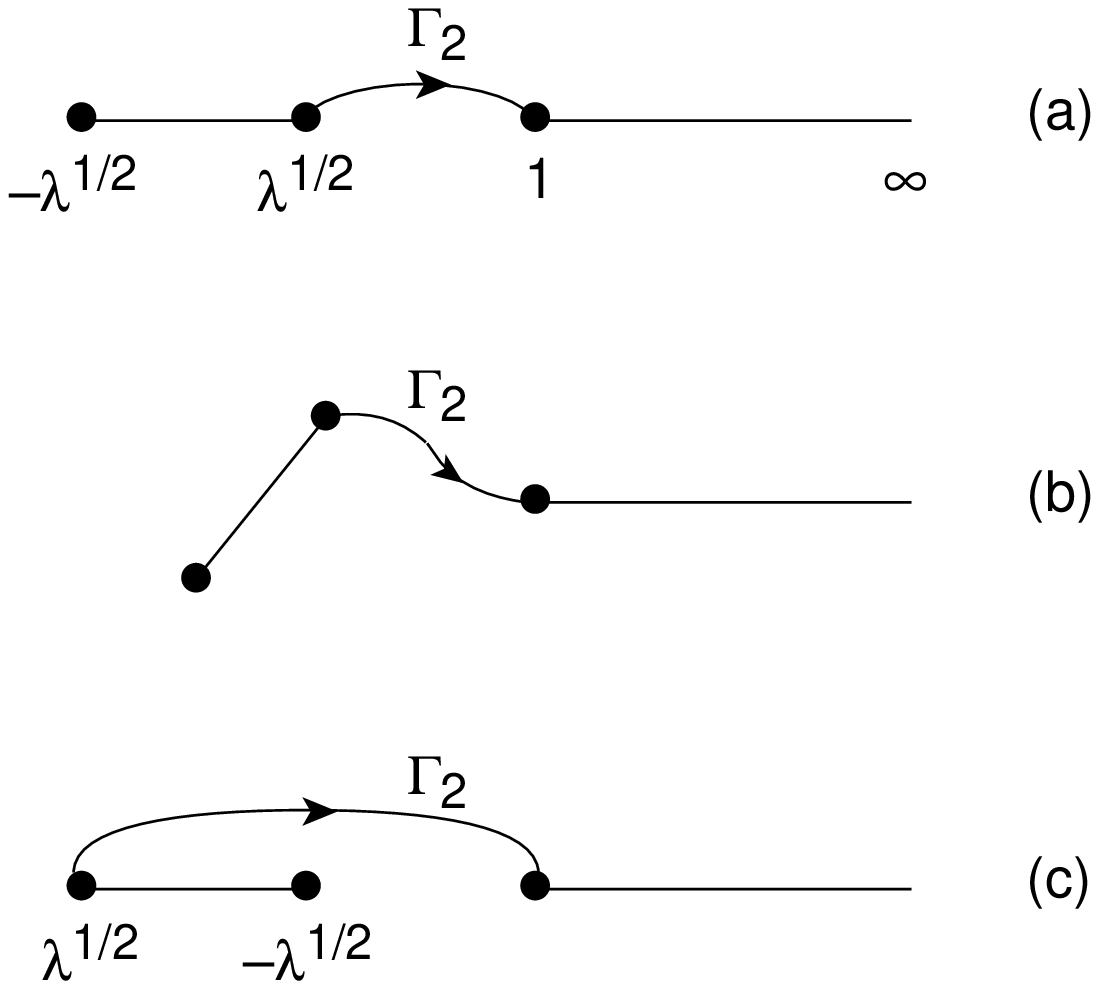}}
\begin{center}Figure 7\end{center}

Given any polynomial, a useful quantity is its discriminant defined as
$$
\Delta = \prod_{i<j} (e_i - e_j)^2,
$$
where the $e_i$'s are the roots.  $\Delta$ can be expressed in terms of
the coefficients of the polynomial. Clearly, at a singularity when two
branch points coincide, $\Delta = 0$ (except for a singularity at
$\infty$). For the example $y^2 = (x-1)(x^2-u^n)$, $\Delta \sim u^n$
near $u = 0$ where the monodromy is conjugate to $T^n$. Hence, in the
stable case, the exponent of the monodromy is the order of zero of
$\Delta$. For instance, for the family of curves described by
(\ref{f2}), $\Delta$ has first-order zeros at $u=\pm\Lambda^2$, and
the monodromies around these points are conjugate to $T$. 
Let us now look at $u\rightarrow \infty$. For large $u$ the branch
points are approximately at $x=0,\,\Lambda^4/4u,\,u,\,\infty$. Thus
the singularity at $u=\infty$ is not stable as more than two branch
points coincide in this limit. This is cured by a change of variables 
$$
x = x'u\,, \qquad y = y' u^{3/2}\,,
$$
which shifts the branch points to $0,\,\Lambda^4/4u^2,\,1,\,\infty$.
The discriminant now behaves like $\Delta \sim u^{-4}$ corresponding
to a monodromy conjugate to $T^{-4}$. Due to the presence of $\sqrt u$
in $y \rightarrow y'$, the monodromy in terms of the original
varialbles $(x,y)$ has an extra minus sign and is conjugate to
$PT^{-4}$. Thus we see that the curve (\ref{f2}) produces the correct
monodromies of the $N_f = 0$ theory.

To obtain the curves for theories with non-zero $N_f$, we should know
the monodromies which arise in these theories. This can be easily
worked out (as in the $N_f=0$ case) since we already know the charge
spectrum of the particles which become massless at the singularities
on the $u$-plane. First note that, as in the $N_f=0$ theory, the
monodromy at $u=\infty$ can be obtained form the perturbative
$\beta$-function (\ref{betanf}), or equivalently (\ref{betanf-mon}),
and is given by 
$$
{\cal M}_{\infty} = PT^{N_f -4}.
$$
The singularities at finite points in the $u$-plane, in general,
correspond to massless magnetic monopoles with one unit of magnetic
charge and $n_e$ units of magnetic charge. To calculate the
corresponding monodromy, we have to go to a dual description of the
theory in which the monopole couples to the dual gauge field the way
an electron couples to the usual gauge field. 
In this frame, the monodromy is determined by the one-loop QED 
$\beta$-function (\ref{beta-hyper}) with $k$ hypermultiplets and is
given by $T^k$. This has to be conjugated with $T^{n_e}S$ which
converts a hypermultiplet of charge $(0,1)$ into a monopole of charge
$(1,n_e)$. Hence the monodromy around a point with $k$ magnetic
monopoles with $(n_m = 1, n_e)$ is $(T^{n_e}S) T^k (T^{n_e}S)^{-1}$.  
Similar arguments can be applied to calculate the monodromy for the
$(2,1)$ state in the $N_f = 3$ theory. In the following we list
all the monodromies around the singularities described in the previous
subsection: 
$$     
\begin{array}{rl}
N_f=0:& STS^{-1},\,(T^2 S)T(T^2 S)^{-1}\rightarrow M_{\infty}=PT^{-4}\,,\\
N_f=1:& STS^{-1},\,(TS)T(TS)^{-1},\, (T^2S)T(T^2S)^{-1} \rightarrow
M_{\infty}=PT^{-3}\,, \\
N_f=2:& ST^2 S^{-1},\,(TS)T^2(TS)^{-1}\rightarrow M_{\infty}=PT^{-2}\,,\\
N_f=3:& (ST^2S)T(ST^2S)^{-1},\,ST^4 S^{-1}\rightarrow
M_{\infty}=PT^{-1}\,. \\
\end{array}
$$
Using $S^2=-1$ and $(ST)^3=1$, it is easy to check that, for each
$N_f$, the product of the monodromy matrices at finite $u$ yields
$M_{\infty}$.  Note that in the first three cases ($N_f =
0,1,2$) the electric charges assigned to the singularities differ in
sign from those determined by our previous arguments.  This is
consistent with the inherent ambiguities in relating the universal
charges with the transformation properties under the centre of the
group. Based on the knowledge obtained so far, in the following we
sketch the arguments of Seiberg and Witten to obtain the explicit form
of the curves for theories with non-zero $N_f$.

\noindent\underline{Properties of the $N_f =0$ Curve:}
Before proceeding further, it is very useful to enumerate some
properties of the $N_f=0$ curve which are expected to remain valid
even for non-zero $N_f$: 
\begin{enumerate}
\item
The equation describing the family of curves is of the generic form
$y^2=F(x,u,\Lambda)$, where $F$ is a polynomial at most cubic in $x$
and $u$.
\item
The part of $F$ cubic in $x$ and $u$ is $F_0 = x^2 (x-u)$.
\item
If we assign $U(1)_{\cal R}$ charge $4$ to $u$ and $x$ and charge $2$
to $\Lambda$, then $F$ has charge $12$. If $y$ is assigned charge $6$,
then the curve is invariant under $U(1)_{\cal R}$ transformations. 
\item
$F$ can be written as $F = F_{0} + \Lambda^4 F_1$ where $F_1 = x/4$.
\end{enumerate}

The property (1) remains an ansatz for $N_f > 0$. It will lead to
the correct monodromy matrices, and it can also be justified in part
when considering the $N_f = 4$ theory.  Property (2) is a consequence
of the fact that as $u \rightarrow \infty$, we must obtain the
monodromy at infinity associated with a logarithm in $a_D$ coming
from the one-loop $\beta$-function. This means that one of the branch
points at finite $x$ should move to infinity as $u\rightarrow\infty$. 
For large $u$, the cubic part can be written as $(x-e_1 u)(x-e_2
u)(x-e_3 u)$. Then the desired behaviour can be obtained if two
$e_i$'s coincide and the other is different, say $e_1=e_2\not=e_3$.
In this case, by a redefinition of $x$, we can bring $F_0$ to the
form $x^2 (x-u)$. Property (4) in the $N_f = 0$ theory is a
consequence of the $U(1)_R$ charge assignments.  Note that $F$ has
only a classical contribution plus a one-instanton term : $\Lambda^4
x/4$.  Now we use these properties to determine the curves for 
$N_f = 1,2,3$.

\noindent\underline{The Curves for Massless $N_f=1$ Theory:}
In this theory, from equation (\ref{inst-nf}), the instanton amplitude
is proportional to $\Lambda^3_{1}$. However, for $N_f \geq 1$, the
instanton factor is odd under the $\rho$ parity and, therefore, only
even powers of it can appear in $F$. Since $\Lambda_{1}$ has $U(1)_R$
charge $2$ and $F$ has charge $12$, the only possibility is 
$$
y^2 = x^2 (x-u) + t \Lambda^{6}_{1} \,.
$$
The constant $t$ can be absorbed in a redefinition of $\Lambda_{1}$:
$t\Lambda^{6}_{1}=\tilde{\Lambda}^6$. The discriminant of this family
of curves is
$$
\Delta = \tilde{\Lambda}^{6}_{1} (4 u^3 - 27 \tilde{\Lambda}^{6}_{1})\,.
$$
This has three zeros which are interchanged under the ${\bf Z}_3$
transformation acting on the $u$-plane and the associated monodromies
are conjugate to $T$. Similarly, the monodromy at large $u$ can be
worked out to be $PT^{-3}$. This is consistent with what we should
have for the $N_f=1$ theory.

\noindent\underline{The Curves for Massless $N_f=2$ Theory:}
In this case the instanton factor is $\Lambda^{2}_{2}$ and, again, in
the absence of bare masses, only even powers of it can appear in $F$. 
Since $\Lambda_2$ has $U(1)_R$ charge 2, the general form of the curve
is
$$
y^2 = x^2 (x-u) + (ax + bu) \Lambda^{4}_{2}\,.
$$
From our discussion of the $N_f=2$ theory, we expect two singularities,
each with two massless monopoles.  Hence the monodromy at each
singularity is conjugate to $T^2$ which means that the discriminant at
each finite singularity should have a double zero.  This condition
determines $a$ and $b$.  After a rescaling of $\Lambda_2$, the
family of $N_f=2$ curves can be written as
$$
y^2 = (x^2 - \tilde{\Lambda}^{4}_{2} ) (x-u)\,,
$$
which has the expected ${\bf Z}_2$ symmetry.

\noindent\underline{The curves for Massless $N_f=3$ Theory:} In this
case, there are two singularities on the $u$-plane with monodromies
conjugate to $T^4$ and $T$ respectively, and there is no symmetry
acting on the $u$-plane. Take the $T^4$ singularity to be at $u =
0$. The discriminant then should have a fourth-order zero at
$u=0$. This, together with the usual $U(1)_{\cal R}$ charge and $\rho$
parity assignments, leads to
$$
F = a \Lambda^{2}_{3} x^2 + bu^2 x + cu x^2 + x^3 \,.
$$
Here, $b \not= 0$ since otherwise the curve is singular for all $u$.
Requiring that the cubic part of $F$ have the expected classical
behaviour, and after some rescaling, one gets
$$
y^2 = x^2 (x-u) + \tilde {\Lambda}^{2}_{3} (x-u)^2\,.
$$

Note that in the above we have only considered theories without a bare
mass term. Since a mass term has odd $\rho$-parity, in a theory with
non-zero bare masses, odd powers of the instanton factor can also
contribute to the equation for the curve. We will not discuss these
cases here, but for the sake of completeness, will simply reproduce
the final results:
\bea
N_f=1:\qquad y^2 &=& x^2(x-u)+ \frac{1}{4}m\Lambda_1^3 x -
\frac{1}{64} \Lambda_1^6\,, \nonumber \\
N_f=2:\qquad y^2 &=& (x^2-\frac{1}{64}\Lambda_2^4) (x-u) + m_1 m_2
\Lambda_2^2 x - \frac{1}{64}(m_1^2+m_2^2)\Lambda_2^4\,,\nonumber \\
N_f=3:\qquad y^2 &=& x^2(x-u)-\frac{1}{64} \Lambda_3^2(x-u)^2
-\frac{1}{64}(m_1^2+m_2^2+m_3^2) \Lambda_3^2(x-u) \nonumber \\
      && +\frac{1}{4}m_1m_2m_3 \Lambda_3 x - \frac{1}{64}
(m_1^2 m_2^2 + m_2^2 m_3^2 + m_1^2 m_3^2)\Lambda_3^2\,. \nonumber
\eea
For more details and discussions (including the $N_f=4$ case), as well
as a more rigorous method of obtaining the equations for the curves,
the reader is referred to the original work of Seiberg and Witten
\cite{SW2}.   

Once the curves are determined, the rest of the procedure is very
similar to the $N_f=0$ case. We define the quantities $a$ and $a_D$ by
the contour integrals
$$
a=\int_{\gamma_1} \, \lambda\,,\quad a_D=\int_{\gamma_2}\lambda\,,
$$
and the metric is given by ${\rm Im}(\tau)$ with $\tau=\frac{da_D}{du}/
\frac{da}{du}$. Here $\lambda$ is a holomorphic one-form such that
$$ 
\frac{d\lambda}{du}=\frac{\sqrt 2}{8\pi}\frac{dx}{y}\,.
$$
The difference with the $N_f=0$ case is that now $\lambda$ can have
poles with non-zero residue, however, the residues should be $u$
independent. If this is the case, then $\tau$ still undergoes an
$SL(2,Z)$ transformation while $(a_D, a)$ undergoes an $SL(2,Z)$
transformation plus a shift. As we saw in the case of QED with quarks
of non-zero bare masses, this shift is actually needed and is
consistent with the modified form of the BPS mass formula
(\ref{bps-m}). From this it follows that the residues of $\lambda$,
which are responsible for the shifts, should be proportional to the
bare quark masses. This was checked in \cite{SW2} for the $N_f=2$
curve and found to be the case. In fact, the existence of residues is
very restrictive and this information was used by Seiberg and Witten
to drive the curve for the $N_f=4$ theory in a rigorous way. The
curves for other theories can then be determined by renormalization
group flow and the results agree with what is listed above.

\noindent{\Large\bf Acknowledgments}

We have benefited from numerous conversations with our colleagues on
the subjects covered in these lectures. We would like to thank in
particular E. Alvarez, S. Randjbar-Daemi, W. Lerche, A. Klemm, S.
Yankielowicz, C. Gomez, K. Kounnas, M. Mari\~no, J. Distler, E.
Verlinde, A. Giveon, E. Rabinovici and G. L. Cardoso for discussions.
We would also like to thank the Trieste Theory Group for the
opportunity to present this material in a stimulating environment.

\end{document}